\documentclass[11pt]{article}
\usepackage{comment,soul}
\usepackage[most]{tcolorbox}
\usepackage[hidelinks]{hyperref}
\usepackage{enumitem}
\usepackage{amsmath,amssymb,epsf,cite,graphicx}
\usepackage{subcaption}   
\usepackage{array}        
\usepackage{amsmath,braket,tikzsymbols,upgreek,booktabs}
\usepackage[bbgreekl]{mathbbol}
\usepackage{comment}
\usepackage[export]{adjustbox}
\usepackage{dsfont}
\usepackage{faktor}
\usepackage{mathrsfs}
\usepackage[normalem]{ulem}
\usepackage{float}

\usepackage{empheq}
\usepackage{simpler-wick}
\usepackage{mathtools}
\usepackage{tikz}
\usepackage{tikzsymbols}

\definecolor{tqftblue}{RGB}{47,93,138}
\definecolor{tqftred}{RGB}{153,67,58}
\definecolor{tqftgray}{RGB}{105,112,119}
\definecolor{tqftlight}{RGB}{238,243,247}
\tikzset{
  boundary/.style={draw=tqftblue,very thick,fill=white},
  cobordism/.style={draw=tqftgray,very thick,fill=tqftlight,rounded corners=13pt},
  ribbon/.style={draw=tqftred,very thick,-{Latex[length=2.2mm]}},
  flow/.style={draw=tqftblue,thick,-{Latex[length=2mm]}},
  labelbox/.style={fill=white,inner sep=2pt,font=\small},
  paneltitle/.style={font=\small\bfseries,text=tqftgray}
}

\usepackage[english]{babel}
\usetikzlibrary{decorations.markings,arrows.meta,patterns,calc,decorations.pathreplacing}

\definecolor{cbBlue}{HTML}{0072B2}
\definecolor{cbVerm}{HTML}{D55E00}
\definecolor{cbGreen}{HTML}{009E73}
\definecolor{cbPurple}{HTML}{CC79A7}
\definecolor{cbGrey}{HTML}{5A5A5A}

\tikzset{
  sec3/.style={line cap=round,line join=round,font=\small},
  bdyphys/.style   ={draw=cbBlue,line width=1.5pt},
  bdytop/.style    ={draw=cbGrey,line width=1.5pt,double,double distance=1.2pt},
  bdyasym/.style   ={draw=cbBlue,line width=1.5pt},
  bdygeo/.style    ={draw=cbGrey,line width=1.1pt,dash pattern=on 3pt off 2pt},
  wall/.style      ={draw=cbGrey,line width=0.7pt},
  ident/.style     ={draw=cbGrey,line width=0.6pt,dash pattern=on 2pt off 2pt},
  bdyline/.style   ={draw=cbBlue,line width=1.4pt},
  anyon/.style     ={draw=cbVerm,line width=1.4pt},
  anyonhid/.style  ={draw=cbVerm,line width=1.0pt,dash pattern=on 2.5pt off 2pt},
  loopx/.style     ={draw=cbGrey,line width=1.1pt,dash pattern=on 3pt off 2pt},
  ribx/.style      ={draw=cbPurple,line width=1.4pt},
  fuse/.style      ={draw=cbVerm,line width=0.9pt,-{Latex[length=1.8mm]}},
  orient/.style    ={draw=cbGrey,line width=1.0pt,-{Latex[length=2.0mm]}},
  jn/.style        ={circle,fill=cbVerm,inner sep=0pt,minimum size=3.2pt},
  jnx/.style       ={circle,fill=cbPurple,inner sep=0pt,minimum size=3.2pt},
  tag/.style       ={font=\footnotesize,inner sep=1.5pt},
  tagbox/.style    ={font=\footnotesize,fill=white,inner sep=1.5pt},
  tile/.style      ={draw=cbGrey,line width=0.35pt},
  tilezero/.style  ={draw=cbBlue,line width=1.1pt},
  tlab/.style      ={font=\fontsize{6}{7}\selectfont,text=cbGrey,inner sep=0.6pt},
  axisline/.style  ={draw=black,line width=0.7pt},
  gridline/.style  ={draw=white,line width=0.5pt,step=1},
  frame/.style     ={draw=black,line width=0.7pt},
  crtbox/.style    ={draw=cbVerm,line width=1.4pt},
  fluxsep/.style   ={draw=black,line width=0.9pt},
  fluxbrace/.style ={draw=cbGrey,line width=0.7pt,
                     decorate,decoration={brace,amplitude=3pt,raise=0pt}},
  flab/.style      ={font=\fontsize{6.5}{7.5}\selectfont,text=cbGrey,anchor=south},
  mval/.style      ={font=\fontsize{6}{7}\selectfont},
  xlab/.style      ={font=\fontsize{6}{7}\selectfont,rotate=45,anchor=north east},
  ylab/.style      ={font=\fontsize{6}{7}\selectfont,anchor=east,xshift=-2pt},
}
\providecommand{\Zc}{Z(\cC)}

\usepackage{amsthm}
 \usepackage{amsfonts}
 \usepackage{cases}
 \usepackage{cancel}
 \usepackage{here}
\definecolor{linkblue}{RGB}{52,86,128}
\hypersetup{colorlinks=true, linkcolor=linkblue, citecolor=linkblue, urlcolor=linkblue, filecolor=linkblue}
\colorlet{darkred}{red!70!black}

\numberwithin{equation}{section}

\definecolor{airforceblue}{rgb}{0.36, 0.54, 0.66}
\newcommand{\beq}{\begin{equation}}
\newcommand{\eeq}{\end{equation}}

\newcommand{\Htop}{\Hil_{\rm top}}
\newcommand{\Zgrav}{Z_{\rm grav}}

\newcommand{\Zwh}{Z_{\rm wh}}

\newcommand{\vr}{\varrho}

\begin{document}
\baselineskip=15.5pt
\pagestyle{plain}
\setcounter{page}{1}
\newcommand{\one}{\mathbb{1}}
\newcommand{\Htot}{\mathcal{H}_{\rm tot}}
\newcommand{\Hil}{\mathcal{H}}
\newcommand{\cat}{\mathcal{C}}
\newcommand{\Tube}{\mathrm{Tube}}
\newcommand{\Irr}{\mathrm{Irr}}
\newcommand{\End}{\mathrm{End}}
\newcommand{\Hom}{\mathrm{Hom}}
\newcommand{\VecG}{\mathrm{Vec}}
\newcommand{\Fib}{\mathrm{Fib}}
\newcommand{\sgn}{\mathrm{sgn}}
\newcommand{\Mult}{\mathcal{M}}
\newcommand{\Cc}{\mathsf{C}}
\newcommand{\C}{\mathcal C}
\newcommand{\Z}{\mathbb Z}
\newcommand{\Mat}{\operatorname{Mat}}
\newcommand{\Rep}{\operatorname{Rep}}
\newcommand{\Tr}{\operatorname{Tr}}
\newcommand{\id}{\mathbf 1}
\newcommand{\cH}{\mathcal H}
\newcommand{\cA}{\mathcal A}
\newcommand{\cC}{\mathcal C}
\newcommand{\cN}{\mathcal N}
\newcommand{\qdim}{\operatorname{qdim}}
\newcommand{\rmean}{\langle r\rangle}
\newcommand{\DS}{D(S_3)}
\newcommand{\cL}{\mathcal{L}}
\newcommand{\CRT}{\mathsf{CRT}}
\newcommand{\Spec}{\operatorname{Spec}}
\providecommand{\Zc}{Z(\cC)}
\begin{center}
{\LARGE \bf Categorical RMT}
\vskip 1cm

\textbf{Chuanxin Cui$^{1,a}$, Kristan Jensen$^{2,b}$, Sridip Pal$^{3,4,c}$, and Moshe Rozali$^{1,d}$}

\vspace{0.5cm}

{\it ${}^1$ Department of Physics and Astronomy, University of British Columbia, Vancouver, BC V6T 1Z1, Canada \\}

{\it ${}^2$ Department of Physics and Astronomy, University of Victoria, Victoria, BC V8W 3P6, Canada\\}

{\it ${}^3$Institut des Hautes Etudes Scientifiques, 91440 Bures-sur-Yvette, France}

{\it ${}^4$Department of Theoretical Physics, Tata Institute of Fundamental Research, Homi Bhabha Rd, Mumbai 400005, India}

\vspace{0.3cm}

{\tt ${}^a$ccxhans@student.ubc.ca, ${}^b$kristanj@uvic.ca, ${}^c$sridip@theory.tifr.res.in, ${}^d$rozali@phas.ubc.ca\\}

\medskip

\end{center}

\vskip1cm

\begin{center}
{\bf Abstract}
\end{center}

Random matrix theory (RMT) describes the level statistics of chaotic systems one symmetry sector at a time. We propose its analogue for local quantum systems on a circle with ordinary or non-invertible symmetries. Symmetry defects define twisted Hilbert spaces, and the symmetry sectors are irreducible representations of the tube algebra, which acts within and between these spaces. Under our proposal, the symmetry fixes exact degeneracies and identifications among sectors, and ordinary RMT describes the rest, with unrelated blocks treated as statistically independent. We call this \emph{categorical RMT}. For a non-invertible symmetry it goes beyond ordinary RMT, since Hilbert spaces with different twists can share a sector, and so part of their spectrum, though no invertible symmetry relates them.

We confirm this in a chaotic Fibonacci anyon chain, and check the framework in chaotic $S_3$ chains. We extend categorical RMT to primary states at fixed spin in $1+1$-dimensional CFTs, where modular covariance requires additional correlations. For continuous symmetries, the Sugawara construction and spectral flow isolate a neutral spectrum, which can have symplectic statistics even in a bosonic theory. In holographic CFTs, torus wormholes decorated by the SymTFT or by Chern--Simons theory realize the predicted ramp and its modular completion.

\hspace{.3cm} 

\newpage

\tableofcontents

\vspace{10pt}
\section{Introduction and summary}

Consider a chaotic quantum mechanical system with a global symmetry $G$. There is a standard way to model its level statistics with random matrix theory (RMT)~\cite{Dyson:1962es,Dyson:1962eeu,Bohigas:1983er,Mehta:2004}. The Hilbert space, symmetry operators, and the Hamiltonian decompose into superselection, or symmetry, sectors as
\begin{equation}
\label{E:theStart}
    \mathcal{H}
    \simeq
    \bigoplus_{\lambda} V_{\lambda}\otimes \Mult_{\lambda}\,,
    \qquad
    U(g)
    =
    \bigoplus_{\lambda}
    \rho_{\lambda}(g)\otimes\mathbf{1}_{\Mult_{\lambda}}\,,
    \qquad
    H
    =
\bigoplus_{\lambda}
\mathbf{1}_{V_{\lambda}}\otimes H_{\lambda}\,.
\end{equation}
Here $\lambda$ labels an irreducible representation $V_{\lambda}$ of the symmetry $G$, and $\Mult_{\lambda}$ is its multiplicity space. The symmetry acts on $V_{\lambda}$, while $\Mult_{\lambda}$ is the unconstrained space on which the effective Hamiltonian $H_{\lambda}$ acts. An RMT ansatz models the full block-diagonal Hamiltonian by the ensemble appropriate to its symmetry class, the Gaussian orthogonal, unitary or symplectic ensemble (GOE, GUE or GSE). Equivalently, each unconstrained $H_{\lambda}$ is drawn from the corresponding random-matrix ensemble, with different blocks statistically independent unless symmetry relates them. The ansatz does not predict the individual energy levels, but captures their universal correlations, including short- and long-range level repulsion within a sector.\footnote{These correlations follow from a universal nonlinear sigma model~\cite{Efetov:1983xg,Andreev:1996tw} associated with causal symmetry breaking. Random-matrix ensembles provide particular microscopic realizations of this effective theory. See especially~\cite{Altland:2020ccq,Altland:2021rqn}.}

In this manuscript we propose an answer to the question of how this prescription generalizes to a local quantum system on a spatial circle.\footnote{We expect a similar prescription for a local quantum system on a spatial strip, with the tube algebra replaced by the strip algebra~\cite{Cordova:2024iti}.}

Spatial locality makes the answer richer than in ordinary quantum mechanics, because it brings in twisted sectors. In 1+1 dimensions, global symmetries are implemented by topological defect lines~\cite{Frohlich:2004ef,Gaiotto:2014kfa,Bhardwaj:2017xup,Chang:2018iay}. A defect line running in the time direction defines a twisted Hilbert space $\mathcal{H}_a$, while networks of topological lines and junctions between them define operators on and between these Hilbert spaces~\cite{Petkova:2000ip,Chang:2018iay,Lin:2022dhv}. The twisted sectors are therefore not generally independent. In a relativistic theory, $\CRT$ can exchange two twisted sectors. More generally, networks of lines can act as intertwiners between twisted sectors, so that a single symmetry multiplet can be spread across several of them~\cite{Lin:2022dhv,Bartsch:2023wvv,Choi:2024tri}. 

We call our proposal \textbf{categorical RMT}. It applies both to ordinary group symmetries and to genuinely non-invertible symmetries, and reduces to conventional symmetry-resolved RMT in the untwisted sector of an ordinary group.

The central ingredient is the \textit{tube algebra} $\operatorname{Tube}(\mathcal C)$, where $\mathcal C$ is the category of symmetry defects~\cite{Izumi:2000qa,Lin:2022dhv,Bartsch:2023wvv,Benini:2025lav}. The tube algebra collects the topological operations acting within and between all twisted Hilbert spaces. For the finite semisimple symmetries considered here, the total Hilbert space and Hamiltonian decompose as
\begin{equation}
\label{eq:catrmtintro}
    \mathcal H_{\mathrm{tot}}
    =
    \bigoplus_a\mathcal H_a
    \simeq
    \bigoplus_\mu V_\mu\otimes \Mult_\mu\,,
    \qquad
    H
    =
    \bigoplus_\mu
    \mathbf{1}_{V_\mu}\otimes H_\mu\,.
\end{equation}
Now $\mu$ labels an irreducible representation $V_\mu$ of the tube algebra, and $\Mult_\mu$ is the corresponding multiplicity space. Categorical RMT models the full block-diagonal Hamiltonian by the ensemble appropriate to its symmetry class. Equivalently, each unconstrained $H_\mu$ is an ordinary random matrix of the corresponding class, and distinct blocks are statistically independent unless symmetry relates them. The full symmetry structure determines the exact degeneracies and relations among twisted sectors, while ordinary random matrix theory governs the remaining dynamics on the multiplicity spaces $\Mult_\mu$.\footnote{Our proposal is consistent with the results of O’Dea~\cite{ODea:2024tkt}, who showed that generalized symmetries, including non-invertible ones such as Kramers-Wannier duality, must be resolved before non-integrable chains show level repulsion.}

How much of this goes beyond ordinary RMT? Suppose one applied ordinary symmetry-resolved RMT to each twisted Hilbert space separately, using every operator that commutes with its Hamiltonian. For a group, together with the conjugation that maps $\mathcal H_h$ to $\mathcal H_{ghg^{-1}}$, this gives the same blocks as categorical RMT. For groups the tube algebra organizes the blocks rather than predicting new relations among them. The new prediction arises when a sector is shared between twisted Hilbert spaces that no invertible symmetry relates. Those Hilbert spaces then have part of their spectrum in common, level by level. This can happen only for a non-invertible symmetry.

We develop and test this proposal in three settings. We first study finite lattice models, then specialize the proposal to $1+1$-dimensional CFTs, and finally construct holographic realizations. These settings play different roles. The lattice models allow direct numerical tests of both local level statistics and long-range spectral correlations, while conformal invariance adds new kinematic and modular constraints. In the holographic examples, the observable available to us is the spectral form factor (SFF) in the ramp regime. The ramp captures long-range RMT correlations, and its slope is constrained by the underlying statistics. Our calculation does not reach the Heisenberg-time crossover to the plateau and therefore does not test short-range level repulsion.

\vspace{-.2cm}
\paragraph{Discrete and categorical symmetries.}

We begin with two classes of periodic lattice models: Potts-like qutrit chains with $S_3$ symmetry (like those considered in~\cite{Sinha:2025jhh}) and anyon chains with categorical Fibonacci symmetry generalizing those considered in~\cite{Feiguin:2006ydp,Chandran:2016vow}. In both cases we add symmetry-preserving deformations that break integrability and study the resulting spectra in the different symmetry sectors by exact diagonalization.

For the $S_3$ chain, we use spatially disordered couplings that preserve $S_3$ and time reversal but break reflection. We resolve the untwisted and twisted Hilbert spaces into the eight irreducible sectors of the tube algebra. Six sectors exhibit orthogonal statistics. Time reversal exchanges the remaining two, so that each exhibits unitary statistics while their spectra coincide. Since $S_3$ is a group, this example checks the framework rather than makes a new prediction. The classes come out as predicted, and apart from the exact pairing, correlations between distinct sectors are very small, including sectors from different twisted Hilbert spaces that share the same disordered couplings.

The Fibonacci chain has a non-invertible symmetry along with time reversal and reflection. Its tube algebra has four irreducible sectors. Two of them appear in the untwisted Hilbert space and three in the Fibonacci-twisted one, which we construct in~\cite{Jensen:2026cat} (see related discussions in~\cite{Buican:2017rxc,Aasen:2020jwb}), so one sector is shared between them. This is the prediction that ordinary RMT, applied to each Hilbert space separately, would miss. The untwisted and twisted chains are different Hamiltonians, acting on spaces of different dimension, yet the shared sector has the same energy levels in both, as the tube algebra requires. We confirm this numerically, resolving both Hilbert spaces into tube-algebra sectors and by lattice momentum. All four sectors have orthogonal statistics, because reflection combined with time reversal acts within each of them. Time reversal alone exchanges the two unshared sectors of the twisted Hilbert space, and their levels coincide at opposite momenta.

We go on to consider spectral correlations in $1+1$-dimensional CFTs on a circle, beginning with CFTs that possess discrete or categorical symmetry. We have in mind irrational and ``chaotic'' CFTs with no extended chiral algebra and an irregular spectrum of Virasoro primaries in the heavy regime $h,\bar h>(c-1)/24$. This introduces three new ingredients relative to the lattice models. First, the infinite-dimensional conformal symmetry organizes states into primaries and their descendant towers, whose quantum numbers are fixed by representation theory. Second, translations around the circle give a conserved momentum, or spin. It is an integer in the untwisted Hilbert space; in a twisted one its fractional part is fixed by the tube-algebra sector. Finally, torus partition functions with fugacity corresponding to the symmetry turned on are modular covariant.

Accordingly, we strip off the universal descendant contributions and study the primary spectrum at fixed spin. At fixed spin $s$ and tube-algebra sector $\mu$, categorical RMT models the unconstrained primary spectrum by the ensemble appropriate to its full symmetry class, including the ever-present $\CRT$ symmetry. Distinct $(s,\mu)$ blocks are statistically independent unless this symmetry class relates them. Modular covariance then requires a modular completion of these RMT correlations, producing additional correlations within and among different blocks. 

Although coupled minimal models provide promising candidates \cite{Antunes:2022vtb,Antunes:2024mfb}, no chaotic irrational CFT is currently under sufficient analytic or numerical control to test these detailed spectral predictions directly. Holography offers a complementary route, where bulk wormhole amplitudes provide predictions for smoothed, long-range spectral correlations in a putative dual description.\footnote{In $1+1$ dimensions another body of work~\cite{Haehl:2023tkr,DiUbaldo:2023qli,Haehl:2023xys,Haehl:2023mhf,Boruch:2025ilr} uses modular invariance to constrain the level statistics of a ``chaotic'' 2d CFT. Wormholes in higher-dimensional Einstein gravity also appear to encode long-range spectral correlations, through a ramp in the spectral form factor~\cite{Saad:2018bqo,Chen:2023hra}. There is a version of these statistics that persists even when the boundary systems exist at slightly different coupling constants~\cite{Cotler:2022rud}, matching the predictions of ``parametric correlations''~\cite{Simons:1993prb,Szafer:1993zz,Simons:1993zza}, that is naturally discussed using the parameterization of wormhole amplitudes appearing in~\cite{Cotler:2020lxj,Cotler:2021cqa}. See also \cite{Perlmutter:2026trb,Perlmutter:2026ujr}. We also note~\cite{Kudler-Flam:2026nzz,Liu:2026fnd} which explored the possibility that these higher-dimensional amplitudes describe the spectral statistics of a single CFT, not necessarily of an ensemble. See also~\cite{Pollack:2020gfa} for earlier discussion along these lines. } For the readers, who may find it surprising that conformal symmetry and quantum chaos can coexist, such examples are known: most notably non-arithmetic hyperbolic surfaces exhibit conformal symmetry\cite{Kravchuk:2021akc,Bonifacio:2023ban,Gesteau:2023brw} and quantum chaos \cite{marklof2006arithmetic,sarnak1993arithmetic}.

When the bulk admits a semiclassical description of Einstein gravity coupled to a sparse spectrum of matter, the dual is expected to be a large-$c$, irrational CFT with a chaotic spectrum of heavy states dual to black hole microstates. This is a working assumption, although aspects of it have been sharpened through quantitative sparseness and universality criteria developed in \cite{Hartman:2014oaa,Pal:2023cgk,Dey:2024nje}, see also \cite{Perlmutter:2026trb,Perlmutter:2026ujr}. Recent work beginning with~\cite{Belin:2020hea,Cotler:2020ugk,Maxfield:2020ale} and continuing with~\cite{Cotler:2020hgz,Belin:2021ryy,Chandra:2022bqq,Collier:2023fwi,Belin:2023efa,Jafferis:2025vyp} (and many more) has given the topological expansion of pure AdS$_3$ gravity an increasingly concrete ensemble interpretation, analogous to the matrix-integral description of JT gravity~\cite{Saad:2019lba}. We do not yet know whether these constructions define a complete nonperturbative theory, and whether the ensemble consists of exact CFTs rather than ``approximate" CFT data.

Motivated by top-down constructions of AdS$_3$ string theory~\cite{Kraus:2006wn}, we consider three-dimensional gravity with negative cosmological constant coupled to the SymTFT that encodes the boundary global or categorical symmetry. This construction is the AdS$_3$ analogue of JT gravity coupled to a topological gauge theory with gauge group $G$, whose boundary description is a random-matrix ensemble graded by $G$~\cite{Kapec:2019ecr}.\footnote{Here the SymTFT is a topological sector of an ordinary bulk theory of AdS quantum gravity. This should be distinguished from proposals in which the SymTFT itself plays the role of a complete holographic description.}

The connected torus wormhole of pure three-dimensional gravity captures long-range spectral correlations in the dual description through the ramp in the spectral form factor~\cite{Cotler:2020ugk}. It does not describe the Heisenberg-time crossover to the plateau. The calculation begins with a contribution that we call the ``seed.'' After accounting for the sum over gluings, including those with a relative Euclidean rotation of $\pi$~\cite{Yan:2023rjh}, the seed reproduces the class-appropriate RMT ramp for primary states at fixed spin, including its correct slope. We obtain the full amplitude by summing the seed over relative modular transformations of the two boundary tori, in a construction analogous to the modular sum defining the Maloney-Witten-Keller~\cite{Maloney:2007ud,Keller:2014xba} partition function. The nontrivial modular images introduce additional correlations among spin sectors as required by modular invariance. These corrections are subleading at low temperature or at late Lorentzian times within the ramp regime, when the ordinary RMT ramp dominates.

We extend this construction by decorating the seed wormhole with a SymTFT cobordism.\footnote{A version of this construction where the bulk has a $(-1)^{\mathsf{F}}$ symmetry has been studied in~\cite{Boruch:2026hbr}. See also~\cite{SUGRA} where an analogous study will appear for minimal bulk AdS$_3$ supergravity, with similar results to those found here.} This new seed amplitude correlates the two twisted Hilbert spaces on the two boundaries precisely when they contain the same tube algebra sector $\mu$. Alone, its gravitational part supplies the RMT ramp of $H_\mu$ with the GUE coefficient, while the SymTFT supplies the exact degeneracies and identifications among twisted sectors. Boundary $\CRT$ is the $\pi$ rotation of the boundary torus, whose importance for 3d gravity computations was emphasized in~\cite{Yan:2023rjh,Harlow:2023hjb}. In the bulk it acts on the SymTFT anyons by charge conjugation. The $\CRT$ image doubles the ramp of every self-dual sector to the GOE value, and makes the spectra of conjugate sectors coincide. Summing over relative modular transformations, including the $\pi$ rotations, then produces the long-range predictions of categorical RMT along with their required vector-valued modular completion. 

\vspace{-.2cm}
\paragraph{Continuous global symmetries.}

We finally consider CFTs with continuous global symmetries. In $1+1$ dimensions, every continuous global symmetry comes with an affine Kac-Moody algebra generated by the modes of its conserved currents. Together with the Virasoro generators, these form a Virasoro-Kac-Moody algebra. For $U(1)_k$, the integer spectral-flow operators further extend the chiral algebra and relate charge sectors that would naively appear independent. More generally, spectral flow relates the untwisted Hilbert space to its continuously twisted counterparts~\cite{Schwimmer:1986mf,Benjamin:2020swg}.

The Sugawara construction cleanly separates the symmetry data from the unconstrained dynamics. It decomposes the stress tensor into a current-algebra contribution, whose spectrum is fixed by representation theory, and a neutral one that commutes with the currents. Once we specify a representation of the full extended chiral algebra and fix the neutral spin, all remaining spectral data reside in the neutral multiplicity space, and it is this neutral spectrum that we model by RMT. This is our proposal in its plainest form: once a symmetry sector is completely specified, the symmetry has fixed everything it can, and ordinary random matrix theory governs what remains.

For holographic CFTs, this decomposition has a simple bulk realization. We study three-dimensional gravity coupled to $U(1)_k\times U(1)_{-k}$ and $SU(2)_k\times SU(2)_{-k}$ Chern-Simons theories, so that there are holomorphic and antiholomorphic symmetry currents. On the torus wormhole, the Chern-Simons amplitude reproduces the current-algebra characters, while large gauge transformations realize spectral flow. The gravitational seed supplies the long-range RMT correlations of the neutral states in the ramp regime. Some neutral sectors have symplectic statistics even though the theory is bosonic. These are the $SU(2)$ modules where the sum of the $SU(2)$ spins $j_L+j_R$ is half-integer, for odd $k$, the $U(1)$ modules of charge $(m_L,m_R)=(0,k)$ or $(k,0)$ mod $2k$. In these modules $\CRT$ squares to $-1$ on the current-algebra part. Since $\CRT^2=+1$ in a bosonic theory, it must also square to $-1$ on the neutral multiplicity space. So these sectors have GSE statistics, and every neutral level is exactly doubly degenerate as a Kramers pair. The torus wormhole gives these sectors a ramp with the GOE slope. That is exactly the GSE prediction when the trace counts each Kramers pair twice, so the torus wormhole alone cannot tell the two classes apart. The modular completion organizes the remaining correlations in the language of Jacobi forms, which we work out explicitly for $U(1)$ and $SU(2)$.

\vspace{-.2cm}
\paragraph{Outline.}

The rest of this manuscript is organized as follows. We describe the categorical RMT proposal in more detail in Section~\ref{sec:categorical-RMT}, including a self-contained review of the relevant mathematical physics, and test it in periodic lattice models. In Section~\ref{S:holoCat} we show how the long-range predictions of categorical RMT are realized for holographic CFTs with ordinary discrete or non-invertible categorical symmetry. Here the spectral-form-factor ramp is encoded by torus wormholes decorated by a bulk topological field theory. We discuss the symmetry-resolved level statistics of two-dimensional CFTs with continuous global symmetries in Section~\ref{sec:continuous-CFT}. There, spectral flow fixes the spectra of all twisted sectors in terms of the untwisted one. We then show how their long-range spectral correlations are realized holographically in Section~\ref{sec:continuous-holography}, through torus wormholes decorated by Chern-Simons theory. We end with a Discussion in Section~\ref{S:discuss}. The Appendices collect technical results.


\section{Categorical RMT}
\label{sec:categorical-RMT}

In this section, we set out categorical RMT in detail and test it in two chaotic lattice models.
In a spatially extended system, a topological line wrapped on the Euclidean time circle produces a twisted Hilbert space,
and networks of such lines act both within and between these spaces. We review how the algebra of
these networks, the tube algebra, fixes the exact degeneracies and the exact coincidences between
twisted Hilbert spaces. The antiunitary symmetries of the model fix the remaining exact relations
and the Wigner--Dyson classes. We then propose that ordinary random matrices govern what both leave
free, which is the content of \eqref{eq:catrmtintro}. The exact diagonalization of an $S_3$ qutrit chain and
of a Fibonacci anyon chain gives the predicted Wigner--Dyson statistics in every resolved sector.

Throughout, a twisted sector carries the label $a$, recording which line wraps the Euclidean time
circle, while a tube algebra sector carries the label $\mu$, the analog of the charge $\lambda$ in
\eqref{E:theStart}. Nothing below in this section assumes conformal invariance, so every statement in this section holds for any local
system on a circle.\footnote{Our review of the categorical formalism follows
\cite{Lin:2022dhv,Bartsch:2023wvv}, whose ideas were used for a closely related purpose in
\cite{Ferragatta:2026ejs}.} In Subsection~\ref{sec:lines} we assemble the tube algebra and derive
\eqref{eq:catrmtintro}. In Subsection~\ref{sec:statement} we state the proposal and draw out its
consequences for spectral densities and their correlations. We then specialize to a finite group in
Subsection~\ref{sec:group} and to $S_3$ in Subsection~\ref{sec:S3}, and we treat the non-invertible
Fibonacci case in Subsection~\ref{sec:fibonacci-elaborated}.

\subsection{Lines, twisted sectors and the tube algebra}
\label{sec:lines}

Let $\cat$ be the collection of topological line defects of a $1+1$-dimensional theory. That these
lines form a fusion category means the following. There are simple line types $a, b, c, \dots$,
and bringing two of them together gives a direct sum of simple lines with non-negative integer
multiplicities. Lines meet at junctions, which are local topological operators. Because the lines
are topological, we may deform them freely without changing correlation functions as long as the
deformation does not cross any operator insertion. The word ``category'' packages these fusion and
junction rules together with the associativity data relating them.

In $1+1$ dimensions a topological line plays two roles. Wrapped on the spatial circle it acts as
an operator on a Hilbert space. Wrapped on the Euclidean time circle it instead defines a twisted
Hilbert space $\Hil_a$, with $a=\mathbf 1$ the untwisted case. The natural arena for a categorical
symmetry is therefore not a single Hilbert space but the direct sum over twisted sectors,
\begin{equation}
  \Htot \;=\; \bigoplus_{a \in \Irr(\cat)} \Hil_a \,.
  \label{eq:Htot}
\end{equation}
Here $\Irr(\cat)$ is the set of simple objects of the fusion category $
\cat$. 
This is already a departure from \eqref{E:theStart}, where there was one Hilbert space to
decompose. A spatially extended theory has many, one for each defect boundary condition around the
circle. The symmetry can act between them as well as inside them.

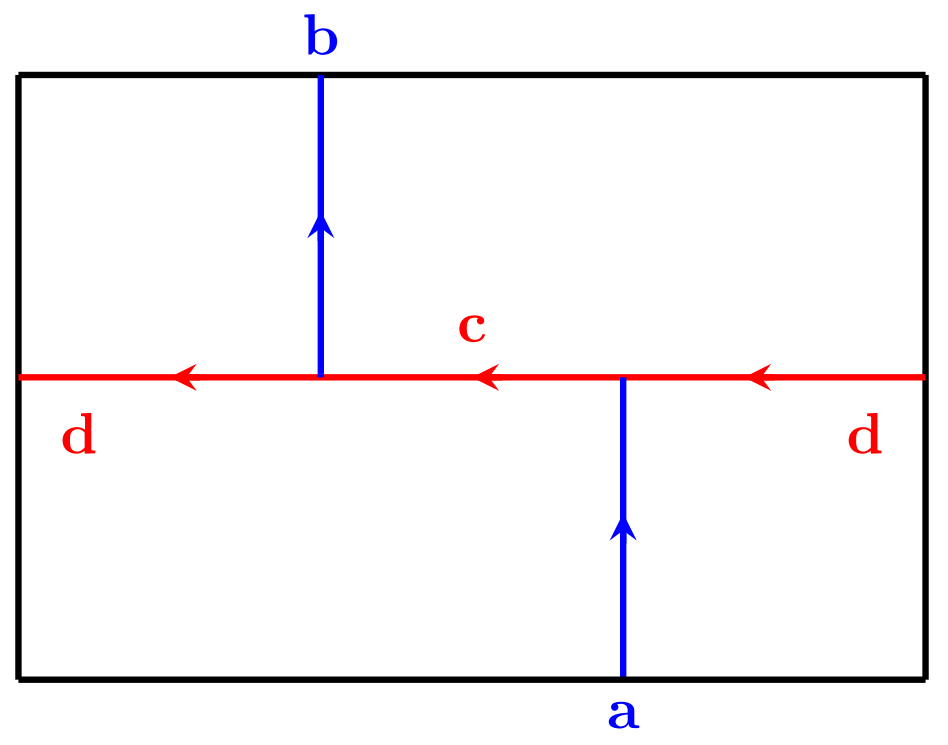
\begin{figure}[H]
  \centering
\begin{tikzpicture}
    \draw[black, very thick] (-3,0)--(3,0);
    \draw[black, very thick]  (-3,0)--(-3,-4);
    \draw[black, very thick]  (3,0)--(3,-4);
    \draw[red, very thick] (-3, -2) --(3 ,-2) ;
    \draw [-stealth, red, very thick](.2,-2) -- (0,-2);
    \draw [-stealth, red, very thick](2,-2) -- (1.8,-2);
    \draw [-stealth, red, very thick](-1.8,-2) -- (-2,-2);
    \draw [red] (2.6, -2.1) node[anchor=north]{\bf d};
    \draw [red] (-2.6, -2.1) node[anchor=north]{\bf d};
    \draw [red] (0, -1.9) node[anchor=south]{\bf c};
    \draw[blue, very thick]  (-1,0)--(-1,-2);
    \draw[blue, very thick]  (1,-2)--(1,-4) node[anchor=north]{\bf a};
    \draw [-stealth, blue, very thick](-1,-1.1) -- (-1,-.9);
    \draw [-stealth, blue, very thick](1,-3.1) -- (1,-2.9);
    \draw [blue] (-1, 0) node[anchor=south]{\bf b};
    \draw[black, very thick] (-3,-4)--(3, -4); 
\end{tikzpicture}
  \caption{A tube algebra element on the cylinder, with the two vertical edges identified. The
  crossing of the horizontal and vertical symmetry lines is resolved in the most general way
  allowed by the fusion rules, which introduces the junction labels $d$ and $c$. The resulting
  operator $\cL^{d,c}_{b \leftarrow a}$ takes the $a$-twisted sector to the $b$-twisted one. Elements
  with $a=b$ act inside one twisted sector and build the projectors of
  Subsections~\ref{sec:group} to~\ref{sec:fibonacci-elaborated}. Elements with $a \neq b$ are intertwiners between
  distinct twisted sectors.}
  \label{fig:tube}
\end{figure}

The algebra organizing these actions is the tube algebra $\Tube(\cat)$. An element of
$\Tube(\cat)$ is a topological operation on a cylinder: a horizontal line together with junction
data, wrapped around a vertical defect line, as in Fig.~\ref{fig:tube}. For a group-like theory
this is the familiar operation of acting on a twisted sector by a symmetry element, which maps $\Hil_h$ to $\Hil_{ghg^{-1}}$. For a
genuinely non-invertible symmetry the line crossing must be resolved through allowed junctions,
and the resulting operator can move between twisted sectors that no invertible line relates,
\begin{equation}
  L : \Hil_a \longrightarrow \Hil_b \,.
  \label{eq:intertwiner}
\end{equation}
The tube algebra is thus an algebra of operators on $\Htot$, not on a single fixed $\Hil_a$. It is
generated by all topological operations available on the cylinder network: sliding lines, fusing
parallel lines, resolving junctions, and applying the associativity moves of the category.

For a unitary fusion category, $\Tube(\cat)$ is finite-dimensional and semisimple. This is what
gives a well-defined decomposition into tube algebra sectors:
\begin{equation}
  \Tube(\cat) \;\simeq\; \bigoplus_{\mu \in \Irr \Tube(\cat)} \End(V_\mu) \,,
  \label{eq:algdecomp}
\end{equation}
where $\mu$ runs over the irreducible modules and $V_\mu$ is the corresponding
finite-dimensional module. The total Hilbert space decomposes accordingly,
\begin{equation}
  \Htot \;\simeq\; \bigoplus_\mu V_\mu \otimes \Mult_\mu \,.
  \label{eq:Hdecomp}
\end{equation}
Representation theory fixes the first factor, while the multiplicity space $\Mult_\mu$ is the
large space on which chaotic dynamics acts.

The same decomposition can be written with the twisted-sector label displayed:
\begin{equation}
  \Hil_a \;\simeq\; \bigoplus_\mu W^\mu_a \otimes \Mult_\mu \,,
  \qquad
  V_\mu \;\simeq\; \bigoplus_a W^\mu_a \,,
  \label{eq:Wdecomp}
\end{equation}
where the finite-dimensional space $W^\mu_a$ records how much of the tube algebra sector $\mu$ is
visible in the $a$-twisted sector. One multiplicity space $\Mult_\mu$ can therefore appear in
several twisted sectors. Those appearances are components of a single irreducible tube algebra
module, so they must not be modeled by independent random matrices.

An element of $\Tube(\cat)$ acts on the categorical index in $V_\mu$, where it may move between
the components $W^\mu_a$ and $W^\mu_b$, and as the identity on $\Mult_\mu$. The large
multiplicity space is therefore invisible to purely topological operations.

Let $H$ be the block-diagonal Hamiltonian on $\Htot$, with one block on each twisted sector.
Preserving the line network means that $H$ commutes with every element of $\Tube(\cat)$. Since
$\Tube(\cat)$ acts irreducibly on $V_\mu$, any such $H$ is proportional to the identity there.
Its commutant is therefore $\bigoplus_\mu 1_{V_\mu} \otimes \End(\Mult_\mu)$, so the
Hamiltonian takes the block form:
\begin{equation}
  H \;=\; \bigoplus_\mu 1_{V_\mu} \otimes H_\mu \,.
  \label{eq:Hblocks}
\end{equation}
The factor $1_{V_\mu}$ produces exact degeneracies and exact relations among twisted sectors.
The genuinely dynamical part is $H_\mu$, acting on $\Mult_\mu$.

\subsection{The categorical RMT statement}
\label{sec:statement}

We can now state our proposal:\footnote{Earlier work on quantum chaos
with categorical symmetry includes \cite{ODea:2024tkt}. The role of the tube
algebra in fixing the Cardy form of the density of states is emphasized in \cite{Lin:2022dhv}. For a group, the Cardy growth in each irreducible sectors, including the twisted sectors, was first derived in \cite{Pal:2020wwd}.}

\begin{tcolorbox}[
  colback=gray!10,
  colframe=blue,
  boxrule=0.6pt,
  arc=2pt,
  left=6pt,
  right=6pt,
  top=4pt,
  bottom=4pt]
  In a chaotic $1+1$-dimensional theory with a preserved categorical line network, the Hamiltonian
  $H_\mu$ on each multiplicity space $\Mult_\mu$ of the tube algebra decomposition \eqref{eq:Hblocks},
  restricted to a sector of any further unitary symmetry such as lattice momentum,
  is an ordinary random matrix, drawn from the Wigner--Dyson class fixed by the antiunitary
  symmetries that map $\mu$ to itself. Distinct blocks $H_\mu$ and $H_\nu$ are statistically
  independent unless a symmetry relates them.
\end{tcolorbox}

Since the independent RMT labels are the tube algebra sectors $\mu$ rather than the simple lines $a$,
two features follow that have no analog in the quantum-mechanical rule \eqref{E:theStart}. First, one tube algebra sector can
have components in several twisted Hilbert spaces. The operators moving between those spaces force
the components to share a single random matrix, so the spectra of different twisted Hilbert spaces
coincide level by level on that shared piece. Second, the number of independent blocks is the
number of irreducible modules of the tube algebra, rather than the number of twisted Hilbert spaces
times a charge count. The first is a sharp prediction that distinguishes categorical RMT from ordinary
symmetry-resolved RMT applied to each twisted Hilbert space separately. For a group, the second also follows from that ordinary procedure, once conjugation by $G$ is used to identify the blocks of conjugate twists. We test them in the two
examples below.

Let $\rho_\mu(E)$ be the spectral density of $H_\mu$. Tracing over the whole of $\Htot$, or
restricting to one twisted sector, gives
\begin{equation}
  \rho(E) \;=\; \sum_\mu \dim(V_\mu)\, \rho_\mu(E) \,,
  \qquad
  \rho_a(E) \;=\; \sum_\mu \dim(W^\mu_a)\, \rho_\mu(E) \,.
  \label{eq:densities}
\end{equation}
The coefficients are exact degeneracies fixed by the symmetry. The second relation is how one
random matrix block contributes to more than one twisted Hilbert space. Write
$\rho_{a,\mu}(E) = \dim(W^\mu_a)\,\rho_\mu(E)$ for the individual summand there.
Connected spectral correlations then inherit the same structure,
\begin{equation}
  \big\langle \rho_{a,\mu}(E)\, \rho_{b,\nu}(E') \big\rangle_c
  \;=\; \delta_{\mu\nu} \dim(W^\mu_a) \dim(W^\mu_b)\, \mathcal K_\mu(E,E') \,,
  \label{eq:corr}
\end{equation}
with $\mathcal K_\mu$ the connected RMT kernel of the appropriate class. Distinct tube algebra sectors carry
no universal correlation unless a symmetry relates them. We turn next to antiunitary symmetries.

\paragraph{Antiunitary symmetries.}
The ensemble modeling $H_\mu$ depends on the antiunitary symmetries of the system, such as time
reversal or $\CRT$. Each antiunitary symmetry that preserves the tube algebra permutes the tube
algebra sectors, and in doing so it plays one of two independent roles. An antiunitary symmetry
that maps $\mu$ to itself acts on the multiplicity space $\Mult_\mu$, and the class of $H_\mu$ is
then GOE or GSE, according to whether that action squares to $+1$ or $-1$. If no antiunitary
symmetry fixes $\mu$, the class is GUE. An antiunitary symmetry that maps $\mu$ to another sector $\nu$ instead makes
$H_\nu$ the complex conjugate of $H_\mu$ in a suitable basis, so the two spectra coincide level by
level whatever their class. This is the exception to the $\delta_{\mu\nu}$ in \eqref{eq:corr}.

Since the two roles are independent, an exchanged pair can belong to either class, and our two
examples realize both. In Subsection~\ref{sec:S3} a single antiunitary symmetry exchanges a
pair of sectors and fixes neither member, so the pair is GUE with coincident spectra. In
Subsection~\ref{sec:fibonacci-elaborated} one antiunitary symmetry exchanges a pair while a second fixes each member, so
the pair is GOE and its spectra still coincide. In either case, the Wigner--Dyson class is assigned sector by
sector, after the decomposition by the tube algebra.

To summarize, the object to compare against a random matrix model is the sector-resolved partition function:
\begin{equation}
  Z_\mu(\beta) \;=\; \mathrm{Tr}_{\Htot}\!\left( P_\mu\, e^{-\beta H}\right) ,
  \label{eq:Zsector}
\end{equation}
where $P_\mu$ is the projector onto the tube algebra sector $\mu$, known as the primitive central
idempotent. 

\subsection{The group case}
\label{sec:group}

\begin{figure}[H]
  \centering
\includegraphics[width=0.40\textwidth]{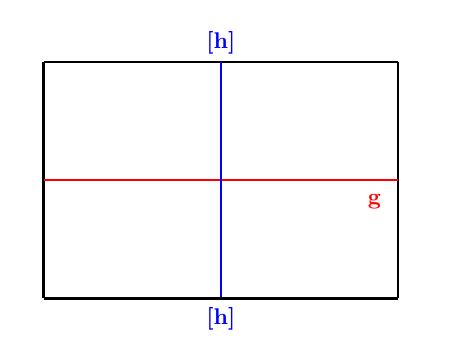}
  \caption{The symmetry action $g$ on the $h$-twisted sector, on the cylinder with its two vertical
  edges identified. A group element $g$ takes the twist $h$ to $ghg^{-1}$. It therefore acts within
  the $h$-twisted sector, through a unitary $U_h(g)$, only when it lies in the centralizer $C(h)$.
  These are the operators used to project onto tube algebra sectors. The vertical defect is labeled
  here by its conjugacy class $[h]$, since that is what labels the tube algebra sector.}
  \label{fig:group}
\end{figure}

In this subsection, we consider the symmetry lines to be invertible and labeled by elements of a finite group $G$, so that
$\cat = \VecG_G$ in the anomaly-free case. Just like before, wrapping the line $h$ on the Euclidean time circle
defines the twisted Hilbert space $\Hil_h$. A group element $g$ acts on $\Hil_h$ by conjugating
the twist. It therefore preserves the twisted sector only when it lies in the centralizer
$C(h) = \{ g \in G : gh = hg\}$. Elements outside the centralizer are intertwiners,
\begin{equation}
  \Hil_h \longrightarrow \Hil_{g h g^{-1}} \,,
  \label{eq:groupintertwiner}
\end{equation}
which is why the tube algebra sector is labeled by the full conjugacy class $[h]$ rather than by a
representative. An
element of the tube algebra in this case is shown in Fig.~\ref{fig:group}.

The projector onto an irreducible representation $R$ of the centralizer is
\begin{equation}
  P^R_h \;=\; \frac{\dim R}{|C(h)|} \sum_{g \in C(h)} \chi_R(g)^* \, U_h(g) \,,
  \label{eq:groupproj}
\end{equation}
where $\chi_R$ is the character of the representation $R$ and $U_h(g)$ is the unitary operator representing $g$. The tube algebra sectors are labeled by pairs $([h],R)$ with $R \in \Irr\, C(h)$. These are exactly the
irreducible representations of the tube algebra of $\VecG_G$, equivalently the simple objects of
the quantum double $D(G)$. In the general notation of Subsection~\ref{sec:lines} we have $W^{([k],R)}_h \simeq V_R$ for
$h \in [k]$ and zero otherwise, with $V_R$ transported from the representative $k$ to the conjugate
element $h$. The dictionary is therefore

\begin{center}
\begin{tabular}{ll}
  \toprule
  General categorical notation & Group-like notation \\
  \midrule
  simple line $a$ & group element $h$ \\
  twisted Hilbert space $\Hil_a$ & $h$-twisted Hilbert space $\Hil_h$ \\
  tube algebra sector $\mu$ & pair $([h],R)$ \\
  finite-dimensional space $W^\mu_a$ & $V_R$ if $a \in [h]$, zero otherwise \\
  multiplicity space $\Mult_\mu$ & $\Mult_{([h],R)}$ \\
  RMT block $H_\mu$ & $H_{([h],R)}$ \\
  \bottomrule
\end{tabular}
\end{center}

\noindent
The independent RMT blocks are labeled by the pairs $([h],R)$, and the same block is seen in every
twisted sector whose twist lies in $[h]$.

Restricting to the untwisted sector $h=1$, for which $C(1)=G$, recovers the ordinary
quantum-mechanical rule \eqref{E:theStart}. There the relevant algebra is the group algebra
$\mathbb{C}[G]$ acting on a single Hilbert space, and the label $\mu$ reduces to an irreducible
representation of $G$. The full $1+1$-dimensional problem instead uses the tube algebra acting on
the direct sum of twisted Hilbert spaces. This is the reduction to conventional symmetry-resolved
RMT promised in the introduction.

\subsection{Example: the group $S_3$}
\label{sec:S3}

Our first example is the smallest non-abelian group. It illustrates the second of the two new
features listed in Subsection~\ref{sec:statement}. The six twisted Hilbert spaces
decompose into eight tube algebra sectors, and those eight should behave as independent random
matrices, apart from one pair that is related by time reversal. We
take
$S_3 = \langle r,\sigma \mid r^3 = \sigma^2 = 1,\ \sigma r \sigma = r^{-1}\rangle$. The six simple lines of $\VecG_{S_3}$
are $L_g$ with $g \in S_3$, all invertible and of quantum dimension one, with fusion
$L_g \otimes L_{g'} \simeq L_{gg'}$. The identity and the three transposition lines are self-dual,
while the two three-cycle lines are dual to one another.

\subsubsection{Twisted sectors and sectors of $D(S_3)$}

Wrapping $L_h$ on the Euclidean time circle gives $\Hil_h$, and the tube algebra acts on
$\Htot = \bigoplus_{h \in S_3} \Hil_h$. The six twists fall into three conjugacy classes:
\begin{equation}
  [e] = \{e\},\quad [t] = \{\sigma, r\sigma, r^2 \sigma\},\quad [c] = \{r, r^2\} \,,
  \label{eq:S3classes}
\end{equation}
with centralizers
\begin{equation}
  C(e) = S_3, \qquad C(\sigma) \simeq \mathbb{Z}_2, \qquad C(r) \simeq \mathbb{Z}_3 \,.
  \label{eq:S3centralisers}
\end{equation}
A convenient basis for the tube algebra is the set of operators
$U_h(g) : \Hil_h \to \Hil_{ghg^{-1}}$, which compose by stacking cylinders,
\begin{equation}
  U_{g_1 h g_1^{-1}}(g_2)\, U_h(g_1) \;=\; U_h(g_2 g_1) \,.
  \label{eq:S3stack}
\end{equation}
The algebra they generate is the quantum double $D(S_3)$.

The simple objects of $D(S_3)$ are pairs $([h],R)$ with $R$ an irreducible representation of the
centralizer. Following the usual terminology we call the class $[h]$ the flux:

\begin{center}
\begin{tabular}{lll}
  \toprule
  Flux class & Centralizer & Tube algebra sectors \\
  \midrule
  $[e]$ & $S_3$ & $([e],1)$, $([e],\sgn)$, $([e],2)$ \\
  $[t]$ & $\mathbb{Z}_2$ & $([t],+)$, $([t],-)$ \\
  $[c]$ & $\mathbb{Z}_3$ & $([c],1)$, $([c],\omega)$, $([c],\omega^2)$ \\
  \bottomrule
\end{tabular}
\end{center}

\noindent
with $\omega = e^{2\pi i/3}$. There are eight tube algebra sectors in all, against six twisted Hilbert
spaces. Their finite-dimensional factors are
supported on the twists in the corresponding class, for instance
\begin{equation}
  V_{([t],+)} \simeq \mathbb{C}^3 , \qquad V_{([c],\omega)} \simeq \mathbb{C}^2 ,
  \label{eq:S3V}
\end{equation}
of dimensions $|[h]| \dim R$. The untwisted sector carries the three irreducible representations
of $S_3$, namely the trivial one, the sign representation and the two-dimensional standard
representation,
\begin{equation}
  \Hil_e \simeq \Mult_{([e],1)} \oplus \Mult_{([e],\sgn)} \oplus \mathbb{C}^2 \otimes \Mult_{([e],2)} \,.
  \label{eq:S3untwisted}
\end{equation}
Each of the three transposition-twisted sectors carries
$\Mult_{([t],+)} \oplus \Mult_{([t],-)}$, and the intertwiners identify the three of them. Each of
the two three-cycle-twisted sectors carries
$\Mult_{([c],1)} \oplus \Mult_{([c],\omega)} \oplus \Mult_{([c],\omega^2)}$, identified with one
another by transport under conjugation.

\paragraph{Projectors.}

Specializing \eqref{eq:groupproj}, the untwisted projectors are
\begin{equation}
  P^1_e = \tfrac{1}{6} \sum_{g \in S_3} U_e(g), \qquad
  P^{\sgn}_e = \tfrac{1}{6} \sum_{g \in S_3} \sgn(g)\, U_e(g), \qquad
  P^2_e = \tfrac{2}{6} \sum_{g \in S_3} \chi_2(g)^*\, U_e(g) \,,
  \label{eq:S3projuntwisted}
\end{equation}
with $\chi_2$ the character of the two-dimensional representation. For a transposition twist and a
three-cycle twist,
\begin{equation}
  P^\pm_\sigma = \tfrac{1}{2}\big( U_\sigma(e) \pm U_\sigma(\sigma) \big), \qquad
  P^{\chi}_r = \tfrac{1}{3}\big( U_r(e) + \chi(r)^* U_r(r) + \chi(r^2)^* U_r(r^2) \big) \,,
  \label{eq:S3projtwisted}
\end{equation}
where $\chi$ runs over the three characters of $\mathbb{Z}_3$. These satisfy
$P^R_h P^{R'}_h = \delta_{RR'} P^R_h$ and $\sum_R P^R_h = 1_{\Hil_h}$. The
primitive central idempotent of the full tube algebra sums them over the class,
\begin{equation}
  P_{([h],R)} \;=\; \sum_{k \in [h]} P^{R_k}_k \,,
  \label{eq:S3central}
\end{equation}
where $R_k$ is obtained from $R$ by conjugation. Comparing \eqref{eq:S3central} with the general
decomposition \eqref{eq:Wdecomp} gives $\dim W^{([h],R)}_k = \dim R$ for $k \in [h]$, and zero
otherwise. Feeding this into \eqref{eq:Zsector} and \eqref{eq:corr} shows that the same random
matrix appears in every twisted Hilbert space whose twist lies in $[h]$.

\subsubsection{Time reversal and the predicted classes}
\label{sec:S3CRT}

The disordered qutrit chain (defined below in Subsection~\ref{sec:S3numerics}) has a single antiunitary symmetry. It is
complex conjugation $\Theta$ in the qutrit basis, under which every $H_h$ is real, and it is the time
reversal referred to in the introduction. The permutation operators $U_h(g)$ are real, so $\Theta$
fixes the twisted sector\footnote{
In a relativistic theory the antiunitary that is always present is $\CRT$, which maps $\Hil_h$ to $\Hil_{h^{-1}}$ and acts on tube algebra sectors by duality. All eight $D(S_3)$ sectors are self-dual.} and conjugates the character in the projector \eqref{eq:groupproj}:
\begin{equation}
  \Theta\, P^{R}_h\, \Theta \;=\; P^{R^*}_h \,, \qquad\text{so}\qquad
  \Theta : ([h],R) \longmapsto ([h],R^*) \,.
  \label{eq:S3crt}
\end{equation}
The three representations of $S_3$, the two of
$\mathbb{Z}_2$ and the trivial representation of $\mathbb{Z}_3$ are real,  so six of the eight sectors are fixed. The two nontrivial characters of
$\mathbb{Z}_3$ are exchanged:
\begin{equation}
  ([c],\omega) \longleftrightarrow ([c],\omega^2) \,.
  \label{eq:S3pair}
\end{equation}
For a real representation $R$, the operator $\Theta$ acts on $\Mult_{([h],R)}$ as an antiunitary
squaring to $+1$, so each fixed sector admits a real basis in which $H_\mu = H_\mu^*$. Those
six sectors are GOE. For the exchanged pair we instead get
\begin{equation}
  H_{([c],\omega^2)} = H_{([c],\omega)}^* \quad \Longrightarrow \quad
  \mathrm{Spec}\, H_{([c],\omega)} = \mathrm{Spec}\, H_{([c],\omega^2)} \,,
  \label{eq:S3degen}
\end{equation}
and no antiunitary symmetry acts within either member. Note that $\Theta$ exchanges two tube algebra
sectors inside the same twisted Hilbert space. It does not exchange two twisted Hilbert spaces.
The predictions are therefore:
\begin{itemize}
  \item The six sectors fixed by $\Theta$ are GOE.
  \item The sectors $([c],\omega)$ and $([c],\omega^2)$ are individually GUE.
  \item $\Theta$ forces their spectra to coincide, so their cross spectral form factor equals the
        corresponding auto spectral form factor.
  \item Apart from the relations imposed by intertwiners and $\Theta$, distinct sectors are
        statistically independent.
\end{itemize}

\subsubsection{Numerical test}
\label{sec:S3numerics}

In this subsection, we test whether the eight tube algebra sectors behave as independent random
matrix blocks with the classes predicted above, using an $S_3$-symmetric qutrit chain with twisted
boundary conditions.

Each of the $L$ sites carries a qutrit with
basis $\{|0\rangle,|1\rangle,|2\rangle\}$, so the many-body space has dimension $3^L$. The group
acts by permuting the on-site labels,
$U(g)|s_{0},\dots,s_{L-1}\rangle = |g(s_{0}),\dots,g(s_{L-1})\rangle$, and we take the generators to be the
transposition $(1\,2)$ and the three-cycle $0 \to 1 \to 2 \to 0$. These realize $\sigma$ and $r$
respectively. The action is on-site and the
Hilbert space is a tensor product, so the symmetry is anomaly-free. The relevant tube algebra is
the quantum double $D(S_3)$ used above, with no cocycle.

For a flux representative $h$ the twisted Hamiltonian is real symmetric. It contains a
nearest-neighbor Potts coupling, the twisted boundary bond, a next-nearest-neighbor coupling, a
three-site interaction, and a permutation-invariant transverse field:
\begin{equation}
  H_h = \sum_{i=0}^{L-2} J_i\, \delta_{s_i,s_{i+1}}
      + J_{L-1}\, \delta_{s_{L-1},\,h(s_0)}
      + \sum_{i=0}^{L-3} J^{(2)}_i\, \delta_{s_i,s_{i+2}}
      + \sum_{i=0}^{L-3} J^{(3)}_i\, \delta_{s_i,s_{i+1}} \delta_{s_{i+1},s_{i+2}}
      + \sum_{i=0}^{L-1} f_i \sum_{a \neq b} |a\rangle_i \langle b|_i \,.
  \label{eq:S3ham}
\end{equation}
Only the nearest-neighbor coupling crosses the twisted bond; the $J^{(2)}$ and $J^{(3)}$ sums stop at $i=L-3$. The couplings carry weak per-site disorder,
\begin{equation}
  J_i \sim 1.00 + 0.15\,\xi, \quad
  f_i \sim 0.82 + 0.12\,\xi, \quad
  J^{(2)}_i \sim 0.32 + 0.08\,\xi, \quad
  J^{(3)}_i \sim 0.20 + 0.06\,\xi \,,
  \label{eq:S3disorder}
\end{equation}
with $\xi$ an independent standard normal variable for each site and term. The disorder preserves
$S_3$ and time reversal but breaks reflection and translation, and both breakings are needed. Were
reflection unbroken, its product with complex conjugation would act inside $([c],\omega)$ and force
that sector to GOE, washing out the prediction of Subsection~\ref{sec:S3CRT}, while unbroken
translation would split each tube algebra sector further by momentum. Both effects appear in the clean chain of
Subsection~\ref{sec:fibonacci-elaborated}.

\paragraph{Diagnostics.}
We unfold each spectrum against a degree-eight polynomial fit to the smooth part of the level
staircase, so that the unfolded levels $\varepsilon_n$ have unit mean density. All diagnostics use
the central $70\%$ of the spectrum. We compare the distribution $P(s)$ of the unfolded gaps
$s_n = \varepsilon_{n+1} - \varepsilon_n$ against the Wigner surmises, and the adjacent-gap ratio
$r_n = \min(\delta_n,\delta_{n+1})/\max(\delta_n,\delta_{n+1})$ built from the raw gaps
$\delta_n = E_{n+1}-E_n$ against the large-matrix benchmarks
$\langle r \rangle_{\rm Poisson} = 0.386$, $\langle r\rangle_{\rm GOE}=0.5307$ and
$\langle r\rangle_{\rm GUE}=0.5996$.

In the time domain we use the unfolded amplitude
$Z(t) = \sum_n e^{-2\pi i \varepsilon_n t}$ and the connected spectral form factor:
\begin{equation}
  \frac{K_c(t)}{N} \;=\; \frac{\langle |Z(t)|^2\rangle - |\langle Z(t)\rangle|^2}{\langle N \rangle} \,,
  \label{eq:sff}
\end{equation}
where the subtraction removes the disorder-averaged density envelope. With time measured in units of the Heisenberg time $t_H$, the inverse of the mean level spacing, RMT predicts a ramp and a plateau at unity. The GOE ramp is the steeper of the two, lying a factor of about two above the GUE ramp at early times. The stronger level repulsion of GUE makes its spectrum more rigid, which suppresses the variance measured by the form factor. For two sectors $A$ and $B$ paired by disorder realization we also use the cross form factor:
\begin{equation}
  \frac{K_c^{AB}(t)}{N} \;=\;
  \mathrm{Re}\, \frac{\langle Z_A Z_B^*\rangle - \langle Z_A\rangle \langle Z_B\rangle^*}
  {\sqrt{\langle N_A\rangle \langle N_B\rangle}} \,,
  \label{eq:crosssff}
\end{equation}
which vanishes for independent sectors and reduces to the auto form factor for identical spectra.
Finally, writing $\delta Z_X$ for the connected part of the density-normalized amplitude
$N_X^{-1/2}\sum_{n \in X} e^{-2\pi i \varepsilon_n t}$, we form the correlation matrix:
\begin{equation}
  C_{AB} \;=\; \frac{\big\langle \mathrm{Re}[\delta Z_A \delta Z_B^*]\big\rangle}
  {\sqrt{\langle |\delta Z_A|^2\rangle \langle |\delta Z_B|^2\rangle}} \,,
  \label{eq:corrmatrix}
\end{equation}
averaged over realizations and over a band of unfolded times. Independent blocks give $C_{AB} \to 0$, while spectra related by an exact symmetry give $C_{AB} \to 1$.

\paragraph{Results.}
We diagonalized all eight sectors fully for $120$ disorder realizations at $L=8$.  Each twisted Hilbert space has dimension $3^8=6561$, and the multiplicity spaces range in dimension from $1093$ to
$3281$.

\begin{table}[H]
  \centering
  \begin{tabular}{lccl}
    \toprule
    sector & flux & $\langle r \rangle$ & class \\
    \midrule
    $([e],1)$ & $[e]$ & $0.5257 \pm 0.0010$ & GOE \\
    $([e],\sgn)$ & $[e]$ & $0.5262 \pm 0.0010$ & GOE \\
    $([e],2)$ & $[e]$ & $0.5289 \pm 0.0008$ & GOE \\
    $([t],+)$ & $[t]$ & $0.5298 \pm 0.0006$ & GOE \\
    $([t],-)$ & $[t]$ & $0.5288 \pm 0.0006$ & GOE \\
    $([c],1)$ & $[c]$ & $0.5294 \pm 0.0008$ & GOE \\
    $([c],\omega)$ & $[c]$ & $0.5956 \pm 0.0007$ & GUE \\
    $([c],\omega^2)$ & $[c]$ & $0.5956 \pm 0.0007$ & GUE \\
    \bottomrule
  \end{tabular}
  \caption{Disorder-averaged adjacent-gap ratio per tube algebra sector at $L=8$ over $120$
  realizations. The
  quoted uncertainty is the standard error of the mean over realizations. }
  \label{tab:S3r}
\end{table}

\begin{figure}[!htbp]
\centering
\includegraphics[width=0.99\textwidth]{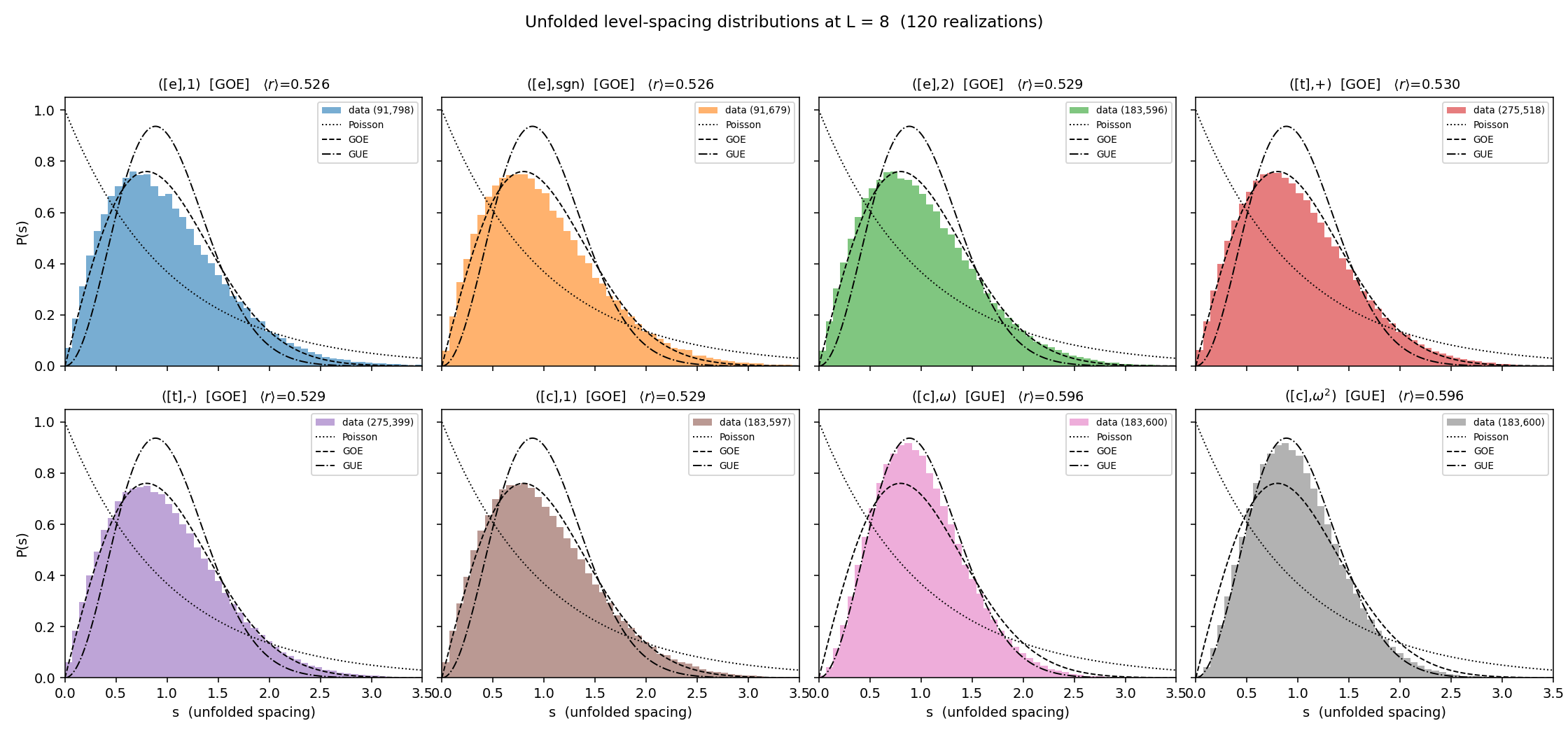}
\caption{Unfolded level-spacing distributions $P(s)$ for the eight $\DS$ sectors ($L=8$, $120$ realizations). Six sectors follow GOE, while the pair $([c],\omega)$, $([c],\omega^2)$, exchanged by time reversal $\Theta$, follows GUE.}
\label{fig:spacing}
\end{figure}

\begin{figure}[!htbp]
\centering
\includegraphics[width=0.70\textwidth]{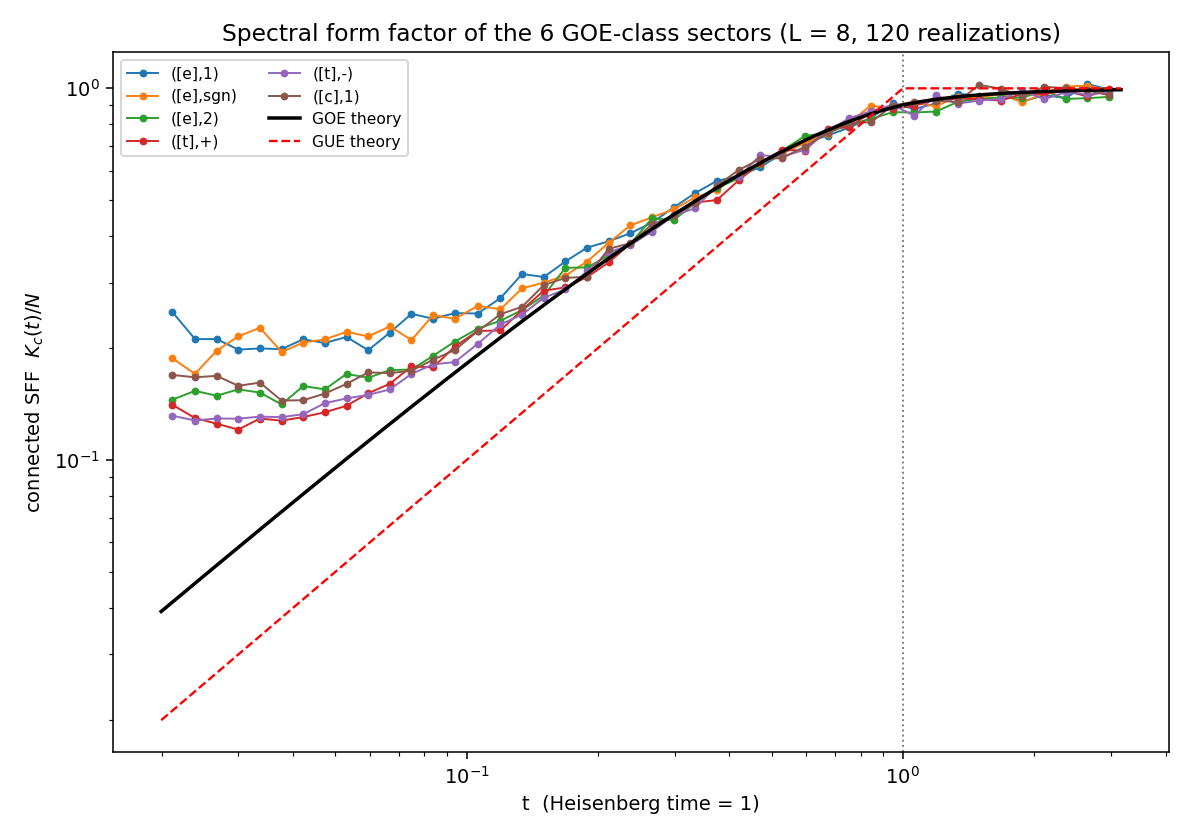}
\caption{Connected SFF of the six GOE-class sectors ($L=8$, $120$ realizations),
matching the GOE ramp$\to$plateau for $t\gtrsim0.2\,t_H$ (GUE shown dashed for contrast).}
\label{fig:sffGOE}
\end{figure}

\begin{figure}[!htbp]
\centering
\includegraphics[width=0.70\textwidth]{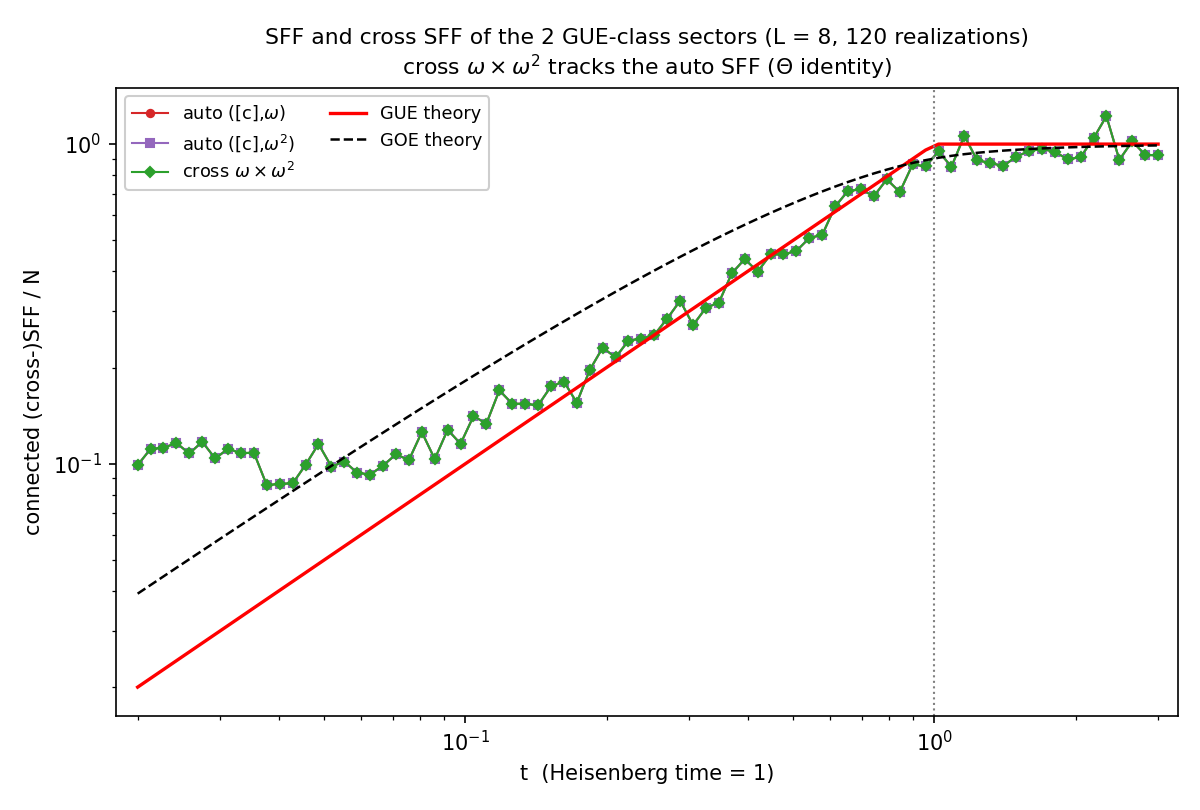}
\caption{Connected SFF of the two GUE-class sectors $([c],\omega)$ and $([c],\omega^2)$ ($L=8$, $120$ realizations). Both auto SFFs follow the GUE ramp for $t\gtrsim0.2\,t_H$, and the cross form factor ($\omega\times\omega^2$ in the legend) tracks the auto SFF, since time reversal $\Theta$ makes the two spectra coincide.}
\label{fig:sffGUE}
\end{figure}

\begin{figure}[!htbp]
\centering
\includegraphics[width=0.62\textwidth]{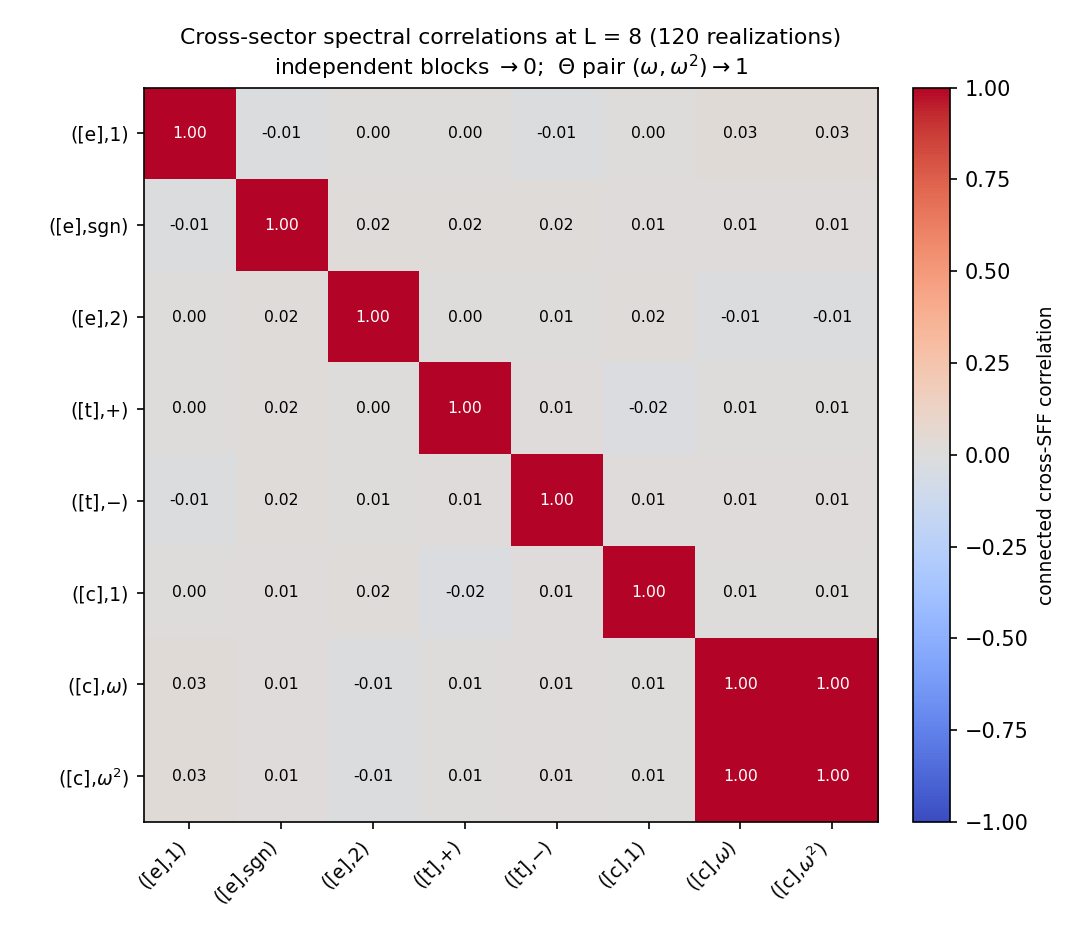}
\caption{Connected cross-sector correlation matrix $C_{AB}$ ($L=8$, $120$ realizations). Distinct sectors give $C_{AB}\approx 0$, while the pair $([c],\omega)$, $([c],\omega^2)$, exchanged by time reversal $\Theta$, gives $C_{AB}=1$. The diagonal is unity by construction.}
\label{fig:corr}
\end{figure}

All four diagnostics confirm the predictions of Subsection~\ref{sec:S3CRT}. The six fixed
sectors are GOE and the pair $([c],\omega)$, $([c],\omega^2)$ is GUE, both in the gap ratio
(table~\ref{tab:S3r})\footnote{The gap ratios in table~\ref{tab:S3r} lie below RMT benchmarks by $1.5$ to $5.7$ standard deviations. This does not affect the class assignment, since the GOE and GUE benchmarks differ by more than ten times the largest offset. A plausible source is the weak disorder, which breaks reflection only slightly. An approximate reflection symmetry almost splits each GOE sector into two independent blocks, lowering $\langle r\rangle$ towards the Poisson value.} and in the spacing distribution (Fig.~\ref{fig:spacing}). The same assignment holds in the time domain (Figs.~\ref{fig:sffGOE} and~\ref{fig:sffGUE}), where the
cross form factor of the exchanged pair coincides with their common auto form factor. The form factors follow RMT for $t\gtrsim0.2\,t_H$. At earlier times they lie above the ramp, most for the smallest blocks; this is the non-universal regime before the Thouless time. The
correlation matrix (see Fig.~\ref{fig:corr}) gives off-diagonal entries close to $0$ 
for every
distinct pair of sectors, with the sole exception of the exchanged pair at $C_{AB}=1$. The eight
$D(S_3)$ sectors are therefore consistent with independent random matrix blocks, up to the exact degeneracy forced
by time reversal which leaves seven statistically independent blocks. Since we diagonalize one representative twist per conjugacy class, this tests the predicted classes and the absence of correlations between blocks from different twisted Hilbert spaces. The identification of conjugate twists is exact by symmetry.

\subsection{Example: the Fibonacci category}
\label{sec:fibonacci-elaborated}

We now spell out the preceding abstract discussion in one of the simplest
non-invertible examples: the Fibonacci fusion category. It is where the first of the two new features listed in
Subsection~\ref{sec:statement} appears in its sharpest form. For a group, a tube algebra sector spans
several twisted Hilbert spaces only when their twists are conjugate, so that an invertible symmetry
relates them. Here, by contrast, the untwisted and the $\tau$-twisted Hilbert spaces share a sector
although no invertible symmetry relates them\footnote{One can gauge the $S_3$ symmetry of the previous mentioned lattice model to see similar features, and the relevant symmetry category becomes $\rm{Rep}(S_3)$.The new features appear in this model as well, with the independent sectors again given by sectors of $D(S_3)$, because $\rm{Rep}(S_3)$ and $\rm{Vec}(S_3)$ share the same Drinfeld center. See \cite{Jensen:2026cat} for details.}. If one first decomposes each twisted Hilbert space
separately, one obtains a seemingly natural list of five blocks. However, the full tube algebra
contains intertwiners between the two twisted Hilbert spaces, and these identify two of those
blocks as components of a single irreducible sector. The RMT sectors are therefore the four
irreducible representations of the tube algebra, equivalently the four simple objects of the Drinfeld center
\(Z({\rm Fib})\).

The Fibonacci category has two simple lines $\mathbf 1$ and $\tau$, with
\begin{equation}
  \tau \otimes \tau \simeq \mathbf 1 \oplus \tau \,.
  \label{eq:fibfusion}
\end{equation}
The line $\tau$ is self-dual, so orientation reversal produces no new simple object. The quantum
dimensions follow from $d_\tau^2 = 1 + d_\tau$,
\begin{equation}
  d_{\mathbf 1} = 1, \qquad d_\tau = \varphi := \tfrac{1+\sqrt 5}{2} \,.
  \label{eq:fibdims}
\end{equation}
The tube algebra acts on $\Htot = \Hil_{\mathbf 1} \oplus \Hil_\tau$. A basis of tube algebra elements is
\begin{equation}
  \cL^{d,c}_{b \leftarrow a} : \Hil_a \longrightarrow \Hil_b \,, \qquad a,b,c,d \in \{\mathbf 1,\tau\} \,,
  \label{eq:fibtube}
\end{equation}
where $a$ and $b$ record the incoming and outgoing twisted sectors. The labels $d$ and $c$ record
the junction data used to resolve the network of Fig.~\ref{fig:tube}. Products are obtained by
stacking cylinders and reducing with \eqref{eq:fibfusion} and the Fibonacci $F$-moves.

\paragraph{From five naive blocks to four sectors.}

Suppose first that we use only tube algebra elements preserving a chosen twisted sector. The
untwisted sector has one nontrivial internal operator $\cL^{\tau,\tau}_{\mathbf 1 \leftarrow \mathbf 1}$, the $\tau$
line wrapped on the spatial circle. Diagonalizing it splits $\Hil_{\mathbf 1}$ into the eigenspaces
$\Hil^{(1)}_{\mathbf 1}$ and $\Hil^{(2)}_{\mathbf 1}$, with eigenvalues $\varphi$ and $-1/\varphi$ respectively. The
$\tau$-twisted sector has two internal operators $\cL^{\tau,\mathbf 1}_{\tau \leftarrow \tau}$ and
$\cL^{\tau,\tau}_{\tau \leftarrow \tau}$, whose common representation theory splits $\Hil_\tau$ into
three pieces $\Hil^{(1)}_\tau$, $\Hil^{(2)}_\tau$ and $\Hil^{(3)}_\tau$.

These five blocks are not the tube algebra sectors, because the tube algebra also contains the
intertwiners with $a \neq b$ in \eqref{eq:fibtube}. The nontrivial intertwiner identifies
\begin{equation}
  \Hil^{(2)}_{\mathbf 1} \longleftrightarrow \Hil^{(2)}_\tau \,,
  \label{eq:fibglue}
\end{equation}
so the two are components of one irreducible module $\mu$, identified below as $(\tau,\tau)$, with
$W^\mu_{\mathbf 1} \simeq W^\mu_\tau \simeq \mathbb{C}$ and $V_\mu \simeq \mathbb{C}^2$. They share one multiplicity space, and by
\eqref{eq:Hblocks} the Hamiltonian on their direct sum is $1_2 \otimes H_\mu$. The five naive blocks therefore collapse to four tube algebra sectors,
\begin{equation}
  \Hil^{(1)}_{\mathbf 1}, \qquad \Hil^{(1)}_\tau, \qquad \Hil^{(3)}_\tau, \qquad
  \Hil^{(2)}_{\mathbf 1} \oplus \Hil^{(2)}_\tau \,,
  \label{eq:fibsectors}
\end{equation}
of which the last straddles both twisted Hilbert spaces, with one component in each.

\paragraph{Drinfeld-center description.}\label{sec:fibcenter}

The irreducible representations of the tube algebra are labeled by the simple objects of the
Drinfeld center. The Fibonacci category is modular, meaning that its braiding is non-degenerate. Its center is
therefore $\Fib \boxtimes \overline{\Fib}$. The four simple objects are $(\mathbf 1,\mathbf 1)$, $(\mathbf 1,\tau)$, $(\tau,\mathbf 1)$ and
$(\tau,\tau)$. Write $F(\mu)$ for the line of $\Fib$, or sum of such lines, that underlies the
center object $\mu$ once its braiding data are dropped. This map is the forgetful functor, and
for a modular category it is simply $F(c,d) = c \otimes d$. Its decomposition into simple lines
gives the multiplicities $\dim W^\mu_a$ directly:

\begin{center}
\begin{tabular}{cccc}
  \toprule
  $\mu$ & $\dim W^\mu_{\mathbf 1}$ & $\dim W^\mu_\tau$ & $F(\mu)$ \\
  \midrule
  $(\mathbf 1,\mathbf 1)$       & 1 & 0 & $\mathbf 1$ \\
  $(\mathbf 1,\tau)$    & 0 & 1 & $\tau$ \\
  $(\tau,\mathbf 1)$    & 0 & 1 & $\tau$ \\
  $(\tau,\tau)$ & 1 & 1 & $\mathbf 1 \oplus \tau$ \\
  \bottomrule
\end{tabular}
\end{center}

\noindent
Importantly, the object $(\tau,\tau)$ in the last row has support in both twisted Hilbert spaces,
because $\tau \otimes \tau = \mathbf 1 \oplus \tau$:
\begin{equation}
  \Hil_{\mathbf 1} \simeq \Mult_{(\mathbf 1,\mathbf 1)} \oplus \Mult_{(\tau,\tau)} \,,
  \qquad
  \Hil_\tau \simeq \Mult_{(\mathbf 1,\tau)} \oplus \Mult_{(\tau,\mathbf 1)} \oplus \Mult_{(\tau,\tau)} \,.
  \label{eq:fibdecomp}
\end{equation}
The two appearances of $\Mult_{(\tau,\tau)}$ are the center version of the intertwiner statement
\eqref{eq:fibglue}. In terms of blocks,
\begin{equation}
  H = H_{(\mathbf 1,\mathbf 1)} \oplus H_{(\mathbf 1,\tau)} \oplus H_{(\tau,\mathbf 1)} \oplus 1_2 \otimes H_{(\tau,\tau)} \,.
  \label{eq:fibblocks}
\end{equation}
Specializing \eqref{eq:corr} to $\mu = \nu = (\tau,\tau)$ and $a=\mathbf 1$, $b=\tau$ leaves the
bare kernel $\mathcal K_{(\tau,\tau)}(E,E')$. Concretely, the
$-1/\varphi$ eigenspace of the untwisted chain and the $(\tau,\tau)$ block of the $\tau$-twisted
chain should have the same spectrum, level by level. { We confirm this in Subsection~\ref{sec:fibnum}, together with the coincidence of the spectra of $(\mathbf 1,\tau)$ and $(\tau,\mathbf 1)$ at opposite lattice momenta that Subsection~\ref{sec:fibcrt} predicts.}

\subsubsection{Antiunitary symmetries and the predicted classes}
\label{sec:fibcrt}

The clean golden chain (defined below in Subsection~\ref{sec:fibnum}) keeps two antiunitary symmetries, and they act
differently on the center labels. The first is complex conjugation $\Theta$ in the fusion path basis,
under which the Hamiltonian is real because the $F$-symbols are. The second is $\mathcal{R}\Theta$, with $\mathcal{R}$ the lattice reflection, which the clean chain also preserves. {This holds on the untwisted chain. The $\tau$-twisted chain will be described in detail in~\cite{Jensen:2026cat}.} Since $\tau$ is self-dual, charge
conjugation plays no role for either.

Consider $\Theta$ first. It reverses time but not space, so it conjugates the topological spin of every
center object and reverses the lattice momentum.\footnote{The topological spin of a center object
is its eigenvalue under the Dehn twist, the $2\pi$ rotation of the spatial circle.} The spins of $(\mathbf 1,\mathbf 1)$ and $(\tau,\tau)$ are real
and those objects are fixed. The spins of $(\mathbf 1,\tau)$ and $(\tau,\mathbf 1)$ are $e^{-4\pi i/5}$ and $e^{4\pi i/5}$ respectively,
complex conjugates of one another, so $\Theta$ exchanges them:
\begin{equation}
  \Theta : (\mathbf 1,\tau) \longleftrightarrow (\tau,\mathbf 1) \,, \qquad k \longmapsto -k \,.
  \label{eq:fibpair}
\end{equation}
Now consider $\mathcal{R}\Theta$. Reflection reverses spatial orientation, which inverts the Dehn twist, and
antilinearity conjugates its eigenvalue. The two reversals cancel on the spin, so $\mathcal{R}\Theta$ fixes all
four center objects and preserves momentum:
\begin{equation}
  \mathcal{R}\Theta : (c,d) \longmapsto (c,d) \,, \qquad k \longmapsto k \,, \qquad (\mathcal{R}\Theta)^2 = +1 \,.
  \label{eq:fibcrt}
\end{equation}
Within a fixed momentum sector, every tube algebra sector therefore carries an antiunitary that
squares to $+1$, and all four are GOE. The operator $\Theta$ adds an exact relation on top: the sector
$(\mathbf 1,\tau)$ at momentum $k$ and the sector $(\tau,\mathbf 1)$ at momentum $-k$ share a spectrum, level by
level. The predictions are therefore:
\begin{itemize}
  \item All four sectors $(\mathbf 1,\mathbf 1)$, $(\tau,\tau)$, $(\mathbf 1,\tau)$ and $(\tau,\mathbf 1)$ are GOE at fixed
        momentum.
  \item The spectra of $(\mathbf 1,\tau)$ at $k$ and of $(\tau,\mathbf 1)$ at $-k$ coincide.
\end{itemize}
Note that the contrast with $S_3$ is entirely a matter of which antiunitary the lattice keeps. The
disorder of Subsection~\ref{sec:S3numerics} removed reflection, leaving only $\Theta$, and the pair it
exchanged was GUE. Here reflection survives and $\mathcal{R}\Theta$ restores GOE, exactly as anticipated in
Subsection~\ref{sec:S3numerics}. The numerics below reach all five blocks, the two GOE sectors of the untwisted Hilbert space and
the three sectors of $\Hil_\tau$, including the mixed pair.

\subsubsection{Numerical results}\label{sec:fibnum}
We can substantiate the above discussion by numerically calculating the  level spacing distribution for a lattice Hamiltonian exhibiting a Fibonacci symmetry. An integrable model with Fibonacci symmetry is the Fibonacci golden chain \cite{Feiguin:2006ydp}, where we place a Fibonacci anyon $\tau$ on each lattice site $i$ and an anyon $a_i \in \{\mathbf 1,\tau\}$ on each link between the sites $i,i+1$, such that $a_{i+1} \in a_i \otimes \tau$. Any such configuration $a_0, a_1, \cdots, a_L=a_0$ is called a fusion path. The dimension of the untwisted Hilbert space (with periodic boundary conditions) for $L$ sites is the Lucas number $F_{L-1}+F_{L+1}$ where $F_L$ is the Fibonacci number. The Fibonacci numbers $F_{L-1},F_{L+1}$ are the dimensions of the sectors $(\mathbf 1,\mathbf 1)$ and $(\tau,\tau)$, respectively, in the untwisted Hilbert space, which we study below. A segment of the fusion path is depicted below. Note that the lines in the Fibonacci case are unoriented.

\begin{center}
\begin{tikzpicture}[scale=0.95, every node/.style={font=\small}]
   \draw[thick] (-1.7, 0) -- (5.7, 0);
   \node[left] at (-1.7, 0) {$\cdots$};
   \node[right] at (5.7, 0) {$\cdots$};
   \foreach \x in {-1, 1, 3, 5} {
      \fill (\x, 0) circle (0.07);
      \draw[thick] (\x, 0) -- (\x, 1.1);
   }
   \node[above] at (-1, 1.15) {$\tau$};
   \node[above] at (1, 1.15) {$\tau$};
   \node[above] at (3, 1.15) {$\tau$};
   \node[above] at (5, 1.15) {$\tau$};
   \node[below] at (-1.5, -0.05) {$a_{i-2}$};
   \node[below] at (0, -0.05) {$a_{i-1}$};
   \node[below] at (2, -0.05) {$a_i$};
   \node[below] at (4, -0.05) {$a_{i+1}$};
   \node[below] at (5.5, -0.05) {$a_{i+2}$};
\end{tikzpicture}
\end{center}

The integrable Hamiltonian and its perturbation can be written in terms of a projector $\Pi^{(1)}_{i}$ which fuses two adjacent $\tau$ anyons at sites $i,i+1$ to the identity. Schematically the matrix elements of the projector in the fusion path basis are
\begin{center}
\begin{tikzpicture}[scale=0.8, every node/.style={font=\small}]
   \draw[thick] (-0.7, -1.6) -- (5.7, -1.6);
   \node[left] at (-0.7, -1.6) {$\cdots$};
   \node[right] at (5.7, -1.6) {$\cdots$};
   \fill (1, -1.6) circle (0.07);
   \fill (3, -1.6) circle (0.07);
   \node[below] at (0, -1.65) {$a_{i-1}$};
   \node[below] at (2, -1.65) {$a_i$};
   \node[below] at (4, -1.65) {$a_{i+1}$};
   \draw[thick] (1, -1.6) -- (2, -0.6);
   \draw[thick] (3, -1.6) -- (2, -0.6);
   \fill (2, -0.6) circle (0.07);
   \node[left] at (1.45, -1.1) {$\tau$};
   \node[right] at (2.55, -1.1) {$\tau$};
   \draw[thick] (2, -0.6) -- (2, 0.6);
   \node[right] at (2.05, 0) {$m=\mathbf 1$};
   \fill (2, 0.6) circle (0.07);
   \draw[thick] (1, 1.6) -- (2, 0.6);
   \draw[thick] (3, 1.6) -- (2, 0.6);
   \node[left] at (1.45, 1.1) {$\tau$};
   \node[right] at (2.55, 1.1) {$\tau$};
   \draw[thick] (-0.7, 1.6) -- (5.7, 1.6);
   \node[left] at (-0.7, 1.6) {$\cdots$};
   \node[right] at (5.7, 1.6) {$\cdots$};
   \fill (1, 1.6) circle (0.07);
   \fill (3, 1.6) circle (0.07);
   \node[above] at (0, 1.65) {$a_{i-1}$};
   \node[above] at (2, 1.65) {$a'_i$};
   \node[above] at (4, 1.65) {$a_{i+1}$};
\end{tikzpicture}
\end{center}

These matrix elements can be calculated by { F-moves} to go back to the original fusion path basis (with $\tau$ anyons { connecting} the bra and ket at each site). The result can be written in terms of the F-symbols of the Fibonacci category. Concretely, in the basis of three adjacent link variables $(a_{i-1},a_i,a_{i+1})$ the projector is diagonal in $a_{i-1}, a_{i+1}$ (and all other $a_{j \neq i}$) but not in $a_i$,
\begin{equation}
   \Pi^{(1)}_i(a'_i, a_i\,;\, a_{i-1},\, a_{i+1}) \;=\;
   \begin{cases}
      \delta_{a_i,\tau}\,\delta_{a'_i,\tau}
         & (a_{i-1},a_{i+1})=(\mathbf 1,\mathbf 1)\\[2pt]
      0
         & (a_{i-1},a_{i+1})\in\{(\mathbf 1,\tau),(\tau,\mathbf 1)\}\\[2pt]
      \bigl(F^{\tau\tau\tau}_{\tau}\bigr)_{a'_i,\mathbf 1}\,
      \bigl(F^{\tau\tau\tau}_{\tau}\bigr)_{a_i,\mathbf 1}
         & (a_{i-1},a_{i+1})=(\tau,\tau)
   \end{cases}
   \label{eq:pi-kernel}
\end{equation}
 In terms of these projectors the integrable Hamiltonian is  
\begin{equation}
    H_0=-  \sum_i \Pi^{(1)}_{i} \,,  
\end{equation}
and the integrability breaking symmetric perturbation is
\begin{equation}
    H_1= \frac{\lambda}{2} \sum_i \left\{\Pi^{(1)}_{i}\, , \Pi^{(1)}_{i+1}\right\}\,,
\end{equation}
which involves next-nearest-neighbor interaction. For large enough perturbation we expect to obtain chaotic Hamiltonians with Fibonacci symmetry. We choose $\lambda =1.25$  which can be seen to be chaotic in exhibiting Wigner--Dyson statistics in symmetry sectors as we now show.

For the untwisted sector one can construct the conserved charge $\cL^{\tau,\tau}_{\mathbf 1\leftarrow \mathbf 1}$, corresponding to an insertion of horizontal $\tau$ line, and verify numerically that it commutes with the Hamiltonian to machine precision. The resulting spectral statistics {are} presented for the two eigenspaces of the symmetry charge. In the $\tau$-twisted Hilbert space the three blocks are resolved by their topological spin, the eigenvalue of the $2\pi$ rotation of the spatial circle, implemented on the lattice by the $L$-th power of the translation of the twisted chain{~\cite{Jensen:2026cat}}. We further resolve the eigenstates of the Hamiltonian by lattice momentum and drop the self-conjugate momentum blocks, in which reflection acts as an additional unitary symmetry. {Since $\Theta$ maps $k$ to $-k$ within $(\mathbf 1,\mathbf 1)$ and within $(\tau,\tau)$, we keep one block from each pair $\pm k$ for these sectors. For $(\mathbf 1,\tau)$ and $(\tau,\mathbf 1)$, which $\Theta$ exchanges together with $k\to-k$, we keep all momenta; these are shifted by the fractional spin, so none is self-conjugate.} To match with RMT, each block is truncated near its edges, keeping $90\%$ of the eigenstates. Unfolding the spectrum follows by fitting the density of states to a degree 7 polynomial.

\begin{figure}[t]
    \centering
    \includegraphics[width=1\linewidth]{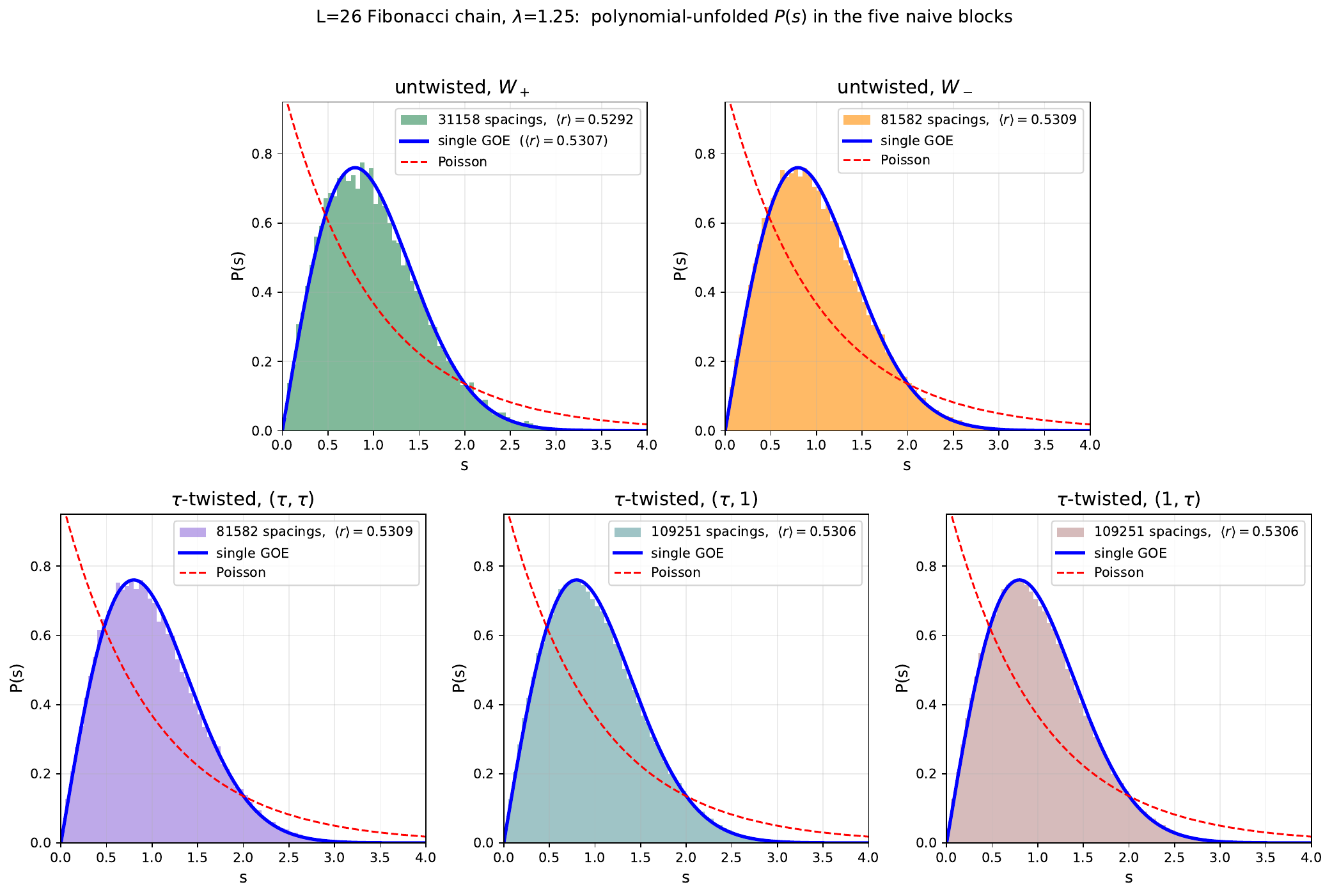}
    \caption{Symmetry-resolved level spacing statistics of the perturbed Fibonacci chain at $\lambda=1.25$ for a chain of $26$ anyons, one panel for each of the five naive blocks. Top row: the two blocks of the untwisted Hilbert space, $W_+=\Hil^{(1)}_{\mathbf 1}$ and $W_-=\Hil^{(2)}_{\mathbf 1}$, the eigenspaces of $\cL^{\tau,\tau}_{\mathbf 1\leftarrow \mathbf 1}$ with eigenvalues $\varphi$ and $-1/\varphi$, which carry the center labels $(\mathbf 1,\mathbf 1)$ and $(\tau,\tau)$. Bottom row: the three blocks of the $\tau$-twisted Hilbert space, labeled by their center labels. The block $(\tau,\tau)$ is the twisted component of the same tube algebra sector as $W_-$ and is isospectral to it, while $(\tau,\mathbf 1)$ and $(\mathbf 1,\tau)$ are exchanged by an antiunitary symmetry and thus have identical spectra. Spectra are resolved by lattice momentum, with self-conjugate momentum blocks dropped, each block is unfolded separately with a degree-7 polynomial after discarding $5\%$ of the levels at each edge, and the spacings are pooled. All five blocks follow the GOE Wigner surmise (solid) rather than Poisson (dashed). The data are taken from an upcoming paper \cite{Jensen:2026cat} which explains the tube-algebra resolution on the lattice and discusses further symmetry-resolved signatures of quantum chaos.}
    \label{numerics_fib}
\end{figure}

The results for a chain of 26 anyons are presented in Fig.~\ref{numerics_fib}, for the two blocks of the untwisted Hilbert space and the three blocks of the $\tau$-twisted Hilbert space. A conventional diagnostic for the emergence of RMT is the adjacent gap ratio $r$ { of Subsection~\ref{sec:S3numerics}, which is insensitive to unfolding}.  The five blocks give $\langle r\rangle=0.5292$ for $W_+$, $0.5309$ for $W_-$, $0.5309$ for the twisted $(\tau,\tau)$, and $0.5306$ for each of $(\tau,\mathbf 1)$ and $(\mathbf 1,\tau)$, all close to $\langle r\rangle_{\rm GOE}=0.5307$. Additionally, the adjacent level statistics plotted in Fig.~\ref{numerics_fib} accurately {follow} the Wigner-Dyson expectations. {Beyond statistics, the $W_-$ block and the twisted $(\tau,\tau)$ block have the same spectrum level by level, to within $2\times10^{-14}$, and so do $(\mathbf 1,\tau)$ and $(\tau,\mathbf 1)$ at opposite momenta~\cite{Jensen:2026cat}.}\footnote{Further spectral quantities will be discussed in \cite{Jensen:2026cat}, in particular any cross-sector spectral form factor fluctuates about zero, showing the independence of tube algebra sectors. }

In summary, in both lattice models every resolved tube algebra sector shows the Wigner--Dyson statistics predicted by its antiunitary symmetries, and the spectra that the tube algebra or an antiunitary symmetry relates coincide level by level.

\section{Holographic CFTs with categorical symmetries}
\label{S:holoCat}

Our discussion thus far has been applicable to any spatially extended system. We now specialize to two-dimensional chaotic CFTs of large central charge. The spectral statistics of scaling dimensions in two-dimensional CFTs have been discussed in \cite{Haehl:2023tkr,Haehl:2023xys,Haehl:2023mhf,DiUbaldo:2023qli,Boruch:2025ilr}. 
The best-studied candidates for chaotic CFTs are those with a weakly coupled gravity dual. For such theories the two-boundary torus wormhole encoding spectral correlations has been calculated in~\cite{Cotler:2020ugk}. Of particular interest are the {\it modular corrections}: a set of cross-correlations between symmetry sectors (in this case sectors of fixed spin) mandated by modular invariance. 

In this section we show that this story generalizes to holographic CFTs with a general categorical symmetry. In holographic models the categorical symmetry is naturally represented by the SymTFT, a bulk topological field theory realizing the symmetry content of the boundary theory, and we refer to this construction as topological holography. It is natural to conjecture that the bulk description of the full theory consists of gravity coupled to the corresponding SymTFT, so that the torus wormhole of~\cite{Cotler:2020ugk} is decorated by the wormhole of the SymTFT. We regard the emergence of the correlations predicted by categorical RMT (Section~\ref{sec:categorical-RMT}) as evidence for this conjecture.

Concretely, we establish three results, which hold for a general categorical symmetry $\C$ and which we illustrate with $\Rep D(S_3)$ and doubled Fibonacci. First, we compute the simplest wormhole, which we call the ``seed'': the torus wormhole of~\cite{Cotler:2020ugk}, decorated by the SymTFT and summed over relative Dehn twists. It reproduces the correlations of categorical RMT sector by sector, including the exact degeneracies, albeit with the GUE coefficient in every sector (Subsection~\ref{sec:seed}). Second, completing the seed into a modular covariant amplitude adds the image of the central element $-\mathbf 1\in SL(2,\Z)$, which implements $\CRT$. This image doubles the coefficient of every self-dual sector, which is the GOE value when $\CRT^2=+1$, and makes the spectra of each pair of conjugate sectors coincide exactly. Orientation reversal of the bulk, when it is a symmetry, similarly pairs the statistics of sectors at opposite spins (see Subsection~\ref{sec:crt}). Third, the remaining modular images, weighted by vector-valued Kloosterman sums, correlate sectors that the seed leaves independent (see Subsection~\ref{sec:fourier}). Modular invariance is therefore incompatible with exactly independent symmetry sectors, and we expect categorical RMT to apply where these corrections are suppressed. Throughout this section we count Virasoro primaries.

\paragraph{Topological holography.} A categorical symmetry $\C$ is encoded in a three-dimensional topological field theory, the SymTFT~\cite{Gaiotto:2020iye,Apruzzi:2021nmk,Freed:2022qnc,Kaidi:2022cpf,Ji:2019jhk,Moradi:2022lqp}. Its bulk lines, or anyons, are the simple objects of the Drinfeld center $Z(\C)$, and they label the tube algebra sectors $\mu$ of Subsection~\ref{sec:lines}. In the sandwich construction the SymTFT lives on an interval, with the physical boundary condition at one end and, at the other, the topological boundary on which the lines of $\C$ live (see e.g.~\cite{Freed:2022qnc,Kaidi:2022cpf}).

\definecolor{figorange}{HTML}{D55E00}
\definecolor{figblue}{HTML}{0072B2}
\definecolor{figgrey}{HTML}{555555}
\colorlet{figfill}{figblue!6}
\begin{figure}[t]
  \centering
  \begin{minipage}[b]{0.42\textwidth}
    \centering
    \begin{tikzpicture}[x=1bp,y=1bp]
      \node[anchor=south west,inner sep=0] at (0,0) {\includegraphics{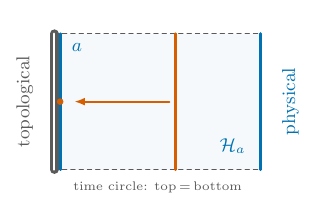}};
      \node[text=figorange] at (92.1,73.6) {\footnotesize $\mu$};
      \node at (59.6,55.4) {\footnotesize $W^\mu_a$};
    \end{tikzpicture}\\[2pt]
    \footnotesize (a) The sandwich. Fusing the anyon $\mu$ into the topological boundary along $a$
    produces the junction space $W^\mu_a$ of \eqref{eq:forget}.
  \end{minipage}\hfill
 \begin{minipage}[b]{0.50\textwidth}
    \centering
    \begin{tikzpicture}[x=1cm,y=1cm]
      \fill[figfill] (0,0) ellipse (1.84 and 0.88);
      \draw[figblue,line width=1.3pt] (0,0) ellipse (2.0 and 1.0);
      \draw[figgrey,double=white,double distance=1.1pt,line width=0.45pt] (0,0) ellipse (1.84 and 0.88);
      \fill[white] (-0.78,0.04) .. controls (-0.3,-0.26) and (0.3,-0.26) .. (0.78,0.04)
        .. controls (0.3,0.2) and (-0.3,0.2) .. (-0.78,0.04);
      \draw[figgrey] (-0.9,0.1) .. controls (-0.35,-0.3) and (0.35,-0.3) .. (0.9,0.1);
      \draw[figgrey] (-0.7,-0.04) .. controls (-0.25,0.2) and (0.25,0.2) .. (0.7,-0.04);
      \draw[figgrey,thick] (0,-0.2) arc (90:-90:0.12 and 0.33);  
      \draw[figgrey,thick,dashed] (0,-0.86) arc (270:90:0.12 and 0.33); \node[text=figgrey,anchor=west] (ctr) at (2.15,-0.75) {\footnotesize contractible};
      \draw[->,figgrey] (ctr.west) -- (0.16,-0.58);
      \node at (0,0.62) {\footnotesize thermal AdS};
      \node[text=figblue] at (0,-1.3) {\footnotesize asymptotic boundary, $Z_a(\tau)$};
    \end{tikzpicture}\\[2pt]
    \footnotesize (b) A one-boundary geometry, the solid torus of thermal AdS. The dashed spatial
    circle is contractible. The SymTFT fills it inside the topological boundary and prepares the vacuum
    state $|1\rangle$. Its modular images prepare $\varrho(\gamma)^{-1}|1\rangle$.
  \end{minipage}\\[10pt]
  \begin{minipage}{0.9\textwidth}
    \centering
    \begin{tikzpicture}[x=1cm,y=1cm]
      \fill[figfill] plot[domain=-2.98:2.98,samples=60] (\x,{0.32*cosh(\x/1.65)})
        -- plot[domain=2.98:-2.98,samples=60] (\x,{-0.32*cosh(\x/1.65)}) -- cycle;
      \draw[figgrey] plot[domain=-3.2:3.2,samples=60] (\x,{0.32*cosh(\x/1.65)});
      \draw[figgrey] plot[domain=-3.2:3.2,samples=60] (\x,{-0.32*cosh(\x/1.65)});
      \draw[figorange,thick] (0,-0.32) arc (-90:90:0.09 and 0.32);
      \draw[figorange,thick,dashed] (0,0.32) arc (90:270:0.09 and 0.32);
      \node at (0,1.25) {\footnotesize $\sum_\mu |\mu\rangle\langle\mu|$};
      \draw[->,figgrey] (0,1.05) -- (0,0.42);
      \foreach \s in {-1,1} {
        \draw[figgrey,double=white,double distance=1.1pt,line width=0.45pt] (\s*2.98,0) ellipse (0.27 and 1.0);
        \draw[figblue,line width=1.3pt] (\s*3.2,0) ellipse (0.3 and 1.13);
      }
      \node at (-1.75,0) {\footnotesize $\dim W^\mu_a$};
      \node at (1.75,0) {\footnotesize $\dim W^\mu_b$};
      \node at (-1.9,1.05) {\footnotesize $M$};
      \node[anchor=north] at (-3.2,-1.25) {\footnotesize $T^2_{\rm out},\ Z_a(\tau_1)$};
      \node[anchor=north] at (3.2,-1.25) {\footnotesize $\overline{T^2_{\rm in}},\ Z_b(-\bar\tau_2)^*$};
    \end{tikzpicture}\\[2pt]
    \footnotesize (c) The seed wormhole. The SymTFT fills the cobordism $M = T^2 \times I$ between the two
    topological boundaries, and its kernel is $K = \sum_\mu |\mu\rangle\langle\mu| = \one$.
    Each end contributes the multiplicity of $\mu$ in its twisted sector. The bar on $T^2_{\rm in}$
    marks its reversed orientation, which conjugates its
    partition function.
  \end{minipage}
  \caption{From the sandwich to the seed wormhole. In panel (a) the horizontal direction is the
  SymTFT interval and the vertical direction is Euclidean time, with its ends identified. In all three
  panels the gray double line is the topological boundary. In panels (b) and (c) the shaded region
  inside it carries the SymTFT, and only gravity lies between it and the asymptotic boundary. The heavy
  blue line is the physical boundary in panel (a) and the asymptotic boundary in panels (b) and (c).}
  \label{fig:sandwich}
\end{figure}

The spaces $W_a^\mu$ of Subsection~\ref{sec:lines} have a direct bulk interpretation, which follows from the forgetful functor $F$ of Subsection~\ref{sec:fibcenter} describing how an anyon decomposes into lines on the topological boundary:
\begin{equation}\label{eq:forget}
F(\mu)\;\simeq\;\bigoplus_{a\in{\rm Irr}(\C)}\dim\big(W_a^\mu\big)\,a\,,\qquad W_a^\mu\;=\;\Hom_\C\big(a,F(\mu)\big)\,.
\end{equation}
The space $W_a^\mu$ is therefore the space of junctions at which $\mu$ ends on the line $a$ (Fig.~\ref{fig:sandwich}). Each twisted partition function then decomposes into sectors,
\begin{equation}\label{eq:Za}
Z_a\;=\;\sum_\mu \dim\big(W_a^\mu\big)\,z_\mu\,,
\end{equation}
where $z_\mu$ is the partition function on $\Mult_\mu$, so that $Z_\mu=\dim(V_\mu)\,z_\mu$ in the notation of~\eqref{eq:Zsector}. This is the decomposition~\eqref{eq:Wdecomp}, written for partition functions.

On a torus the SymTFT has a finite-dimensional state space $\Htop(T^2)$, whose basis state $|\mu\rangle$ is prepared by a solid torus carrying the anyon $\mu$ along its core. The mapping class group $SL(2,\Z)$ acts on this space through a unitary representation $\vr$,\footnote{Since a Drinfeld center has vanishing chiral central charge, $\vr$ carries no framing phase.} in which the Dehn twist $\hat T$ is diagonal,
\begin{equation}\label{eq:theta}
\vr(\hat T)\,|\mu\rangle=\theta_\mu\,|\mu\rangle\,,\qquad\theta_\mu=e^{2\pi is_\mu}\,.
\end{equation}
The topological spin $\theta_\mu$ determines the spins in the sector $\mu$, which lie in $s_\mu+\Z$. For our two examples the topological spins are
\begin{equation}\label{eq:spins}
\theta_{([h],R)}=\frac{\chi_R(h)}{\dim R},\qquad\theta_{(\tau,\mathbf 1)}=\overline{\theta_{(\mathbf 1,\tau)}}=e^{4\pi i/5},\qquad\theta_{(\mathbf 1,\mathbf 1)}=\theta_{(\tau,\tau)}=1,
\end{equation}
for $\Rep D(S_3)$ and doubled Fibonacci, respectively.\footnote{In the Fibonacci labels, $\tau$ is the nontrivial simple line of the Fibonacci category, not the modular parameter of the torus.}

\paragraph{The proposal.} We propose to combine gravitational and topological holography by placing the sandwich inside the bulk:

\begin{tcolorbox}[
  colback=gray!10,
  colframe=blue,
  boxrule=0.6pt,
  arc=2pt,
  left=6pt,
  right=6pt,
  top=4pt,
  bottom=4pt]
A holographic CFT with categorical symmetry $\C$ is dual to three-dimensional gravity coupled to the SymTFT of $\C$. The sandwich sits inside the bulk, with its physical boundary at the asymptotic boundary and its topological boundary in the interior. A one-boundary amplitude is then a gravitational solid torus filled with the SymTFT. A two-boundary amplitude is a torus wormhole of~\cite{Cotler:2020ugk}, with the SymTFT filling the region between its two topological boundaries. In each geometry the gravitational and topological factors multiply.
\end{tcolorbox}

As Fig.~\ref{fig:sandwich} illustrates, the geometry now closes off the SymTFT and gravity takes over the role of the physical boundary. On thermal AdS the SymTFT prepares the vacuum state $|1\rangle$, whereas in a wormhole an anyon label propagates from one boundary to the other, correlating the two twisted sectors.

\subsection{The seed wormhole}\label{sec:seed}

In this subsection we compute the seed and show that it reproduces the correlations of categorical RMT sector by sector. Before turning to the calculation, we first clarify an orientation subtlety that arises in the wormhole geometry.
 
A wormhole has two boundaries, with moduli $\tau_1$ and $\tau_2$, on which the bulk induces opposite orientations. Following~\cite{Cotler:2020ugk}, we flip the orientation of boundary 2, so that both boundaries carry the same orientation and each enters $\langle Z_a(\tau_1)Z_b(\tau_2)\rangle$ as an ordinary partition function. Since boundary 2 would have modulus $-\bar\tau_2$ in the orientation induced from the bulk, the identity
\begin{equation}\label{eq:conj}
Z(\tau)=Z(-\bar\tau)^*\,,
\end{equation}
then gives $\langle Z_a(\tau_1)Z_b(\tau_2)\rangle=\langle Z_a(\tau_1)Z_b(-\bar\tau_2)^*\rangle$, which is the cross form factor~\eqref{eq:crosssff} in the bulk orientation.

We now proceed to calculate  $\langle Z_a(\tau_1)Z_b(\tau_2)\rangle$, which consists of a topological factor and a gravitational factor.

\paragraph{The topological factor.}
The simplest wormhole has the topology $T^2\times I$, and the SymTFT assigns to it a kernel $K$ pairing the state spaces $\Htop(T^2)$ at its two ends. Since $T^2\times I$ is a product, this kernel is the identity,
\begin{equation}\label{eq:Kdiag}
K=\sum_\mu|\mu\rangle\langle\mu|=\one.
\end{equation}
Note that \eqref{eq:Kdiag} does not describe an anyon line stretched across the wormhole. Rather, it states that the anyon label is conserved. Since the anyon $\mu$ meets the topological boundary at each end through the junction space $W_a^\mu$, the sector $\mu$ contributes the topological factor $\dim W_a^\mu\dim W_b^\mu$ for twists $a$ and $b$.

\paragraph{The gravitational factor.}
For a single geometry, the gravitational factor is that of~\cite{Cotler:2020ugk},
\begin{equation}\label{eq:Zgrav}
\Zgrav(\tau_1,\tau_2)
= \frac{1}{2\pi^2}Z_0(\tau_1) Z_0(\tau_{2})\frac{\text{Im}(\tau_{1}) \text{Im}(\tau_2)}{|\tau_1+\tau_2|^2},
\end{equation}
where $Z_0(\tau)=\frac{1}{\sqrt{{\rm Im}\,\tau}\,|\eta(\tau)|^2}$ takes the form of the non-compact boson partition function. The function~\eqref{eq:Zgrav}  satisfies simultaneous modular invariance:
\begin{equation}\label{eq:gravinv}
\Zgrav(\gamma\!\cdot\!\tau_1,\tilde\gamma\!\cdot\!\tau_2)=\Zgrav(\tau_1,\tau_2)\,,\qquad\gamma\in SL(2,\Z)\,,
\end{equation}
where $\tilde\gamma$ denotes $\gamma$ with the signs of its off-diagonal entries reversed, which accounts for the flipped orientation of boundary 2.

On the torus wormhole, the  gravitational factor should be summed over relative Dehn twists, which are relative $\hat T^{\ell}$ transformations of the boundaries \cite{Cotler:2020ugk}. In the presence of topological kernel $K$,  these twists  multiply the sector $\mu$ by $\vr(\hat T)^{-\ell}=\theta_{\mu}^{-\ell}$. Summing over all such twists gives the seed in the sector $\mu$,
\begin{equation}\label{eq:seedGamma}
Z_\mu^{\rm seed}(\tau_1,\tau_2)=\sum_{\ell\in\Z}\theta_\mu^{-\ell}\,\Zgrav(\tau_1,\tau_2+\ell)\,.
\end{equation}
The phases select exactly the spins in $s_\mu+\Z$ on both boundaries. The connected seed amplitude in Fig.~\ref{fig:sandwich}(c) is then
\begin{equation}\label{eq:seedamp}
\big\langle Z_a(\tau_1)\,Z_b(\tau_2)\big\rangle_{\rm seed}=\sum_\mu\dim\big(W_a^\mu\big)\dim\big(W_b^\mu\big)\,Z_\mu^{\rm seed}(\tau_1,\tau_2)\,.
\end{equation}

\paragraph{Comparison with categorical RMT.}
The seed alone reproduces the structure of categorical RMT, with correlations diagonal in spin and anyon labels, and a ramp in each sector separately. Let us review how this is achieved in \cite{Cotler:2020ugk}. We first  get rid of the descendant contributions by multiplying \eqref{eq:Zgrav} with $\prod_{n \geq 1}|1-q_1^n|^2|1-q_2^n|^2$ to obtain the primary-counting gravitational factor 
\begin{align}
 Z^P_{\rm grav}(\tau_1,\tau_2)=\frac{1}{2\pi^2}\frac{|q_1q_2|^{-\frac{1}{12}}}{\sqrt{\text{Im}(\tau_{1}) \text{Im}(\tau_2)}}\frac{\text{Im}(\tau_{1}) \text{Im}(\tau_2)}{|\tau_1+\tau_2|^2}.
 \label{equ:CJ_primary_sec3}
\end{align}
Then replacing $Z_{\mathrm{grav}}$ in \eqref{eq:seedGamma} by \eqref{equ:CJ_primary_sec3}, and fixing the spins $s_i$ by  Fourier transforming $x_i=\mathrm{Re}\tau_i$ at fixed $y_i={\rm Im}\,\tau_i$, the connected seed amplitude \eqref{eq:seedamp} becomes
\begin{align}
    \delta_{s_1s_2} \sum_{\mu\,:\,s_1\in s_\mu+\Z}  \dim\big(W_a^\mu\big)\dim\big(W_b^\mu\big)\, \frac{1}{2\pi} \frac{\sqrt{y_1 y_2}}{y_1+y_2}e^{-2\pi(|s_1|-\frac1{12})(y_1+y_2)}\,.
    \label{equ:RMT_result_cate}
\end{align}
The result \eqref{equ:RMT_result_cate} reproduces the Categorical RMT prediction \eqref{eq:corr} in three respects:

\begin{itemize}
\item Distinct $\mu$-sectors are uncorrelated.
\item Each sector carries the weight $\dim W_a^\mu\dim W_b^\mu$, which reproduces the degeneracy~\eqref{eq:densities} forced by the tube algebra.
\item Two twisted sectors $\Hil_a$ and $\Hil_b$ are correlated through those sectors with $a,b\subset F(\mu)$. For Fibonacci this is the sector $(\tau,\tau)$, in line with the identification~\eqref{eq:fibglue}.
\end{itemize}
The coefficient $1/2\pi$ is the prediction for a GUE value. The $\CRT$ image of Subsection~\ref{sec:crt} doubles it for self-dual sectors.

\subsection{Modular corrections}\label{sec:modular}

The seed \eqref{eq:seedamp} contains only the relative Dehn twists, whereas every other relative mapping class gives a further geometry. In this subsection we sum over all of them, denoting the modular $S$-matrix by $S=\vr(\hat S)$. The outcome is~\eqref{eq:preK}, in which the seed and its $\CRT$ image reproduce the ramp of each sector with the slope appropriate to its class, while the remaining modular images correlate sectors that the seed leaves independent.

\subsubsection{The modular sum}\label{sec:completion}

A simultaneous change of parametrization acts by $\gamma$ on boundary 1 and by $\tilde\gamma$ on boundary 2. Its action on the kernel is
\begin{equation}\label{eq:Kconj}
K\longmapsto\vr(\gamma)^{T}K\,\vr(\tilde\gamma)\,.
\end{equation}
Since $\vr(\hat T)$ is diagonal and $S$ is symmetric and unitary, $\vr(\tilde\gamma)=\overline{\vr(\gamma)}$, and the identity kernel~\eqref{eq:Kdiag} is therefore invariant under~\eqref{eq:Kconj}. Combined with~\eqref{eq:gravinv}, this makes each single geometry invariant under simultaneous transformations, so that only relative mapping classes produce new geometries. Note that a single projector $|\nu\rangle\langle\nu|$, by contrast, is not invariant, because $S$ mixes the anyons, and the sum over images will therefore mix different sectors.

We collect the seeds into the diagonal matrix denoted by $Z^{\mu\nu}_{\rm seed}=\delta^{\mu\nu}Z^{\rm seed}_\nu$. The full amplitude must transform covariantly under each boundary separately:
\begin{align}
\Zwh^{\mu\nu}(\gamma\!\cdot\!\tau_1,\tau_2)&=\vr(\gamma)^{\mu}{}_{\kappa}\,\Zwh^{\kappa\nu}(\tau_1,\tau_2),\label{eq:equiv1}\\
\Zwh^{\mu\nu}(\tau_1,\gamma\!\cdot\!\tau_2)&=\vr(\gamma)^{\nu}{}_{\kappa}\,\Zwh^{\mu\kappa}(\tau_1,\tau_2).\label{eq:equiv2}
\end{align}
Since both boundaries carry the same orientation, they transform in the same representation. The seed already contains the sum over the group $\Gamma_\infty$ generated by $\hat T$ (which does not contain $-\mathbf 1$), so that, as in~\cite{Cotler:2020ugk}, the remaining sum acts on boundary 2:
\begin{equation}\label{eq:Zwh}
\Zwh^{\mu\nu}(\tau_1,\tau_2)=\sum_{\gamma\in\Gamma_\infty\backslash SL(2,\Z)}\big(\vr(\gamma)^{-1}\big)^{\nu\kappa}\,Z^{\mu\kappa}_{\rm seed}(\tau_1,\gamma\!\cdot\!\tau_2).
\end{equation}
Each term is independent of the choice of coset representative, and the twisted amplitudes are recovered as
\begin{align}
    \langle Z_aZ_b\rangle=\sum_{\mu\nu}\dim W_a^\mu\dim W_b^\nu\,\Zwh^{\mu\nu}.
\end{align}
 We show covariance on boundary 2 in Appendix~\ref{app:VVMF}, and covariance on boundary 1 then follows from~\eqref{eq:gravinv} and the invariance of $K$. As in~\cite{Cotler:2020ugk}, the sum requires regularization, and since adding any weight-zero vector-valued modular form preserves~\eqref{eq:equiv1}--\eqref{eq:equiv2}, \eqref{eq:Zwh} is only one completion among many.

Because $\Gamma_\infty$ does not contain $-\mathbf 1$, the coset of $-\mathbf 1$ appears as a separate term in~\eqref{eq:Zwh}, which contributes even when $\vr(-\mathbf 1)$ is trivial. We now show that this term implements $\CRT$.

\subsubsection{$\CRT$ and parity}\label{sec:crt}

Recall from Subsection~\ref{sec:statement} that antiunitary symmetries play two roles: one that maps a sector to itself fixes its Wigner--Dyson class, and one that maps it to another sector makes the two spectra coincide exactly. In a CFT two kinds of symmetry are relevant: $\CRT$ preserves the spin, so it maps a sector either to itself or to another sector at the same spin, whereas parity and time reversal reverse the spin and pair sectors at opposite spins. These are the continuum counterparts of $\mathcal R\Theta$ and $\Theta$ in Subsection~\ref{sec:fibcrt}, and both have a natural bulk origin, which we now describe.

\paragraph{$\CRT$.} In Euclidean signature $\CRT$ rotates the spacetime torus by $\pi$, which is the central element $-\mathbf 1\in SL(2,\Z)$ and leaves $\tau$ unchanged. On the SymTFT it acts as charge conjugation,
\begin{equation}\label{eq:CRTbulk}
\vr(-\mathbf 1)=S^2=\Cc\,,\qquad \Cc\,|\nu\rangle=|\bar\nu\rangle\,.
\end{equation}
On the boundary, $\CRT$ reverses the orientation of a line along the time circle, mapping $\Hil_a$ to $\Hil_{\bar a}$, and, being antilinear, it also conjugates the charge. For a group this sends $([h],R)$ to $([h^{-1}],R^*)$, which becomes $\bar\nu$ once the flux is transported back to the class representative, in agreement with~\eqref{eq:CRTbulk}. In particular, since  $[\Cc,\vr(\hat{T})]=0$,  $\CRT$ preserves the spin: $\theta_{\bar\nu}=\theta_\nu$.

The coset of $-\mathbf 1$ contributes $\Cc\,Z_{\rm seed}$, and together with the identity coset it yields
\begin{equation}\label{eq:seedCRT}
\big(\delta^{\mu\nu}+\delta^{\mu\bar\nu}\big)\,Z^{\rm seed}_\nu(\tau_1,\tau_2)\,.
\end{equation}
Covariance~\eqref{eq:equiv1}--\eqref{eq:equiv2} under $-\mathbf 1$ then makes the pairing exact, $\Zwh^{\mu\nu}=\Zwh^{\bar\mu\nu}=\Zwh^{\mu\bar\nu}$, and the two possible cases reproduce precisely the two outcomes of categorical RMT. The diagonal coefficient of every self-dual sector doubles, whatever the sign of $\CRT^2$ on $\Mult_\nu$. When $\CRT^2=+1$, as in every sector of our two examples, this is the GOE ramp, twice the GUE ramp at early times (see Subsection~\ref{sec:S3numerics}). When $\CRT^2=-1$ the sector is GSE, and counting each Kramers pair twice gives the same slope (as occurs in Section~\ref{sec:continuous-holography}). For a pair $\nu\neq\bar\nu$, by contrast, each diagonal coefficient keeps its GUE value and the off-diagonal entry carries the same weight, so that the two spectra coincide \emph{exactly}.

Notably, the same factor of two fixes the normalization. For trivial $\C$, \eqref{eq:seedCRT} doubles the amplitude of~\cite{Cotler:2020ugk}, as it must: that amplitude carries the GUE coefficient~\eqref{equ:RMT_result_cate}, whereas a bosonic relativistic theory always has $\CRT$ and must therefore be GOE~\cite{Yan:2023rjh,DiUbaldo:2023qli}.

Since every anyon of $\Rep D(S_3)$ and of doubled Fibonacci is self-dual, $\CRT$ fixes every sector, and all sectors are GOE at fixed spin, in agreement with the golden chain, where $\mathcal R\Theta$ fixes all four sectors~\eqref{eq:fibcrt}. The simplest example of a $\CRT$ pair occurs for $\C=\VecG_{\Z_N}$ with $N>2$, where the dual of $(a,k)$ is $(-a,-k)$, so that in the untwisted sector the pair consists of the charges $k$ and $-k$, the familiar GUE pair of symmetry-resolved RMT.

\paragraph{Parity.} A second pairing arises when the bulk is invariant under orientation reversal, as Einstein gravity is.\footnote{A gravitational Chern--Simons term would break this invariance.} Orientation reversal conjugates the modular data of the SymTFT, so it acts on anyons by an involution $\sigma$ with
\begin{equation}\label{eq:sigma}
\theta_{\sigma(\nu)}=\overline{\theta_\nu},\qquad S_{\sigma(\mu)\sigma(\nu)}=\overline{S_{\mu\nu}}.
\end{equation}
Such an involution exists in both of our examples, and since it satisfies $s_{\sigma(\nu)}=-s_\nu$ mod 1, it can fix only sectors with $s_\nu\in\{0,\tfrac12\}$ mod 1. In both examples it fixes all of them. The dual ensemble is then invariant under a spin-reversing symmetry, the counterpart of $\Theta$ in Section~\ref{sec:categorical-RMT}, which at the level of amplitudes means
\begin{equation}\label{eq:Zsigma}
\Zwh^{\mu\nu}(\tau_1,\tau_2)=\Zwh^{\sigma(\mu)\sigma(\nu)}(-\bar\tau_1,-\bar\tau_2).
\end{equation}
In other words, the sector $\nu$ at spin $s$ and the sector $\sigma(\nu)$ at spin $-s$ have the same statistics.

For a group, the involution is
\begin{equation}\label{eq:sigmaG}
\sigma:([h],R)\longmapsto([h],R^*),
\end{equation}
which is the action~\eqref{eq:S3crt} of $\Theta$. 

Both roles of the antiunitary symmetries thus have a bulk origin: the element $-\mathbf 1$ fixes the class and the pairs at equal spin, while orientation reversal relates the statistics of sectors at opposite spin.

\subsubsection{Fourier coefficients and the pre-Kuznetsov formula}\label{sec:fourier}

Following Appendix~\ref{app:VVMF}, we compute Fourier coefficients
after stripping the Virasoro descendants. As in~\eqref{equ:CJ_primary_sec3},
we define the primary-counting amplitude by
\begin{equation}
Z_{\rm wh}^{P,\mu\nu}(\tau_1,\tau_2)
:=\left[\prod_{i=1}^2\prod_{r\geq1}|1-q_i^r|^2\right]
\Zwh^{\mu\nu}(\tau_1,\tau_2),
\qquad q_i=e^{2\pi i\tau_i}.
\end{equation}
Its Fourier coefficients have frequencies shifted by the topological spins:
\begin{equation}\label{eq:Fourier}
\begin{split}
Z_{nm}^{P,\mu\nu}(iy_1,iy_2)
:={}&\int_0^1\!\!\int_0^1
Z_{\rm wh}^{P,\mu\nu}(x_1+iy_1,x_2+iy_2)\\
&\qquad\times
e^{-2\pi i[(s_\mu+n)x_1+(s_\nu+m)x_2]}\,dx_1\,dx_2.
\end{split}
\end{equation}

To unfold the modular sum, it is convenient to introduce the reduced
gravitational kernel
\[
\widehat Z_{\rm grav}(\tau_1,\tau_2)
:=\frac{\Zgrav(\tau_1,\tau_2)}{Z_0(\tau_1)Z_0(\tau_2)}
=\frac{1}{2\pi^2}
\frac{\operatorname{Im}\tau_1\,\operatorname{Im}\tau_2}
{|\tau_1+\tau_2|^2}.
\]
Since $Z_0$ is modular invariant, its factors come out of the sum.
Stripping descendants then replaces them by
\[
\left[\prod_{i=1}^2\prod_{r\geq1}|1-q_i^r|^2\right]
Z_0(\tau_1)Z_0(\tau_2)
=\frac{e^{\pi(y_1+y_2)/6}}{\sqrt{y_1y_2}}.
\]
This factor is evaluated at the physical boundary moduli and is independent
of $x_1,x_2$, so it stays outside the Fourier integrals.
Applying the Bruhat decomposition~\eqref{eq:Bruhat_decomp} to boundary 2,
the $c=0$ terms are the cosets of $\mathbf1$ and $-\mathbf1$.
Unfolding the remaining translations, together with the sum over $\ell$
in~\eqref{eq:seedGamma}, extends both integrals to the real line and gives
\begin{align}
Z_{nm}^{P,\mu\nu}(iy_1,iy_2)
&=\frac{e^{\pi(y_1+y_2)/6}}{\sqrt{y_1y_2}}
\Biggl[
(\delta^{\mu\nu}+\delta^{\mu\bar\nu})\,\delta_{nm}
\int_{-\infty}^{\infty}
\widehat Z_{\rm grav}(x_1+iy_1,iy_2)
e^{-2\pi i(s_\mu+n)x_1}\,dx_1
\nonumber\\
&\quad+\sum_{c>0}
\Bigl(A^{\mu\nu}(s_\mu+n,s_\nu+m;c)
+A^{\mu\bar\nu}(s_\mu+n,s_\nu+m;c)\Bigr)
\nonumber\\
&\qquad\times
\int_{-\infty}^{\infty}\!\!\int_{-\infty}^{\infty}
\widehat Z_{\rm grav}\!\left(
x_1+iy_1,-\frac{1}{c^2(x_2+iy_2)}\right)
\nonumber\\
&\qquad\qquad\times
e^{-2\pi i[(s_\mu+n)x_1+(s_\nu+m)x_2]}
\,dx_1\,dx_2
\Biggr],
\label{eq:preK}
\end{align}
where $A^{\mu\nu}$ are the vector-valued Kloosterman sums
\begin{align}
   A^{\mu \nu}(s_\mu+n,s_\nu+m;c)\coloneqq \sum_{d(\mathrm{mod}c)\atop (d,c)=1} e^{2\pi i\left((s_\mu+n)\frac{d}{c}+(s_\nu+m)\frac{a}{c} \right)}  \left(\vr\left(\gamma_{cd} \right)^{-1}\right)^{\mu \nu},
   \label{eq:Kloost_VMF_sec3}
\end{align}
 and the terms with $\bar\nu$ come from $-\gamma$, because $\vr(-\gamma)=\Cc\,\vr(\gamma)$. In unfolding the integral, we assumed that the gravitational kernel has been regularized as in~\cite{Cotler:2020ugk}, so that the sum converges. We have also used the symmetry $A^{\nu\mu}(s_\nu+m,s_\mu+n;c)=A^{\mu\nu}(s_\mu+n,s_\nu+m;c)$, which follows from the symmetry of the matrices $S$ and $\vr(\hat T)$. The first line evaluates to the Fourier coefficient of the seed computed in Subsection~\ref{sec:seed}, multiplied by the $\CRT$ factor. Formulae of this type are known as pre-Kuznetsov formulae~\cite{Iwaniec2002SpectralMO}.\footnote{They correspond to the Kuznetsov trace formula before its spectral side is expanded.}

Equation~\eqref{eq:preK} is the main result of this subsection. Its first line, the seed together with its $\CRT$ image, reproduces the RMT ramp of Section~\ref{sec:categorical-RMT} sector by sector, with the slope appropriate to each class. The terms with $c\ge1$ are the modular corrections, which correlate sectors that the seed leaves independent. Modular invariance is therefore incompatible with \emph{exactly} independent symmetry sectors, just as it is incompatible with independent spin sectors~\cite{Haehl:2023mhf,DiUbaldo:2023qli}. Categorical RMT holds where the $c\ge1$ terms are small, namely at low temperature or at late times within the ramp.

In summary, within the proposal that the bulk is gravity coupled to the SymTFT, we have found that the wormhole amplitudes reproduce the structure of categorical RMT. The seed supplies the ramp of every tube algebra sector together with its exact degeneracy. The $\CRT$ image fixes the Wigner--Dyson classes and the exact pairings at equal spin, and orientation reversal relates the statistics at opposite spins. The remaining modular images add cross-sector correlations that lie beyond categorical RMT. In Sections~\ref{sec:continuous-CFT} and~\ref{sec:continuous-holography} we turn to continuous global symmetries, for which the role of the SymTFT is played by a bulk Chern--Simons gauge field.

\section{RMT for CFTs with continuous global symmetries}
\label{sec:continuous-CFT}

In Section~\ref{sec:categorical-RMT}, we saw that in a spatially extended system, the sectors carrying independent RMT statistics need not coincide with the naive ``charge" sectors. The tube algebra contains intertwiners connecting charge sectors that live in different twisted Hilbert spaces, and components related by an intertwiner share a single RMT dynamical block. In this section, we develop the parallel story for 2d CFTs with continuous global symmetries.

In the continuous case, the symmetry comes with conserved currents, whose modes generate an affine Kac-Moody algebra, and this has two important consequences. First, through the Sugawara construction, the currents contribute to the stress tensor, so the Virasoro and affine symmetries organize into a Virasoro-Kac-Moody (VKM) algebra for each chirality. Every holomorphic Virasoro generator now decomposes into a neutral part and a Sugawara part
\begin{align}
    L_n=L_n^N+L_n^{\mathrm{Sug}},
\end{align}
and the same is true for the antiholomorphic part. The spectrum is that of the neutral degrees of freedom, dressed by current-algebra modules. Second, the VKM algebra admits spectral flows, a family of isomorphisms that shift both the Cartan charges and the conformal weights while preserving the algebra. Depending on the flow parameter, spectral flow either maps the untwisted Hilbert space to a twisted Hilbert space, or relates different charge sectors within a single Hilbert space. Spectral flow here plays a role analogous to that of the tube-algebra intertwiners in Section~\ref{sec:categorical-RMT}.

 For the untwisted Hilbert space, we include the local spectral-flow operators in the current algebra where needed and decompose the spectrum into irreducible modules of the resulting algebra. After further removing neutral Virasoro descendants and resolving any additional exact symmetries, we propose ordinary RMT statistics for the remaining primary spectrum at fixed neutral spin in each sector. Distinct sectors are independent in the universal RMT regime unless $\CRT$ relates them. For the twisted Hilbert spaces, each of them is isomorphic to the untwisted one by spectral flow, with the same neutral multiplicity spaces. The twisted sectors therefore introduce no new random-matrix blocks, and their spectra are exact flow images of the untwisted ones.

Subsections~\ref{sec:sugawara-general}--\ref{sec:continuous-RMT-statement} develop these statements in general. We then give explicit examples for the abelian symmetry $U(1)$ and the nonabelian symmetry $SU(2)$ in Subsections~\ref{sec:U1-CFT}--\ref{sec:nonabelian-CFT}.

\subsection{Sugawara dressing of the neutral sector}
\label{sec:sugawara-general}

Consider a unitary bosonic two-dimensional CFT with compact connected symmetry group $G_L\times G_R$, where the two factors have Lie algebra $\mathfrak g$ and are generated by holomorphic and antiholomorphic currents, respectively. We take their current levels to agree and assume $c_L=c_R=c$. For the Sugawara formula below, we take $\mathfrak g$ to be simple or $u(1)$. 

The holomorphic symmetry algebra is the Virasoro-Kac-Moody algebra. The holomorphic currents $J^a(z)$ have modes $J^a_n$ generating the affine Kac--Moody algebra $\hat{\mathfrak g}_k$ at level $k$, which combine with the Virasoro generators $L_n$ into
\begin{align}
[J^a_m,J^b_n]&=km\,\delta^{ab}\delta_{m+n,0}+if^{ab}{}_{c}\,J^c_{m+n}\,,\nonumber\\
    [L_m,J^a_n]&=-n\,J^a_{m+n}\,,
    \label{eq:vkm-general}\\
    [L_m,L_n]&=(m-n)L_{m+n}+\frac{c}{12}(m^3-m)\,\delta_{m+n,0}\,,\nonumber
\end{align}
where $f^{ab}{}_{c}$ are the structure constants of $\mathfrak g$ and $k$ is the level. (For $\mathfrak g=u(1)$, only the first term in the first line survives, which we instead normalize
as $[J_m,J_n]=2km\,\delta_{m+n,0}$].) The currents contribute to the stress tensor through the Sugawara construction,
\begin{equation}
    L^{\mathrm{Sug}}_n=\frac{1}{2(k+h^{\vee})}\sum_{m=-\infty}^{\infty}:J^a_mJ^a_{n-m}:\,,
    \label{eq:sugawara-general}
\end{equation}
with $::$ the normal ordering and $h^\vee$ the dual Coxeter number of $\mathfrak g$. The modes $L^{\mathrm{Sug}}_n$ generate a Virasoro algebra of central charge $c^{\mathrm{Sug}}=\frac{k\dim\mathfrak g}{k+h^\vee}$. The antiholomorphic generators $\bar L_n$ and $\bar J^a_n$ obey an independent VKM algebra at the same level, and their Sugawara generators $\bar L_n^{\mathrm{Sug}}$ have the same central charge $c^{\mathrm{Sug}}$.

The neutral generators, defined by $L^N_n:=L_n-L^{\mathrm{Sug}}_n$, commute with the current, and hence with the Sugawara modes,
\begin{equation}
    [L^N_m,J^a_n]=0\,,
    \qquad
    [L^N_m,L^{\mathrm{Sug}}_n]=0\,.
    \label{eq:neutral-commutes}
\end{equation}
Thus, they generate their own Virasoro algebra $\mathrm{Vir}^N$ with central charge $c^N=c-c^{\mathrm{Sug}}$. Likewise, $\bar L^N_n:=\bar L_n-\bar L_n^{\mathrm{Sug}}$ generate $\overline{\mathrm{Vir}}^N$ with $\bar c^N=c^N$. The untwisted Hilbert space then decomposes as
\begin{equation}
\cH\simeq\bigoplus_{\lambda_L,\lambda_R}V_{\lambda_L}\otimes\bar V_{\lambda_R}\otimes\Mult_{\lambda_L\lambda_R}\,,
    \label{eq:continuous-decomposition}
\end{equation}
where $V_{\lambda_L}$ and $\bar V_{\lambda_R}$ are irreducible highest-weight modules of the two current algebras, and $\Mult_{\lambda_L\lambda_R}$ is a representation of $\mathrm{Vir}^N\oplus\overline{\mathrm{Vir}}^N$. Note that  the neutral spin $L_0^N-\bar L_0^N$ can be fractional even if  the physical spin $L_0-\bar L_0$ is integral.

The Sugawara spectrum in $V_{\lambda_L}\otimes\bar V_{\lambda_R}$ is completely fixed by current-algebra representation theory. In the remaining space $\Mult_{\lambda_L\lambda_R}$, the spectrum of neutral Virasoro descendants is also fixed once the primary spectrum is specified. Therefore, the unconstrained spectral data lie in the neutral-primary spectrum of $\Mult_{\lambda_L\lambda_R}$.
Restricting to the untwisted Hilbert space entails no loss of generality, since spectral flow maps it isomorphically onto every $G_L\times G_R$-twisted Hilbert space, as we show in Subsection~\ref{sec:spectral-flow-general}.

The Sugawara dressing of the neutral sector can also be seen from the partition function. We distinguish the canonical Hilbert space trace from the torus path integral, which differs by a factor depending on the sources. We use independent sources $z_L,z_R$ for the Cartan charges $\mathbf Q_L,\mathbf Q_R$ and write
\begin{equation}
    X=(\tau,\mathbf z),\qquad \mathbf z=(z_L,z_R),\qquad
    \tau_L=\tau,\quad \tau_R=-\bar\tau,\quad
    q_a=e^{2\pi i\tau_a}\quad(a=L,R).
\end{equation}
In our convention, both $z_L$ and $z_R$ appear with the same sign in the canonical trace, which can be written as
\begin{equation}
Z^{\mathrm{can}}(X)
    =\Tr_{\cH}\left[
      q_L^{L_0-\frac c{24}}q_R^{\bar L_0-\frac c{24}}
      e^{2\pi i(z_L\cdot\mathbf Q_L+z_R\cdot\mathbf Q_R)}\right].
\label{eq:CFT-canonical-trace}
\end{equation}

The path integral on a torus differs from this trace by an overall factor. For the two rank-one groups considered below, this factor can be written, using the boundary counterterm convention of~\cite{Kraus:2006wn}, as
\begin{equation}
\begin{aligned}
    Z^{\mathrm{PI}}(X)&=\mathcal G_G(X)Z^{\mathrm{can}}(X),\\
    \mathcal G_G(X)&=\exp\!\left[\frac{\pi\kappa_G}{\operatorname{Im}\tau}
                    (z_L^2+z_R^2)\right],\qquad
    \kappa_{U(1)}=k,\quad \kappa_{SU(2)}=\frac{k}{4}.
\end{aligned}
\label{eq:CFT-PI-canonical}
\end{equation}
Here $\kappa_G$ is the Jacobi index in our normalization. For higher rank groups, $\kappa_G z_a^2$ should be replaced by the corresponding quadratic form on Cartan sources. It is the path integral partition function $Z^{\mathrm{PI}}$ that is invariant under modular transformations with
\begin{equation}
    \tau\longmapsto\frac{a\tau+b}{c\tau+d},\qquad
    z_L\longmapsto\frac{z_L}{c\tau+d},\qquad
    z_R\longmapsto\frac{z_R}{c\bar\tau+d}.
\end{equation}
Bear in mind that to uncover a spectral interpretation from the partition function, we must do so from $Z^{\mathrm{can}}$.

The canonical affine characters in these conventions are
\begin{equation}
\begin{aligned}
    \chi_{\lambda_L}(\tau,z_L)
    &=\Tr_{V_{\lambda_L}}\left[q^{L^{\mathrm{Sug}}_0-\frac{c^{\mathrm{Sug}}}{24}}
      e^{2\pi i z_L\cdot\mathbf Q_L}\right],\\
    \chi_{\lambda_R}(-\bar\tau,z_R)
    &=\Tr_{\bar V_{\lambda_R}}\left[\bar q^{\bar L^{\mathrm{Sug}}_0-\frac{c^{\mathrm{Sug}}}{24}}
      e^{2\pi i z_R\cdot\mathbf Q_R}\right].
\end{aligned}
\label{eq:character-general}
\end{equation}
Since the current algebras have the same level, the two chiral characters above are described by the same function evaluated at $(\tau_L,z_L)$ and $(\tau_R,z_R)$, respectively.
The decomposition \eqref{eq:continuous-decomposition} implies that the canonical partition function can be decomposed as the neutral trace dressed by affine characters:
\begin{equation}
\begin{aligned}
    Z^{\mathrm{can}}(X)
    &=\sum_{\lambda_L,\lambda_R}\chi_{\lambda_L}(\tau,z_L)\,
      \chi_{\lambda_R}(-\bar\tau,z_R)\,
      Z^N_{\lambda_L\lambda_R}(\tau,\bar\tau),\\
    Z^N_{\lambda_L\lambda_R}(\tau,\bar\tau)
    &=\Tr_{\Mult_{\lambda_L\lambda_R}}\left[
      q^{L^N_0-\frac{c^N}{24}}\bar q^{\bar L^N_0-\frac{c^N}{24}}\right].
\end{aligned}
\label{eq:dressed-Z-general}
\end{equation}

To isolate the primary spectrum, define the neutral-primary subspace
\begin{equation}
    \mathcal P_{\lambda_L\lambda_R}
    :=\left\{v\in\Mult_{\lambda_L\lambda_R}:
    L^N_n v=\bar L^N_n v=0\ \text{for all }n>0\right\}.
    \label{equ:neutral_primary_space}
\end{equation}
Tensoring a state in $\mathcal P_{\lambda_L\lambda_R}$ with current-module ground states gives a primary state  under both current algebras and neutral Virasoro algebras. Its neutral spin $\hat s$ is the eigenvalue of $L^N_0-\bar L^N_0$,
\begin{equation}
    \hat s=s-\left(h^{\mathrm{Sug}}_{\lambda_L}-h^{\mathrm{Sug}}_{\lambda_R}\right),
\end{equation}
where $h^{\mathrm{Sug}}_{\lambda_L}$ and $h^{\mathrm{Sug}}_{\lambda_R}$ are the Sugawara weights of the current primaries. The neutral spins can thus be fractional even if the physical spins $s$ are integral.  The neutral-primary subspace can be further decomposed at fixed neutral spin as
\begin{equation}
\mathcal P_{\lambda_L\lambda_R}
 =\bigoplus_{\hat s}\mathcal P^{(\hat s)}_{\lambda_L\lambda_R}.
\label{eq:neutral-spin-blocks}
\end{equation}
A naive RMT ansatz would then apply to these $\mathcal P^{(\hat s)}_{\lambda_L\lambda_R}$ sectors. However, as we discuss next, the current modules must first be regrouped under spectral flow.

\subsection{Spectral flow}
\label{sec:spectral-flow-general}
The VKM algebra admits a family of isomorphisms called spectral flows.
They are labeled by independent flow parameters $\zeta_L,\zeta_R\in\mathbb R^r$ for the two chiralities, where $r=\operatorname{rank}\mathfrak g$.  We first write the holomorphic formulas. Since spectral flow treats the Cartan and root directions differently, we will work in the Cartan-Weyl basis, with Cartan modes $J^i_n$, $i=1,\dots,r$, and root currents
$J^\alpha_n$. Note that Latin indices $i,j$ label Cartan directions and Greek indices $\alpha$
label roots, which should not be confused with orthonormal currents $J^a$ in~\eqref{eq:vkm-general}. In the charge normalization of~\eqref{eq:CFT-canonical-trace}, we have $\mathbf Q=(J^1_0,\ldots,J^r_0)$. Their commutators are
\begin{equation}
[J^i_m,J^j_n]=K_G\,\delta^{ij}\,m\,\delta_{m+n,0},\qquad
  [J^i_m,J^\alpha_n]=\alpha^i\,J^\alpha_{m+n}.
  \label{eq:cartan_weyl}
\end{equation}
The spectral flow acts as
\begin{equation}
\begin{aligned}
  \sigma_\zeta(J^i_n)&=J^i_n+K_G\,\zeta^i\,\delta_{n,0},\\
  \sigma_\zeta(J^\alpha_n)&=J^\alpha_{n+\alpha\cdot\zeta},\\
  \sigma_\zeta(L_n)&=L_n+\zeta\cdot J_n+\tfrac12K_G\,\zeta\cdot\zeta\,\delta_{n,0},
\end{aligned}
\label{eq:general-flow}
\end{equation}
 where $\zeta\cdot J_n=\sum_i\zeta^i J^i_n$. These formulas preserve the VKM algebra.   
On a state of charge $\mathbf Q$ and conformal weight $h$, they give
\begin{equation}
\mathbf Q\longmapsto\mathbf Q+ K_G\zeta,
 \qquad h\longmapsto h+\zeta\cdot\mathbf Q
                  +\frac12  K_G \zeta\cdot\zeta.
\label{eq:general-flow-state}
\end{equation}
The antiholomorphic flow $\bar\sigma_{\zeta_R}$ has the same formulas with barred generators. The explicit rank-one formulas appear in~\eqref{eq:u1-spectral-flow-algebra_1}--\eqref{eq:u1-spectral-flow-algebra_2} for $u(1)$ and in~\eqref{su(2)_J3_SF}-\eqref{su(2)_Jpm_SF} for $su(2)$. 

The flow of $L_n$ comes entirely from the Sugawara generators~\eqref{eq:sugawara-general}. Using $L_n^N=L_n-L_n^{\mathrm{Sug}}$, one can check that the neutral Virasoro algebras are invariant,
\begin{equation}
\sigma_{\zeta_L}(L^N_n)=L^N_n,\qquad
       \bar\sigma_{\zeta_R}(\bar L^N_n)=\bar L^N_n.
\label{eq:neutral-invariant}
\end{equation}
Consequently, while spectral flow reshuffles the charges and physical spin, it leaves all neutral data untouched, including the neutral spin $\hat{s}$.

Depending on the value of $\zeta$, there are two qualitatively different situations for the spectral flow,
which we discuss below.

 \paragraph{Fractional spectral flow: twisted sectors.}
For generic $\zeta_L,\zeta_R\in\mathbb{R}^r$, the flowed generators in~\eqref{eq:general-flow} shift the rotation generator $L_0-\bar L_0$ by $\zeta_L\cdot J_0-\zeta_R\cdot\bar J_0+\text{const}$. As a result, a field with charges $(\mathbf Q_L,\mathbf Q_R)$ obeys the twisted boundary condition
\begin{equation}
  \Phi_{\mathbf Q_L,\mathbf Q_R}(e^{2\pi i}w,e^{-2\pi i}\bar w)
  =e^{2\pi i(\zeta_L\cdot\mathbf Q_L-\zeta_R\cdot\mathbf Q_R)}
   \Phi_{\mathbf Q_L,\mathbf Q_R}(w,\bar w).
  \label{eq:twisted-bc}
\end{equation}
This twist is implemented by inserting a topological symmetry defect,
\begin{equation}
  (g_L,g_R)
  =\left(e^{2\pi i\zeta_L\cdot J_0},\,e^{-2\pi i\zeta_R\cdot\bar J_0}\right)
  \in G_L\times G_R,
  \label{eq:defect}
\end{equation}
extended along the Euclidean time direction, where $H_0,\bar H_0$ are the Cartan charge
operators of the untwisted theory. This defect depends only on the conjugacy class of
$(g_L,g_R)$ and yields the twisted Hilbert space
$\mathcal{H}_{\zeta_L,\zeta_R}$.

We call this generic flow fractional spectral flow. It provides an isomorphism between untwisted and twisted Hilbert spaces,
\begin{equation}
\sigma_{\zeta_L}\otimes\bar\sigma_{\zeta_R}:\quad
 \cH\ \xrightarrow{\ \simeq\ }\ \cH_{\zeta_L,\zeta_R}.
\label{eq:flow-isomorphism}
\end{equation}
Applying $\sigma_{\zeta_L}\otimes\bar\sigma_{\zeta_R}$ to the untwisted decomposition~\eqref{eq:continuous-decomposition}, and using the invariance of the neutral generators, each twisted Hilbert space decomposes as
\begin{equation}
\cH_{\zeta_L,\zeta_R}
 \simeq\bigoplus_{\lambda_L,\lambda_R}
 V^{\zeta_L}_{\lambda_L}\otimes
 \bar V^{\zeta_R}_{\lambda_R}\otimes
 \Mult_{\lambda_L\lambda_R},
\label{eq:twisted-decomposition}
\end{equation}
where $V^{\zeta_L}_{\lambda_L}\otimes
 \bar V^{\zeta_R}_{\lambda_R}$ are flowed current modules, while each neutral space $\Mult_{\lambda_L\lambda_R}$ is the same as in the untwisted decomposition. This is the continuous counterpart of the twisted-sector decompositions~\eqref{eq:Wdecomp} of Section~\ref{sec:categorical-RMT}. There is, however, one important distinction: a tube-algebra sector $\mu$ need not have support $W_a^\mu$ in every twisted Hilbert space, whereas here every nonzero neutral space in the untwisted decomposition occurs for every twist. The twists therefore introduce no new independent random-matrix blocks.
 
The corresponding canonical characters satisfy
\begin{equation}
\chi^{\zeta_a}_{\lambda_a}(\tau_a,z_a)
 =e^{2\pi i\left(\frac{\tau_a}{2} K_G\zeta_a^\cdot \zeta_a
                      + K_G z_a\cdot  \zeta_a\right)}
   \chi_{\lambda_a}(\tau_a,z_a+\tau_a\zeta_a),
 \qquad a=L,R,
\label{eq:flowed-character-general}
\end{equation}
with $\tau_L=\tau$ and $\tau_R=-\bar\tau$. Replacing the two characters in~\eqref{eq:dressed-Z-general} by~\eqref{eq:flowed-character-general} gives the canonical trace in the twisted Hilbert space.

\paragraph{Symmetry flow: the extended chiral algebra.}
Spectral flow acts within the untwisted Hilbert space when the defect~\eqref{eq:defect} is trivial. For each chiral factor, this means the flow parameter lies in the cocharacter lattice
\begin{equation}
\Lambda_G=\left\{\zeta\in\mathfrak t:
                 e^{2\pi i\zeta\cdot J_0}=\mathbf1\ \text{in }G\right\}.
\label{eq:cocharacter-lattice}
\end{equation}
In this case, spectral flow maps $\cH$ to itself ~\cite{Schwimmer:1986mf,Benjamin:2020swg}, relabeling states with the charge and weight shifts as in~\eqref{eq:general-flow-state}.
Each such flow can be implemented by a local chiral operator $U_{\zeta}$ in each chiral sector. These operators extend the VKM algebra to a larger
chiral algebra, whose irreducible modules are labeled by spectral-flow
orbits
\begin{equation}
V^{\rm ext}_{[\lambda_a]}
 =\bigoplus_{\lambda'_a\in\Lambda_G\cdot\lambda_a}V_{\lambda'_a},
 \qquad a=L,R,
\label{eq:orbit-module}
\end{equation}
with a bar on $V$ for the right-moving modules.
\footnote{In the $U(1)$ and $SU(2)$ cases considered below,
\eqref{eq:orbit-module} is exact. For other groups, such as non-simply-connected groups, a spectral flow can instead map a module to itself. For example, for $SO(3)_k$, this happens at $j=k/4$. Such a fixed point gives rise to several inequivalent irreducible modules of the extended algebra, so the orbit label must be supplemented by a resolution label.}

Meanwhile, since the neutral generators $L_n^N$ and $\bar{L}_n^N$ are invariant under spectral flow, all multiplicity spaces on the flow orbit are identified $\Mult_{[\lambda_L][\lambda_R]}\simeq\Mult_{\lambda_L\lambda_R}$, and we can choose one of them as the representative on each orbit. Therefore, the decomposition~\eqref{eq:continuous-decomposition} regroups into extended modules as
\begin{equation}
\cH\simeq\bigoplus_{[\lambda_L],[\lambda_R]}
 V^{\rm ext}_{[\lambda_L]}\otimes
 \bar V^{\rm ext}_{[\lambda_R]}\otimes
 \Mult_{[\lambda_L][\lambda_R]}.
\label{eq:extended-decomposition}
\end{equation}
After removing neutral Virasoro descendants, the remaining spectral data lie in the neutral-primary spaces $\mathcal P_{[\lambda_L][\lambda_R]}^{\hat{s}}$ at fixed $\hat s$, defined as in \eqref{eq:neutral-spin-blocks} after regrouping.  These are the spaces to which we apply the RMT ansatz to the neutral Hamiltonian  $H^N=L_0^N+\bar L_0^N-c^N/12$.

\subsection{The RMT statement}
\label{sec:continuous-RMT-statement}
We can now state the continuous-symmetry analogue of our proposal in Subsection~\ref{sec:statement}:
\begin{tcolorbox}[
  colback=gray!10,
  colframe=blue,
  boxrule=0.6pt,
  arc=2pt,
  left=6pt,
  right=6pt,
  top=4pt,
  bottom=4pt]
In a chaotic bosonic $1+1$-dimensional CFT with continuous symmetry $G_L\times G_R$, the universal local spectral statistics of $H^N$ on each neutral-primary block $\mathcal P^{(\hat s)}_{[\lambda_L][\lambda_R]}$ are described by an ordinary random matrix drawn from the Wigner--Dyson class fixed by the antiunitary symmetries preserving that block. Distinct blocks are statistically independent in the RMT regime unless an exact symmetry relates them. Spectral flow identifies these primary spaces across all the twisted Hilbert spaces, which therefore introduce no new independent random matrices.
\end{tcolorbox}

The RMT labels are thus $([\lambda_L],[\lambda_R];\hat s)$, with current-algebra modules and neutral Virasoro descendants dressing these primary spectra to recover the full canonical traces. The dictionary with the categorical discussion is:
\begin{center}
\begin{tabular}{c|c}
Categorical symmetry & Continuous symmetry \\
\hline
fusion category $\cC$ & global symmetry $G_L\times G_R$ \\
twisted Hilbert spaces $\cH_a$ & twisted Hilbert spaces $\cH_{\zeta_L,\zeta_R}$ \\
tube algebra $\Tube(\cC)$ & VKM algebras and spectral flow \\
sector $\mu$ and further symmetry labels & $([\lambda_L],[\lambda_R];\hat s)$ \\
symmetry space $V_\mu$ & $V^{\mathrm{ext}}_{[\lambda_L]}\otimes\bar V^{\mathrm{ext}}_{[\lambda_R]}$ \\
symmetry degeneracies $\dim W_a^\mu$ & $\chi_{[\lambda_L]}(\tau,z_L)\chi_{[\lambda_R]}(-\bar\tau,z_R)$ \\
chaotic space $\Mult_\mu$ & neutral-primary space $\mathcal P^{(\hat s)}_{[\lambda_L][\lambda_R]}$ \\
intertwiners between twisted sectors & spectral flow
\end{tabular}
\end{center}
The characters in the table are canonical characters of the extended current modules. In a twisted sector they are replaced by their flowed versions, while the neutral-primary spaces remain the same.

\paragraph{$\CRT$ action and symmetry classes.}
As in Subsection~\ref{sec:statement}, the Wigner-Dyson class is assigned by antiunitary symmetries.
We assume that the only antiunitary symmetry is $\CRT$. $\CRT$ commutes with $G_L\times G_R$ and, being antilinear, conjugates the current-module labels. It preserves both neutral conformal weights and hence $\hat s$. Its action on the RMT labels is therefore
\begin{equation}
([\lambda_L],[\lambda_R];\hat s)
 \longmapsto([\lambda_L^*],[\lambda_R^*];\hat s).
\label{eq:continuous-CRT-labels}
\end{equation}
If $([\lambda_L],[\lambda_R]) \neq
 ([\lambda_L^*],[\lambda_R^*])$, $\CRT$ exchanges the two sectors, so their spectra coincide and
each is individually GUE.

If $([\lambda_L],[\lambda_R]) =
 ([\lambda_L^*],[\lambda_R^*])$, $\CRT$ then acts within $V^{\rm ext}_{[\lambda_L]}\otimes
 \bar V^{\rm ext}_{[\lambda_R]}\otimes
 \Mult_{[\lambda_L][\lambda_R]}$, where it factorizes as
\begin{equation}
\CRT=\CRT_V\otimes\CRT_{\bar V}\otimes\CRT_M.
\end{equation}
The first two factors act on $V^{\mathrm{ext}}_{[\lambda_L]}$ and $\bar V^{\mathrm{ext}}_{[\lambda_R]}$, while the last acts on the neutral multiplicity space $
 \Mult_{[\lambda_L][\lambda_R]}$ and commutes with $H^N$. Since RMT is applied to its neutral-primary subspace, the symmetry class is determined by $\CRT_M^2$. Write $\CRT_V^2=\epsilon_L$ and $\CRT_{\bar V}^2=\epsilon_R$, with $\epsilon_L,\epsilon_R\in\{+1,-1\}$. On the untwisted bosonic Hilbert space, $\CRT^2=+1$, and we have
\begin{equation}
\CRT^2=\CRT_V^2\,\CRT_{\bar V}^2\,\CRT_M^2=1,
 \qquad \CRT_M^2=\epsilon_L\epsilon_R.
\label{equ:CRTV-CRTM}
\end{equation}
Therefore, the neutral-primary block is GOE when $\epsilon_L\epsilon_R=+1$ and GSE when $\epsilon_L\epsilon_R=-1$. Notably, GSE statistics can arrised from opposite signs from the two current modules,   even though the full theory is bosonic.

Finally, for the twisted Hilbert spaces, $\CRT$ maps $\cH_{\zeta_L,\zeta_R}$ to $\cH_{-\zeta_L,-\zeta_R}$. The symmetry class of $\cH_{\zeta_L,\zeta_R}$ is nevertheless inherited from its untwisted partner by undoing the fractional spectral flow. This is again because spectral flow leaves the neutral-primary space, its Hamiltonian, and so its symmetry class unchanged. 

In the rest of this section, we demonstrate the previous statements through explicit examples with the abelian symmetry $U(1)$ and nonabelian symmetry $SU(2)$.

\subsection{Example: abelian symmetry $U(1)$}
\label{sec:U1-CFT}

We consider a bosonic CFT with $U(1)_L\times U(1)_R$ symmetry, equal levels $k\in\mathbb Z_{>0}$, and $c_L=c_R=c$. We first write the holomorphic algebra, using $w$ for the plane coordinate:
\begin{equation}
J(w)=\sum_{n\in\mathbb Z}\frac{J_n}{w^{n+1}},
\qquad T(w)=\sum_{n\in\mathbb Z}\frac{L_n}{w^{n+2}}.
\label{eq:u1-mode-expansion}
\end{equation}
Our physical charge is normalized as $Q_L=2J_0\in\mathbb Z$, and the current OPE is
\begin{equation}
J(w)J(0)\sim\frac{k/2}{w^2},
\qquad [J_m,J_n]=\frac{k}{2}m\delta_{m+n,0}.
\label{eq:u1-current-ope}
\end{equation}
The antiholomorphic algebra has the same normalization, with $Q_R=2\bar J_0\in\mathbb Z$. Each chiral $U(1)$ factor is separately compact, with integral charges in this normalization. A compact boson with irrational $R^2$ does not satisfy this assumption~\cite{Benjamin:2020swg}. The Sugawara generators are
\begin{equation}
L_n^{\mathrm{Sug}}=\frac1k\sum_{m\in\mathbb Z}:J_mJ_{n-m}:,
\qquad c^{\mathrm{Sug}}=1,
\qquad c^N=c-1,
\end{equation}
and similarly on the right. We use the canonical trace and path-integral conventions of Subsection~\ref{sec:sugawara-general}, with $\tau_L=\tau$, $\tau_R=-\bar\tau$, and independent sources $z_L,z_R$ coupling with the same sign.

\subsubsection{Spectral flow and the torus partition function}
\label{sec:U1-spectral-flow}

With respect to the Cartan–Weyl basis in \eqref{eq:cartan_weyl}, the Cartan current in $U(1)$ case is $J_n^{i=1}=2J_n$, so
\begin{equation}
 K_{U(1)}=2k.
\end{equation}
The spectral flow of the $u(1)_k$ VKM algebra is
\begin{align}
\sigma_\zeta(L_n)&=L_n+2\zeta J_n+k\zeta^2\delta_{n,0},
\label{eq:u1-spectral-flow-algebra_1}\\
\sigma_\zeta(J_n)&=J_n+k\zeta\delta_{n,0}.
\label{eq:u1-spectral-flow-algebra_2}
\end{align}
Writing $Q$ for the charge in either chirality, the corresponding transformation of weights and charges is
\begin{equation}
Q\longmapsto Q^\zeta=Q+2k\zeta,
\qquad h\longmapsto h^\zeta=h+\zeta Q+k\zeta^2.
\label{spectral-flow-state}
\end{equation}

\paragraph{Integer spectral flow.}
For $\zeta\in\mathbb Z$, spectral flow acts within the untwisted Hilbert space. It gives an exact relabeling of states with the shifts~\eqref{spectral-flow-state}, while leaving their neutral weights unchanged. Bosonizing $J=i\sqrt{k/2}\,\partial\phi$, the flow is implemented by the charged vertex operator
\begin{equation}
\mathcal U_\zeta(w)=:e^{i\zeta\sqrt{2k}\phi(w)}:,
\label{equ:SF_vertex_opt}
\end{equation}
which has charge $2k\zeta$ and weight $k\zeta^2$. These vertex operators are local with respect to the integral charge lattice and extends the current algebra.

The extended current algebra has finitely many irreducible modules, labeled by the spectral-flow orbits $m=Q\bmod 2k$, with $m\in\mathbb Z_{2k}$. Their canonical characters collect an entire flow orbit and its current descendants:
\begin{align}
\begin{split}
\chi_m^k(\tau,z)
&=\Tr_{V_{[m]}^{\mathrm{ext}}}
 \left[q^{L_0^{\mathrm{Sug}}-1/24}e^{2\pi izQ}\right]\\
&=\frac1{\eta(\tau)}\sum_{w\in\mathbb Z}
 q^{(m+2kw)^2/(4k)}e^{2\pi iz(m+2kw)},
\label{2k-theta_function}
\end{split}
\end{align}
where the sum over $w$ runs over the flow images of charge $Q=m+2kw$, each with its Sugawara weight $\frac{Q^2}{4k}$, and $\eta(\tau)=q^{1/24}\prod_{n\ge1}(1-q^n)$ is the Dedekind eta function. There are $2k$ extended current modules for each chirality. The dressing formula~\eqref{eq:dressed-Z-general} becomes
\begin{equation}
Z^{\mathrm{can}}(X)
 =\sum_{m_L,m_R=0}^{2k-1}
 \chi_{m_L}^k(\tau,z_L)\chi_{m_R}^k(-\bar\tau,z_R)
 Z^N_{m_L,m_R}(\tau,\bar\tau),
 \qquad X=(\tau,\mathbf z).
\label{eq:U1-dressing}
\end{equation}

\paragraph{Fractional spectral flow.}
For nonintegral flow parameters $\zeta_L,\zeta_R\in(0,1)$, spectral flow instead gives a twisted Hilbert space $\mathcal H_{\zeta_L,\zeta_R}$~\cite{Schwimmer:1986mf,Benjamin:2020swg}. In the conventions of Subsection~\ref{sec:spectral-flow-general}, a charged field obeys the twisted boundary condition
\begin{equation}
\mathcal O_{\mathbf Q}(e^{2\pi i}w,e^{-2\pi i}\bar w)
 =e^{2\pi i(\zeta_LQ_L-\zeta_RQ_R)}
 \mathcal O_{\mathbf Q}(w,\bar w).
\end{equation}
This is implemented by a symmetry defect line along the Euclidean time direction,
\begin{align}
    U_{\zeta_L,\zeta_R}=e^{2\pi i\zeta_L 2J_0}\,e^{-2\pi i\zeta_R2\bar J_0}.
\end{align}
In this case, the twist is trivial when $\zeta_L,\zeta_R\in\mathbb{Z}$, and $\mathcal H_{\zeta_L,\zeta_R}$ reduces to $\mathcal{H}$. 

\paragraph{Torus partition function.}
The canonical partition function in~\eqref{eq:U1-dressing}  is
\begin{equation}
Z^{\mathrm{can}}(X)=\Tr_{\mathcal H}
 \left[q^{L_0-c/24}\bar q^{\bar L_0-c/24}
 e^{2\pi i(z_LQ_L+z_RQ_R)}\right].
\label{eq:U1-Z-def}
\end{equation}
This partition function has two properties that are both related to spectral flow.
First, the canonical trace is invariant under the relabeling of states induced by integer spectral flow in either chirality. For example,
\begin{align}
   \Tr_{\mathcal H}
 \left[q^{L_0-c/24}\bar q^{\bar L_0-c/24}
 e^{2\pi i(z_LQ_L+z_RQ_R)}\right]= \Tr_{\mathcal H}
 \left[q^{\sigma_{\zeta}(L_0)-c/24}\bar q^{\bar L_0-c/24}
 e^{2\pi i(z_L \sigma_{\zeta}(Q_L)+z_RQ_R)}\right], \quad \zeta \in \mathbb{Z}.
\end{align}
This gives the elliptic transformation law
\begin{equation}
Z^{\mathrm{can}}(\tau;\ldots,z_a+u\tau_a+v,\ldots)
 =e^{-2\pi ik(u^2\tau_a+2uz_a)}Z^{\mathrm{can}}(X),
 \qquad u,v\in\mathbb Z,\quad a=L,R.
\label{U(1)_elliptic}
\end{equation}

Second, the canonical trace transforms under modular transformations as
\begin{equation}
Z^{\mathrm{can}}(\gamma\cdot X)
 =\exp\!\left[2\pi ikc
 \left(\frac{z_L^2}{c\tau+d}
       -\frac{z_R^2}{c\bar\tau+d}\right)\right]
 Z^{\mathrm{can}}(X),
\label{U(1)_modular}
\end{equation}
where $\gamma=\left(\begin{smallmatrix}a&b\\c&d\end{smallmatrix}\right)\in SL(2,\mathbb Z)$ and
\begin{align}
    \gamma\cdot X=\left(\frac{a\tau+b}{c\tau+d},\frac{z_L}{c\tau+d},\frac{z_R}{c\bar\tau+d} \right).
    \label{eq:gamma_coord}
\end{align}
Note that $c$ in  \eqref{U(1)_modular} and \eqref{eq:gamma_coord}  is an $SL(2,\mathbb Z)$ matrix entry and should not be confused with the central charge.  
For real $z_L,z_R$, this modular transformation law has a useful interpretation in terms of symmetry lines~\cite{Benjamin:2020swg} and spectral flow. Define $U_a(z)=e^{2\pi izQ_a}$. Then
\begin{equation}
Z^{\mathrm{can}}(X)=\Tr_{\mathcal H}
 \left[U_L(z_L)U_R(z_R)
 q^{L_0-c/24}\bar q^{\bar L_0-c/24}\right].
\label{U(1)_modular_eq1}
\end{equation}
In the $S$-dual interpretation, the symmetry line becomes a defect and the partition function is equivalent to a trace over the twisted Hilbert space $\mathcal{H}_{-z_L,z_R}$  with torus modulus  $-1/\tau$,
\begin{align}
Z^{\mathrm{can}}(X)&=\Tr_{\mathcal H_{-z_L,z_R}}
 \exp\!\left[-\frac{2\pi i}{\tau}(L_0-c/24)
             +\frac{2\pi i}{\bar\tau}(\bar L_0-c/24)\right]
\nonumber\\
&=e^{-2\pi ik(z_L^2/\tau-z_R^2/\bar\tau)}
 Z^{\mathrm{can}}(-1/\tau;z_L/\tau,z_R/\bar\tau)
 .
\label{U(1)_modular_eq2}
\end{align}
The second equality uses fractional spectral flow and reproduces the $S$ transformation in~\eqref{U(1)_modular}.

In the path-integral convention of~\eqref{eq:CFT-PI-canonical},
\begin{equation}
Z^{\mathrm{PI}}(X)=\mathcal G_{U(1)}(X)Z^{\mathrm{can}}(X),
\qquad \mathcal G_{U(1)}(X)
 =\exp\!\left[\frac{\pi k}{\operatorname{Im}\tau}(z_L^2+z_R^2)\right],
\end{equation}
the Gaussian cancels the modular pre-factor in \eqref{U(1)_modular}, so that $Z^{\mathrm{PI}}(\gamma\cdot X)=Z^{\mathrm{PI}}(X)$. 

\subsubsection{RMT sectors and symmetry classes}
\label{sec:U1-RMT-expectations}

We now specialize the RMT ansatz of Subsection~\ref{sec:continuous-RMT-statement} to the $U(1)_k$ case.
The random spectrum is that of the neutral Hamiltonian $H^N$ on the neutral-primary subspace $\mathcal P_{[m_L][m_R]}^{(\hat s)}$, within which primary states have conformal weight
\begin{equation}
\hat h=L_0-\frac{Q_L^2}{4k}=L_0^N,
\qquad \hat{\bar h}=\bar L_0-\frac{Q_R^2}{4k}=\bar L_0^N.
\end{equation}
The neutral spin is therefore
\begin{equation}
\hat s=\hat h-\hat{\bar h}
 =s-\frac{Q_L^2-Q_R^2}{4k},
\qquad s=L_0-\bar L_0.
\label{eq:neutral-spin}
\end{equation}
The RMT sectors of a chaotic $U(1)_k$-symmetric CFT are thus labeled by the flow invariants associated with $\mathcal P_{[m_L][m_R]}^{(\hat s)}$,
\begin{equation}
(m_L,m_R;\hat s)
 =\left(Q_L\bmod2k,Q_R\bmod2k;
 s-\frac{Q_L^2-Q_R^2}{4k}\right).
\label{eq:U1-sector-labels}
\end{equation}

To obtain the predictions for the two-point spectral correlations, we first define partition function on $\mathcal P_{[m_L][m_R]}^{(\hat s)}$ as
\begin{align}
    \mathcal Z^N_{\hat s,\mathbf m}(\beta)=\Tr_{\mathcal P_{[m_L][m_R]}^{(\hat s)}}e^{-2\pi\beta H^N}.
\end{align}
If we instead use the physical generators in the trace, the partition function is given by
\begin{align}
     Z^P_{s,\mathbf Q}(\beta)&=\Tr_{\mathcal P_{[m_L][m_R]}^{(\hat s)}} e^{-2\pi\beta (L_0+\bar L_0-1/12)}\nonumber \\
     &=e^{-2\pi\beta(\epsilon_{\mathbf Q}-1/12)} \mathcal Z^N_{\hat s,\mathbf m}(\beta), \qquad \epsilon_{\mathbf Q}=\frac{Q_L^2+Q_R^2}{4k}
     \label{equ:Z_p_partition}
\end{align}
The trace above recieves contributions from different states related by integer spectral flow. To count only extended primaries instead, we include only the current-module ground states, whose charges in the orbit $[m]$ obey $Q\equiv m\bmod2k$ and $|Q|\le k$. In particular, for $V_{[k]}^{\mathrm{ext}}$, there are two ground states with charges $\pm k$. Let
\begin{equation}
h_m=\min_{Q\equiv m\,\mathrm{mod}\,2k}\frac{Q^2}{4k},
\qquad d_m=\begin{cases}2,&m=k,\\1,&m\ne k,\end{cases}
\qquad D_{\mathbf m}=d_{m_L}d_{m_R}.,
\end{equation}
the RMT statement then gives diagonal correlations in RMT labels,
\begin{equation}
\left\langle Z^P_{s_1,\mathbf m_1}(\beta_1)
 Z^P_{s_2,\mathbf m_2}(\beta_2)\right\rangle_{c,\mathrm{ramp}}
 \ \propto\ D_{\mathbf m_1}^2\delta_{\hat s_1,\hat s_2}
 \left(\delta_{\mathbf m_1,\mathbf m_2}
       +\delta_{\mathbf m_1,-\mathbf m_2}\right).
\label{eq:diagonal-prediction}
\end{equation}
The second term $\delta_{\mathbf m_1,-\mathbf m_2}$ implements $\CRT$, which conjugates the RMT labels as $(m_L,m_R;\hat s) \rightarrow (-m_L,-m_R;\hat s)$, with the minus signs understood modulo $2k$. In terms of fixed physical charges, the same statement is
\begin{equation}
\left\langle Z^P_{s_1,\mathbf Q_1}(\beta_1)
 Z^P_{s_2,\mathbf Q_2}(\beta_2)\right\rangle_{c,\mathrm{ramp}}
 \ \propto\ \delta(s_1,\mathbf Q_1;s_2,\mathbf Q_2)
 +\delta(s_1,\mathbf Q_1;s_2,-\mathbf Q_2),
\label{SFF_untwisted_bdy}
\end{equation}
where
\begin{equation}
\delta(s_1,\mathbf Q_1;s_2,\mathbf Q_2)
 =\begin{cases}
 1,&\hat s_1=\hat s_2\ \text{and }Q_{1,a}\equiv Q_{2,a}\bmod2k
       \ \text{for }a=L,R,\\
 0,&\text{otherwise}.
 \end{cases}
\label{SF_delta_bdy}
\end{equation}
Therefore, although the spectral correlation is actually diagonal in RMT sectors, it might exhibit
off-diagonal correlations between naive charge sectors. This seeming contradiction is resolved by
extending the chiral algebra with spectral-flow operators.

The symmetry classes follow from Subsection~\ref{sec:continuous-RMT-statement}. For a pair with $\mathbf m\ne-\mathbf m$, $\CRT$ exchanges the two sectors, so they are individually GUE with coincident spectra. The self-conjugate sectors have $m_L,m_R\in\{0,k\}$. The vacuum module is real, with $\CRT_V^2=+1$ on $V_{[0]}^{\mathrm{ext}}$. On $V_{[k]}^{\mathrm{ext}}$,  denote its charge-$k$ ground state by $|k\rangle$, one has
\begin{equation}
\CRT_V|k\rangle=\alpha|-k\rangle,
\qquad |-k\rangle=(\mathcal U_{-1})_0|k\rangle,
\qquad |k\rangle=(\mathcal U_1)_0|-k\rangle,
\end{equation}
where $|\alpha|=1$, and $(\mathcal U_{\pm1})_n $ is the $n$-th mode of the spectral flow vertex operator \eqref{equ:SF_vertex_opt}: $\mathcal U_{\pm1}(z)=\sum_n (\mathcal U_{\pm1})_n z^{-n-k}$. The last two equation arises because $|k\rangle$ and $|-k\rangle$  in the same $V_{[k]}^{\mathrm{ext}}$ module, and are thus related by the spectral flow operator.  By $\CRT$ theorem, $\CRT$ maps a bosonic local operator of spin $s$ to $(-1)^s$ times its Hermitian conjugates. The spectral flow operator has spin $k$, and obey $\mathcal U_{\pm1}^{\dagger}=\mathcal U_{\mp1}$. We thus have
\begin{align}
    \CRT(\mathcal U_{\pm1})_n \CRT^{-1}=(-1)^k (\mathcal U_{\mp1})_n.
    \label{equ:CRT_SF_opt}
\end{align}
Since $\mathcal U_{\pm1}$ act only on the current module, the above relation holds with $\CRT$ replaced by $\CRT_V$. Antilinearity  then gives
\begin{align}
    \CRT_V^2 |k\rangle= \alpha^{*} \CRT_V|-k\rangle=(-1)^k \alpha^* \alpha (\mathcal U_{1})_0|-k\rangle= (-1)^k|k\rangle.
\end{align}
The same sign $ \CRT_V^2$ also holds  for every other pair of states in $V_{[k]}$ related by $\CRT$. One can check this by recursively using \eqref{equ:CRT_SF_opt} together with the relation
\begin{align}
    |(2n+1)k\rangle =(\mathcal U_{1})_{-2nk}|(2n-1)k\rangle, \quad  |(2n-1)k\rangle =(\mathcal U_{-1})_{2nk}|(2n+1)k\rangle, \quad n\in\mathbb{Z}.
\end{align}
As a result, $\CRT_V^2=(-1)^k$ on $V_{[k]}^{\mathrm{ext}}$. The same argument applies to $\bar{V}_{[k]}^{\mathrm{ext}}$ with $\CRT_{\bar{V}}=(-1)^k$. Equation~\eqref{equ:CRTV-CRTM} then gives the classes in Table~\ref{table:U1_classes}.

\begin{table}[H]
\centering
\begin{tabular}{lcc}
\toprule
Current-module labels $(m_L,m_R)$ & $\epsilon_L\epsilon_R$ & Neutral class\\
\midrule
$(0,0),(k,k)$ & $+1$ & GOE\\
$(k,0),(0,k)$, $k$ even & $+1$ & GOE\\
$(k,0),(0,k)$, $k$ odd & $-1$ & GSE\\
Other pairs & --- & GUE, paired with $(-m_L,-m_R)$\\
\bottomrule
\end{tabular}
\caption{Wigner--Dyson classes of the neutral-primary blocks for $U(1)_L\times U(1)_R$, assuming $\CRT^2=1$ and no additional antiunitary symmetries beyond those specified in Subsection~\ref{sec:continuous-RMT-statement}.}
\label{table:U1_classes}
\end{table}

Each block in a twisted Hilbert space $\mathcal H_{\zeta_L,\zeta_R}$ is the flow image of an untwisted sector $(m_L,m_R)$ and shares its neutral multiplicity space. Its symmetry class is therefore inherited from the untwisted partner, as given in Table~\ref{table:U1_classes}. 

\subsection{Example: non-abelian symmetry $SU(2)$}
\label{sec:nonabelian-CFT}

We now consider $SU(2)_L\times SU(2)_R$ at equal levels $k\in\mathbb Z_{>0}$. In the physical Cartan normalization $Q_L=J_0^3$, the holomorphic commutators are
\begin{equation}
[J_m^3,J_n^3]=\frac{k}{2}m\delta_{m+n,0},\qquad
 [J_m^3,J_n^\pm]=\pm J_{m+n}^\pm,\qquad
 [J_m^+,J_n^-]=km\delta_{m+n,0}+2J_{m+n}^3.
\end{equation}
The right-moving algebra is identical, with $Q_R=\bar J_0^3$. In this basis the Sugawara generators are
\begin{equation}
L_n^{\mathrm{Sug}}=\frac1{k+2}\sum_{m\in\mathbb Z}
 \left(:J_m^3J_{n-m}^3:
 +\frac12:J_m^+J_{n-m}^-:
 +\frac12:J_m^-J_{n-m}^+:\right),
\label{sug_su(2)}
\end{equation}
with $c^{\mathrm{Sug}}=3k/(k+2)$ and $c^N=c-c^{\mathrm{Sug}}$ for each chirality. The relation to the orthonormal basis $J_n^a$ $(a=1,2,3)$ is given in~\eqref{equ:SU(2)_new_basis}.

\paragraph{Spectral flow.}
The spectral flow acts as 
\begin{align}
\sigma_\zeta(J_n^3)&=\frac{k\zeta}{2}\delta_{n,0}-J_n^3,
\label{su(2)_J3_SF}\\
\sigma_\zeta(J_n^+)&=J_{n+\zeta}^-,\qquad
\sigma_\zeta(J_n^-)=J_{n-\zeta}^+.
\label{su(2)_Jpm_SF}
\end{align}
 With respect to the spectra flow formula in  Subsection~\ref{sec:spectral-flow-general}, we have $K_{SU(2)}=\frac{k}{2}$.
For $\zeta = 0$, this reduces to the action of the Weyl group of $su(2)$. The Virasoro generators transform as
\begin{equation}
\sigma_\zeta(L_n)=L_n-\zeta J_n^3
 +\frac{k\zeta^2}{4}\delta_{n,0}.
\end{equation}
This can be shown using \eqref{sug_su(2)}. The normal ordering in \eqref{sug_su(2)} is subtle under the spectral flow, and we provide the details in Appendix ~\ref{su(2)_vir_SF} for $\zeta\in \mathbb{Z}$.

Unlike the $U(1)$ case, the spectral flow acts within the untwisted Hilbert space only when $\zeta\in2\mathbb Z$. This follows from the cocharacter lattice~\eqref{eq:cocharacter-lattice}. A spatial $SU(2)$ twist is labeled by a conjugacy class, with a representative in the Cartan subgroup,
\begin{align}
    g_{\zeta}=\exp(2\pi i\zeta J_0^3).
\end{align}
Since Cartan charges can be half-integral, the cocharacter lattice of $SU(2)$ is
\begin{align}
     \Lambda_{SU(2)}=\bigl\{\zeta\in\mathbb R:\ e^{2\pi i\zeta J_0^3}=\mathbf1\ \text{in }SU(2)\bigr\}=2\mathbb Z.
    \label{eq:cocharacter-lattice_SU2}
\end{align}
In particular, odd integer flow implements the nontrivial center element $e^{2\pi iJ_0^3}=-\mathbf1$ and maps the untwisted Hilbert space to the center-twisted sector.

For $\zeta\in 2\mathbb{Z}$, the Cartan charge $J^3_0$ and the conformal weight $h$ transform inside the untwisted Hilbert space as
\begin{align}
    &J^3_0 \rightarrow {J_0^3}^\zeta= \zeta\frac{k}{2}- J_0^3, \\
    &h \rightarrow h^{\zeta}= h-\zeta J_0^3+\frac{k\zeta^2}{4}.
\end{align}
For $su(2)_k$, the Cartan charge of a primary state $J_0^3$ and the level $k$ must satisfy an integrability bound,
\begin{align}
    0\le j\le\frac{k}{2},\qquad m=-j,-j+1,\ldots,j.
\end{align}
This comes from the requirement of certain descendant states in the spectrum to have non-negative norm. Naively, the shift of $J_0^3$ under spectral flow seems to suggest that $J_0^3$ can be shifted to values larger than $\frac{k}{2}$. However, the resulting state is no longer a primary state.\footnote{We demonstrate this explicitly for $\zeta=1$ in the Appendix \ref{SF_states}. A brief review of the highest weight representation of $su(2)_k$  is also given in Appendix \ref{HWR_su(2)}.}
Unlike $U(1)$, the spectral flow of $SU(2)$ reorganize states already present in the affine module and require no extension of the current algebra.

For $\zeta\notin 2\mathbb{Z}$, the spectral flow implements a $g_{\zeta}$-twist that gives rise to a twisted Hilbert space. Again, every twisted sector is isomorphic to the untwisted sector by the spectral flow. They thus carry no independent spectral data.

\paragraph{Characters and the partition function.}
Let us first define the level-$k$ theta function,
\begin{equation}
\vartheta_m^{(k)}(\tau,z)
 =\sum_{n\in\mathbb Z+m/(2k)}e^{2\pi i\tau kn^2}e^{2\pi izkn}.
\end{equation}
The canonical character of $V_j$, with $z$ coupling to $J_0^3$, is given by the Weyl-Kac formula
\begin{equation}
\chi_j^k(\tau,z)
 =\frac{\vartheta_{2j+1}^{(k+2)}(\tau,z)-\vartheta_{-2j-1}^{(k+2)}(\tau,z)}
 {\vartheta_1^{(2)}(\tau,z)-\vartheta_{-1}^{(2)}(\tau,z)},
\qquad j=0,\tfrac12,\ldots,\tfrac k2.
\label{su(2)-char}
\end{equation}
Since the allowed untwisted spectral flows do not extend the current algebra, $V_j=V^{\mathrm{ext}}_{[j]}$ in the previous notation. The ground states in $V_j$ form a spin-$j$ multiplet of dimension $2j+1$, with weight
\begin{equation}
h_j=\frac{j(j+1)}{k+2}.
\label{eq:su2-hm}
\end{equation}
The canonical partition function is
\begin{equation}
Z^{\mathrm{can}}(X)
 =\sum_{j_L,j_R=0,\frac12,\ldots,\frac k2}
 \chi_{j_L}^k(\tau,z_L)\chi_{j_R}^k(-\bar\tau,z_R)
 Z^N_{j_L,j_R}(\tau,\bar\tau).
\label{eq:SU2-dressing}
\end{equation}

The path integral partition function again differs by the canonical one with an overall factor. With the present charge normalization, the path integral partition function can be written as
\begin{equation}
Z^{\mathrm{PI}}(X)
 =\exp\!\left[\frac{\pi k}{4\operatorname{Im}\tau}(z_L^2+z_R^2)\right]
 Z^{\mathrm{can}}(X).
\end{equation}
The Gaussian cancels the modular prefactor in the modular transformation as before.

\paragraph{RMT statistics.}
The RMT sectors of a chaotic $SU(2)_k$-symmetric CFT are the untwisted neutral blocks labeled by
\begin{equation}
(j_L,j_R;\hat s),
\qquad \hat s=s-h_{j_L}+h_{j_R},
\qquad j_L,j_R=0,\tfrac12,\ldots,\tfrac k2.
\label{eq:su2-sector-labels}
\end{equation}
At fixed $(j_L,j_R)$, fixing $\hat s$ is equivalent to fixing the physical spin $s$. Removing both affine and neutral Virasoro descendants, and counting all states in each spin-$j$ primary multiplet, the prediction for the two-point spectral correlations can be written as
\begin{equation}
\left\langle Z^P_{s_1,\mathbf j_1}(\beta_1)
 Z^P_{s_2,\mathbf j_2}(\beta_2)\right\rangle_{c,\mathrm{ramp}}
 \ \propto\ 2D_{\mathbf j_1}^2
 \delta_{j_{1,L},j_{2,L}}\delta_{j_{1,R},j_{2,R}}\delta_{s_1,s_2}.
\label{eq:su2-diagonal-prediction}
\end{equation}
Here $\mathbf j=(j_L,j_R)$, and $D_j=(2j_L+1)(2j_R+1)$ counts the spin-$j$ multiplet degeneracy. The primary-counting partition function $ Z^P_{s,\mathbf j}$ is defined as in \eqref{equ:Z_p_partition}. The overall factor of 2 comes from CRT, since every $SU(2)$ representation is self-conjugate by the Weyl symmetry.

The symmetry-class analysis follows the $U(1)$ case. $\CRT$ preserves each sector because every $SU(2)$ representation is self-conjugate. Thus, the class is determined by the sign of $\CRT_M^2=\CRT_V^2\CRT_{\bar V}^2$. We first compute $\CRT_V^2$.
Since $\CRT_V$ reverses $J_0^3$, it acts on each $|j,m\rangle$ state in the multiplet as
\begin{align}
    \CRT_V |j,m\rangle= c_{m} |j,-m\rangle,
\end{align}
with $|c_{m}|=1$. Different states in the multiplets are related by the raising and lowering operators $J_0^{\pm}$ (see Appendix \ref{HWR_su(2)})
\begin{align}
    J_0^{+}|j,m\rangle = N_{\mathrm{CG}} |j,m+1\rangle, \quad   J_0^{-}|j,m\rangle = N_{\mathrm{CG}}^{\prime} |j,m-1\rangle,
\end{align}
where $N_{\mathrm{CG}}$ and $N_{\mathrm{CG}}^{\prime}$ are real ladder coefficients depending on $j,m$. Since $J^\pm$ are spin-one currents, the $\CRT$ theorem gives
\begin{align}
    \CRT_VJ_n^{\pm} \CRT^{-1}_V=-J_n^{\mp}.
\end{align}
We thus have
\begin{align}
    \CRT_V^2 |j,m\rangle = c_{m}^* \CRT_V |j,-m\rangle = c_{m}^* c_{m} (-1)^{2m}|j,m\rangle = (-1)^{2j}|j,m\rangle,
\end{align}
where in the second equality we used $ \CRT_V(J_n^{-})^{2m} \CRT^{-1}_V=(-1)^{2m}(J_n^{+})^{2m}$.  The right-moving sign is the same. Therefore, we have
\begin{align}    \CRT_M^2=\CRT_V^2\CRT_{\bar V}^2
 =(-1)^{2(j_L+j_R)}.
\end{align}
Thus the neutral-primary spectrum is GOE when $j_L+j_R$ is integral and GSE when it is half-integral. Table \ref{table:SU2_classes} summarizes the $SU(2)_k$ symmetry classes.

\begin{table}[H]
\centering
\begin{tabular}{lcc}
\toprule
Current-module labels $(j_L,j_R)$ & $\epsilon_L\epsilon_R$ & Neutral class\\
\midrule
$j_L+j_R\in\mathbb Z$ & $+1$ & GOE\\
$j_L+j_R\in\mathbb Z+\tfrac12$ & $-1$ & GSE\\
\bottomrule
\end{tabular}
\caption{Wigner--Dyson classes of the neutral-primary blocks for $SU(2)_L\times SU(2)_R$, under the same antiunitary-symmetry assumptions as Table~\ref{table:U1_classes}.}
\label{table:SU2_classes}
\end{table}

As a check, $su(2)_1$ coincides with the extended $u(1)_1$ current algebra, with $J^\pm$ identified with $\mathcal U_{\pm1}$. Since the $U(1)$ charge is twice the $SU(2)$ Cartan charge, their characters obey
\begin{equation}
\chi_j^{SU(2)_1}(\tau,z)
 =\chi_{m=2j}^{U(1)_1}(\tau,z/2),
\qquad j=0,\tfrac12.
\end{equation}
The four sectors $(0,0)$, $(\tfrac12,0)$, $(0,\tfrac12)$, and $(\tfrac12,\tfrac12)$ in  $su(2)_1$ have classes GOE, GSE, GSE, and GOE. They correspond to the $(0,0)$, $(1,0)$, $(0,1)$, and $(1,1)$ sectors of $u(1)_1$ with the same respective symmetry classes and  ramp weights $2,8,8,32$.

Finally, twisted sectors contain the same neutral-primary spectra as their untwisted partners. Their classes are assigned using those original labels, as in the $U(1)_k$ case. 

\section{Holographic CFTs with continuous global symmetries}
\label{sec:continuous-holography}

We now specialize to 2d holographic CFTs of large central charge, and identify how the RMT structure of Section~\ref{sec:continuous-CFT} is encoded in the dual gravitational description. 
We consider a holographic CFT with $c_L=c_R$ and a $G_L\times G_R$ symmetry with holomorphic and antiholomorphic  current algebras at level $k$, described in the bulk by gravity coupled to $G_k\times G_{-k}$ Chern-Simons theory. As in our holographic models with boundary categorical symmetry in Section~\ref{S:holoCat}, the torus wormhole is a sum
\begin{equation}
    Z_{\rm wormhole}(X_1,X_2) = \sum_{\gamma\in SL(2,\mathbb{Z})}Z_{\rm seed}(X_1,\gamma\cdot X_2)\,, \qquad X_i = (\tau_i,\mathbf{z}_i)\,,
    \label{eq:PI-modular-sum}
\end{equation}
where $\mathbf{z}_i=(z_{i,L},z_{i,R})$ are the chemical potentials for the left- and right-moving charges on the $i^{\rm th}$ boundary. The seed amplitude is the factorized product of the gravitational seed of~\cite{Cotler:2020ugk} times a $G_k \times G_{-k}$ Chern-Simons path integral $Z_{\rm CS}^{\rm PI}$,
\begin{equation}
    Z_{\rm seed}(X_1,X_2) = Z_{\rm grav}(\tau_1,\tau_2)Z_{\rm CS}^{\rm PI}(\tau_1,\mathbf{z}_1;\tau_2,\mathbf{z}_2)\,.
    \label{eq:seed-factorization}
\end{equation}
Like the gravitational seed, the Chern-Simons amplitude comes from an integral over moduli and edge modes, the latter being the flavor counterparts of the reparameterization modes~\cite{Cotler:2018zff} of pure AdS$_3$ gravity. Each chiral half is a Wess-Zumino-Witten torus partition function. For example, the $G_k$ amplitude is a $G_k$ WZW torus partition function with the left-movers placed on the 1 boundary and the right-movers placed on the 2-boundary, together with an orientation flip so that they become left-moving. 

Note that while the Chern-Simons and gravitational degrees of freedom are locally decoupled, they are effectively correlated through the $SL(2;\mathbb{Z})$ sum over the different gluings that produce torus wormholes.

The path-integral seed is simultaneously modular invariant, which implies that the wormhole is invariant under independent modular transformations. Both the canonical and path-integral seeds are elliptically covariant on each boundary and in each chirality. The Chern-Simons amplitude with twisted boundary conditions, which encodes spectral statistics in twisted sectors of the boundary CFT, can be obtained by taking the amplitude above and acting with fractional spectral flows. 

To extract the statistics of energy levels from the wormhole amplitude, which encodes a contribution to $\langle Z^{\rm PI}(X_1)Z^{\rm PI}(X_2)\rangle_c$, we must pass from the path integral version of the torus amplitude to the canonical Hilbert space trace, strip off VKM descendants, and then project onto fixed spin and representation sectors. In Subsections~\ref{sec:U1-holography} and~\ref{sec:nonabelian-holography} we do this when $G=U(1)$ and $G=SU(2)$ respectively, which gives the expected RMT behavior, along with corrections required by modular invariance which we discuss in Subsection~\ref{sec:jacobi-corrections}. In particular, these modular corrections have a continuous-current analogue of the pre-Kuznetsov structure discussed in Subsection~\ref{sec:fourier}, with labels for both the left- and right-moving representations.

\subsection{Extracting spectral statistics}
\label{sec:CS-kernel}

The Chern-Simons amplitude for $G_k$ Chern-Simons theory is well known \cite{Witten:1988hf, Moore:1989yh, Elitzur:1989nr}. In canonical quantization, it is the kernel of the identity operator on the finite-dimensional Chern-Simons Hilbert space on $\mathbb{T}^2$, expressed in the polarizations selected by the two boundaries. We first construct this canonical amplitude, and then convert it to the path integral $Z_{\rm CS}^{\mathrm{PI}}$ appearing in~\eqref{eq:seed-factorization}. From there we sum over modular images, and interpret the result as a spectral correlator.

\paragraph{Phase space and quantization on $\mathbb{T}^2$.} Consider Chern-Simons theory with compact gauge group $G$ at level $k$ on the torus wormhole 
\begin{equation}
    S=\frac{k}{4\pi}\int \Tr\left(A\wedge dA+\frac{2}{3}\,A\wedge A\wedge A\right).
    \label{eq:CS-action-general}
\end{equation}
The action for $G_k\times G_{-k}$ Chern-Simons theory is $S[A_L]-S[A_R]$. The $G_k$ and $G_{-k}$ factors are decoupled and can be treated separately. We present the quantization of the $G_k$ factor below. For $G=U(1)$ we normalize the trace so that the action is $\frac{2k}{4\pi}\int A\wedge dA$.

In canonical quantization we single out time, $A_0$ enters \eqref{eq:CS-action-general} linearly, so that it becomes a Lagrange multiplier enforcing the flatness of the spatial components of the field strength,\footnote{This remains true in Coulomb gauge.}
\begin{equation}
    F\big|_{\mathbb{T}^2}=0\,.
    \label{eq:flatness-constraint}
\end{equation}
The physical phase space is therefore the moduli space of flat $G$-connections on the spatial torus,\footnote{We assume that if $G$ is non-abelian, then it is simply connected.}
\begin{equation}
    \mathcal P(\mathbb{T}^2)
    =\mathrm{Hom}\bigl(\pi_1(\mathbb{T}^2),G\bigr)/G
    \simeq (T\times T)/W ,
    \label{eq:CS-phase-space}
\end{equation}
which is parametrized by the two commuting holonomies, each valued in the maximal torus $T$ of the gauge group up to the Weyl group $W$ that preserves the maximal torus. Writing a flat connection as $A=\frac{1}{2\pi}\left(\alpha\,dx+\beta\,dy\right)$, with $x,y\in[0,2\pi)$ the coordinates on the two cycles and $\alpha,\beta$ the commuting Cartan-valued holonomy angles, the action \eqref{eq:CS-action-general} reduces to a kinetic term that equips the phase space with the symplectic form proportional to the level
\begin{equation}
    \omega=\frac{k}{2\pi}\,\Tr\left(d\alpha\wedge d\beta\right).
    \label{eq:CS-symplectic}
\end{equation}
 Since the phase space is compact, quantization produces a finite-dimensional Hilbert space $\cH_{\rm CS}(\mathbb{T}^2)$.

In $\cH_{\rm CS}(\mathbb{T}^2)$, there is a natural basis of states $\left\{|\lambda \rangle \right\}$, where $\lambda$ labels integrable representations of the affine $G_k$ current algebra, which is prepared by the path integral on a solid torus $D^2\times S^1$ with a Wilson line in the representation $\lambda$ at the center of the disk. The wavefunction of  $|\lambda\rangle$ can be obtained canonically from the boundary data. The boundary complex structure $\tau$ selects a holomorphic polarization of the phase space, in which we use $z$ as  a complex coordinate on the moduli space of flat connections, constructed from the two commuting holonomy variables $\alpha$ and $\beta$ using $\tau$. Denoting by   $|\tau,z\rangle$ the state associated with this holomorphic polarization, the wavefunction $\Psi_\lambda(\tau,z)= \langle \tau,z|\lambda\rangle$ is a well-known result
\cite{Witten:1988hf, Elitzur:1989nr}
\begin{equation}
    \Psi_\lambda(\tau,z)= \chi_{\lambda}(\tau,z)\,,
    \label{eq:CS-wavefunctions}
\end{equation}
i.e. the character of the boundary extended chiral algebra, with $\lambda$ running over the finitely many integrable representations at level $k$. In our conventions, $\langle \lambda|\tau,z\rangle = \chi_{\lambda}(\tau,z)^* = \chi_{\lambda}(-\bar{\tau},-\bar{z})$. In presenting Eq.~\eqref{eq:CS-wavefunctions} we have fixed a holomorphic frame and written the character as a trace. 

\paragraph{The canonical Chern-Simons amplitude.}
The Chern-Simons amplitude on $\mathbb{T}^2\times I$ is defined as a trivial cobordism from the torus to itself. It is therefore the identity map on $\cH_{\rm CS}(\mathbb{T}^2)$,
\begin{equation}
    K_{G}
    =\mathbf 1
    =\sum_{\lambda}|\lambda\rangle\langle \lambda|.
    \label{eq:CS-identity-kernel}
\end{equation}
The canonical version of the $G_k$ Chern-Simons amplitude is, after using the orientation flip on the second boundary,
\begin{equation}
    \mathcal{K}_G(\tau_1,z_1;\tau_2,z_2) =\langle \tau_1,z_1|\mathbb{1}|-\bar{\tau}_2,-\bar{z}_2\rangle =  \sum_{\lambda} \chi_{\lambda}(\tau_1,z_1)\chi_{\lambda}(\tau_2,z_2)\,.
    \label{eq:CS-kernel-general}
\end{equation}
This is a diagonal sum of extended characters of the boundary current algebra. 

So much for a single chiral half. For the $G_k\times G_{-k}$ amplitude, we label the complex structures and chemical potentials as in Subsection~\ref{sec:sugawara-general},
\begin{equation}
    X_i=(\tau_i,\mathbf z_i),\qquad \mathbf z_i=(z_{i,L},z_{i,R}),\qquad
    \tau_{i,L}=\tau_i,\quad \tau_{i,R}=-\bar\tau_i,\quad
    q_{i,a}=e^{2\pi i\tau_{i,a}}\quad(a=L,R),
\end{equation}
so that the dual CFT partition function is a Hilbert space trace over a factor whose flavor part is $e^{2\pi i(z_{i,L}Q_L+z_{i,R}Q_R)}$, as in~\eqref{eq:CFT-canonical-trace}. Using~\eqref{eq:CS-kernel-general} then gives the canonical version of the $G_k\times G_{-k}$ Chern-Simons amplitude
\begin{equation}
\begin{aligned}
 Z_{\rm CS}(\tau_1,\mathbf z_1;\tau_2,\mathbf z_2)
 &=\prod_{a=L,R}\mathcal K_G(\tau_{1,a},z_{1,a};\tau_{2,a},z_{2,a})=\sum_{\lambda_L,\lambda_R}\prod_{i=1}^2
   \chi_{\lambda_L}(\tau_i,z_{i,L})
   \chi_{\lambda_R}(-\bar\tau_i,z_{i,R}).
\end{aligned}
\label{eq:CS-kernel-paired}
\end{equation}
Note that the labels $\lambda_L$ and $\lambda_R$ are completely independent.

\paragraph{The covariant Chern-Simons amplitude.}
The canonical Chern-Simons partition function $Z_{\rm CS}$ transforms up to a multiplicative factor under simultaneous modular transformations. That factor is a function of the complex structures and holonomies. We convert to a covariant path integral version of the amplitude using the same factor as in our discussion in~\eqref{eq:CFT-PI-canonical}. For $U(1)$ or $SU(2)$ global symmetry it involves 
\begin{equation}
\mathcal G_G(X)=\exp\!\left[\frac{\pi\kappa_G}{\operatorname{Im}\tau}
 (z_L^2+z_R^2)\right],\qquad
 \kappa_{U(1)}=k,\quad \kappa_{SU(2)}=\frac{k}{4}\,.
\label{eq:canonical-PI-conversion}
\end{equation}
Here $\kappa_G$ is the Jacobi index. In terms of $\mathcal{G}_G$ the covariant version of the Chern-Simons amplitude is
\begin{equation}
Z_{\rm CS}^{\mathrm{PI}}(X_1;X_2)
 =\mathcal G_G(X_1)\mathcal G_G(X_2)Z_{\rm CS}(X_1;X_2)\,.
\label{eq:CS-PI-kernel}
\end{equation}
This is simultaneously modular invariant. Writing an $SL(2;\mathbb{Z})$ transformation as $\gamma=\left(\begin{smallmatrix}a&b\\c&d\end{smallmatrix}\right)$ which acts on $X$ as
\begin{equation}
\gamma\cdot X
 =\left(\frac{a\tau+b}{c\tau+d},\frac{z_L}{c\tau+d},
 \frac{z_R}{c\bar\tau+d}\right)\,,
\label{eq:paired-source-modular-action}
\end{equation}
we have
\begin{equation}
Z_{\rm CS}^{\mathrm{PI}}(\gamma\cdot X_1;\gamma^\sharp\cdot X_2)
 =Z_{\rm CS}^{\mathrm{PI}}(X_1;X_2)\,,
\label{eq:CS-PI-invariance}
\end{equation}
with $\gamma^\sharp=\left(\begin{smallmatrix}a&-b\\-c&d\end{smallmatrix}\right)$. The full seed~\eqref{eq:seed-factorization} is then simultaneously modular invariant.

\paragraph{The full amplitude and spectral data.}
With the wormhole amplitude~\eqref{eq:PI-modular-sum} in hand we now show how to extract the spectral statistics it describes. To do so we perform the modular sum over the seed, convert back to a two-point function of Hilbert space traces, strip off VKM descendants, and then Fourier transform to fixed spin and charge. To carry this out, we define a canonical version of the wormhole amplitude, 
\begin{equation}
\begin{aligned}
 Z_{\rm wormhole}^{\mathrm{can}}(X_1;X_2)
 &=\frac{Z_{\rm wormhole}(X_1;X_2)}
 {\mathcal G_G(X_1)\mathcal G_G(X_2)},\\
 Z_{\rm seed}^{\mathrm{can}}(X_1;X_2)
 &=\frac{Z_{\rm seed}(X_1;X_2)}
 {\mathcal G_G(X_1)\mathcal G_G(X_2)}
 =Z_{\rm grav}(\tau_1;\tau_2)Z_{\rm CS}(X_1;X_2)\,,
\end{aligned}
\label{eq:PI-seed-conversion}
\end{equation}
which we interpret as contributing to the connected two-point function of Hilbert space traces,
\begin{equation}
    \langle Z^{\rm can}(X_1)Z^{\rm can}(X_2)\rangle_c \simeq Z^{\rm can}_{\rm wormhole}(X_1;X_2) \,.
\end{equation}
(The $\simeq$ summarizes our interpretation that (a) there is a provisional CFT correlation dual to the two-boundary amplitude, and (b) that the wormhole contributes to it, but other geometries may too.)
We can write this canonical amplitude as
\begin{equation}
Z_{\rm wormhole}^{\mathrm{can}}(X_1;X_2)
 =\sum_{\gamma\in SL(2,\mathbb Z)}
 \frac{\mathcal G_G(\gamma\cdot X_2)}{\mathcal G_G(X_2)}
 Z_{\rm seed}^{\mathrm{can}}(X_1;\gamma\cdot X_2).
\label{eq:canonical-modular-sum}
\end{equation}

Next, we strip off the VKM descendant contributions. Writing $\Lambda = (\lambda_L,\lambda_R)$, we define
\begin{equation}
\begin{aligned}
 \Xi_\Lambda(X)&=\chi_{\lambda_L}(\tau,z_L)
                  \chi_{\lambda_R}(-\bar\tau,z_R)\,,\\
 U^G_{\Lambda_1\Lambda_2}(\gamma)
 &=\rho_G(\gamma)_{\lambda_{1,L}\lambda_{2,L}}\,
   \overline{\rho_G(\gamma)_{\lambda_{1,R}\lambda_{2,R}}}\,,
\end{aligned}
\label{eq:paired-modular-matrix}
\end{equation}
where $\rho_G$ is the modular representation of the chiral characters. With a little (modular) algebra the canonical wormhole amplitude becomes
\begin{equation}
Z_{\rm wormhole}^{\mathrm{can}}(X_1;X_2)
 =\frac{Z_0(\tau_1)Z_0(\tau_2)}{2\pi^2}
 \sum_{\Lambda_1,\Lambda_2}
 \Xi_{\Lambda_1}(X_1)\Xi_{\Lambda_2}(X_2)
 \mathcal M^G_{\Lambda_1\Lambda_2}(\tau_1,\tau_2)\,,
\label{eq:full-character-expansion}
\end{equation}
where
\begin{equation}
\begin{aligned}
 \mathcal M^G_{\Lambda_1\Lambda_2}(\tau_1,\tau_2)
 &=\sum_{\gamma\in SL(2,\mathbb Z)}
 U^G_{\Lambda_1\Lambda_2}(\gamma)\,
 \mathcal W_\gamma(\tau_1,\tau_2)\,,\\
 \mathcal W_\gamma(\tau_1,\tau_2)
 &=\frac{\text{Im}(\tau_1)\text{Im}(\gamma\cdot \tau_2)}{|\tau_1+\gamma\cdot \tau_2|^2}\,,
 \qquad y_i=\operatorname{Im}\tau_i\,.
\end{aligned}
\label{eq:full-modular-coefficient}
\end{equation}

Removing neutral Virasoro descendants replaces $Z_0(\tau_i)$ by $|q_i|^{-1/12}/\sqrt{y_i}$. Removing Kac-Moody descendants replaces $\Xi_\Lambda$ by the primary-counting character $\Xi_\Lambda^P$, which we describe for $G=U(1)$ and $G=SU(2)$ in the next Subsections. From this we obtain 
\begin{equation}
 \big\langle Z^{P,\mathrm{can}}(X_1)Z^{P,\mathrm{can}}(X_2)\big\rangle_c
 \simeq \frac{|q_1q_2|^{-1/12}}{2\pi^2\sqrt{y_1y_2}}
 \sum_{\Lambda_1,\Lambda_2}
 \Xi_{\Lambda_1}^{P}(X_1)\Xi_{\Lambda_2}^{P}(X_2)
 \mathcal M^G_{\Lambda_1\Lambda_2}(\tau_1,\tau_2)\,.
\label{eq:full-spectral-projection}
\end{equation}
From here we must Fourier transform to fixed spin and project onto representations. The terms in the modular sum with $c=0$ encode a ramp as predicted by our categorical RMT proposal, while the $c>0$ terms encode modular corrections.

\subsection{Example: $U(1)$ global symmetry}
\label{sec:U1-holography}

In this section we work out the details of the torus wormhole amplitude and its spectral interpretation when $G=U(1)$. 

\paragraph{Wormhole seed.}
We begin with the canonical $U(1)_k$ Chern-Simons amplitude. It reads
\begin{align}
   \mathcal{K}_{\rm U(1)_k}(\tau_1,z_1;\tau_2,z_2)= \left(\eta(\tau_1)\eta(\tau_{2})\right)^{-1}\sum_{p,w=-\infty}^{\infty} q_1^{\frac{k}{4}(
\frac{p}{k}+w)^2}q_2^{\frac{k}{4}(\frac{p}{k}-w)^2} e^{2\pi i z_1(p+kw)}e^{2\pi i z_2(p-kw)}\,,
\label{U(1)_seed_2sum}
\end{align}
which we can also write in terms of the theta function $\chi_m^k$ defined in ~\eqref{2k-theta_function} as
\begin{align}
    \mathcal K_{\rm U(1)_k}&(\tau_1,z_1;\tau_2,z_2) =\sum_{m=0}^{2k-1}  \chi_m^k(\tau_1,z_1)  \chi_m^k(\tau_2,z_2) 
\label{U(1)_seed_3sum}
\\
    &=\left(\eta(\tau_1)\eta(\tau_{2})\right)^{-1}\sum_{w_1,w_2=-\infty}^{\infty}\sum_{p=0}^{2k-1} q_1^{\frac{k}{4}(\frac{p}{k}+2w_1)^2}e^{2\pi i z_1(p+2kw_1)}~ q_2^{\frac{k}{4}(\frac{p}{k}+2w_2)^2}e^{2\pi i z_2(p+2kw_2)}\,.
    \label{U(1)_seed_Wi}
\end{align}

It is straightforward to verify that this amplitude has the modular transformation law~\eqref{U(1)_modular} under the simultaneous modular transformations of Subsection~\ref{sec:CS-kernel}, which here amounts to
\begin{align}
       \mathcal{K}_{U(1)_k}\left(\gamma\cdot X_1;\gamma^{\sharp}\cdot X_2\right)=e^{ \frac{2\pi i k c z_1^2}{c\tau_1+d}+\frac{2\pi ik c z_2^2}{c\tau_2-d}} \mathcal{K}_{U(1)_k}(X_1;X_2),
       \label{equ:U(1)_simul_modular}
\end{align} 
The simplest way to show this is to use that the $U(1)_k$ characters transform covariantly under modular transformations with symmetric and unitary $T$ and $S$ matrices representing the action of $SL(2;\mathbb{Z})$, with
\begin{align}
\begin{split}
        \chi_{m}^k(\tau+1,z)&=\sum_{\lambda =0}^{2k-1} T_{m\lambda} \chi_{\lambda}^k(\tau,z)\,,
        \\
    \chi_{m}^k\left(-\frac{1}{\tau},\frac{z}{\tau}\right)&=e^{\frac{2\pi ikz^2}{\tau}}\sum_{\lambda =0}^{2k-1} S_{m\lambda} \chi_{\lambda}^k(\tau,z)\,.
\end{split}
\label{eq:U1_ST_matrix}
\end{align}

The canonical amplitude also transforms elliptically upon relabeling $Q_{1} \to Q_{1}+2ku_1$ and $Q_2\to Q_2 + 2ku_2$ with $u_{1},u_2$ valued in the charge lattice of~\eqref{U(1)_seed_2sum}. This preserves the condition $Q_{1}\equiv Q_{2}\pmod{2k}$, so that in terms of $X_1'=(\tau_1,z_1'=z_1 + u_1 \tau_1)$ and $X_2' = (\tau_2,z_2'=z_2+u_2\tau_2)$ we have
\begin{equation}
\mathcal K_{\rm U(1)_k}(X_1';X_2')
 =e^{-2\pi i k(u_1^2\tau_1+2u_1 z_1+u_2^2\tau_2+2u_2z_2)}
 \mathcal K_{\rm U(1)_k}(X_1;X_2)\,.
\label{eq:U1-paired-elliptic}
\end{equation}

\paragraph{Twisted sectors.} Our expression~\eqref{U(1)_seed_3sum} for the $U(1)_k$ seed applies when the dual CFTs are in the untwisted sector. The result when the CFTs are in twisted sectors can be obtained from it by inserting a topological line $U_{\zeta_{i}}$, with $i=1,2$, along the $y$-circle (the Euclidean time) of each torus boundary. In the bulk, this fixes an extra holonomy $2\pi\zeta_i$, with $\zeta_{1,2}\in (0,1)$, for the gauge field around the corresponding boundary $x$-circle. This extra holonomy can be easily encoded into the seed by noting that the sums over $w_{1,2}$ in~\eqref{U(1)_seed_Wi} can be understood as sums over the winding sectors of bulk large gauge transformations at the two boundaries. The extra holonomy introduced by the topological line fractionally shifts these winding modes, $w_{1,2} \rightarrow w_{1,2}+\zeta_{1,2}$. Viewed from the boundary CFT, this shift is precisely the fractional spectral flow that defines the twisted sectors $\mathcal{H}_{\zeta}$ in Subsection~\ref{sec:U1-spectral-flow}. The twisted $U(1)$ seed is then
\begin{equation}
 \mathcal K_{\rm U(1)_k}^{\zeta_1,\zeta_2}(\tau_1,z_1;\tau_2,z_2)=\frac{1}{\eta(\tau_1)\eta(\tau_2)}
 \sum_{w_1,w_2\in\mathbb Z}\sum_{p=0}^{2k-1}
 \prod_{i=1}^2 q_i^{[p+2k(w_i+\zeta_i)]^2/(4k)}
 e^{2\pi i z_i[p+2k(w_i+\zeta_i)]}\,.
\label{U(1)_seed_3sum_twisted}
\end{equation}
Relabeling the three sums into two sums, we have
\begin{align}
\begin{split}
    & \mathcal K_{\rm U(1)_k}^{\zeta_1,\zeta_2}(\tau_1,z_1;\tau_2,z_2)=\left(\eta(\tau_1)\eta(\tau_{2})\right)^{-1}   \\
     &\hspace{1in}\times\sum_{p,w=-\infty}^{\infty}q_1^{\frac{k}{4}(\frac{p}{k}+(w+2\zeta_1))^2}q_2^{\frac{k}{4}(\frac{p}{k}-w+2\zeta_2)^2}e^{2\pi i z_1(p+k(w+2\zeta_1))}e^{2\pi i z_2(p-k(w-2\zeta_2))}\,.
     \label{U(1)_seed_twisted}
\end{split}
\end{align}

Now, for the canonical version of the $U(1)_k\times U(1)_{-k}$ amplitude, let $\boldsymbol\zeta_i=(\zeta_{i,L},\zeta_{i,R})$ denote the fractional spectral-flow parameters, with the convention $Q_{i,a}\to Q_{i,a}+2k\zeta_{i,a}$. This fixes the sign of the right-moving twist. The paired amplitude in a general background is then
\begin{equation}
Z_{\rm CS}^{\boldsymbol\zeta_1,\boldsymbol\zeta_2}
 =\prod_{a=L,R}\mathcal K_{\rm U(1)_k}^{\zeta_{1,a},\zeta_{2,a}}
 (\tau_{1,a},z_{1,a};\tau_{2,a},z_{2,a}).
\end{equation}
So, apart from inducing a fractional spectral flow from the untwisted seed, the twisted seed contains essentially no new information. For this reason, in the following discussion we will focus only on the statistics in the untwisted sector.

\paragraph{Spectral data and the RMT contribution.}
We now write the wormhole amplitude in the form~\eqref{eq:full-spectral-projection}, from which we can more easily transform to fixed spin and charges. The canonical seed is
\begin{equation}
Z_{\rm seed}^{\mathrm{can}}(X_1;X_2)
 =Z_{\rm grav}(\tau_1;\tau_2)
 \prod_{a=L,R}\mathcal K_{\rm U(1)_k}
 (\tau_{1,a},z_{1,a};\tau_{2,a},z_{2,a})\,.
\label{U(1)_full_seed}
\end{equation}
There are infinitely many primary states charged under the left-moving $U(1)$ , from which we build a primary-counting character
\begin{equation}
\chi_m^{k,P}(\tau,z)
 =q^{-1/24}\sum_{\substack{Q\in\mathbb Z\\Q\equiv m\ ({\rm mod}\ 2k)}}
 q^{Q^2/(4k)}e^{2\pi i zQ}\,,
\label{eq:U1-primary-character}
\end{equation}
in terms of which the object $\Xi^P_{\textbf{m}}$ with $\textbf{m}=(m_L,m_R)$ is $\Xi^P_{\textbf{m}}=\chi^{k,P}_{m_L}\chi^{k,P}_{m_R}$. The full primary-counting amplitude is~\eqref{eq:full-spectral-projection} with $\mathcal{M}^{U(1)}$ summed over all of $SL(2;\mathbb{Z})$.

Let us parameterize $\tau_i=x_i+i\beta_i$, $\mathbf Q_i=(Q_{i,L},Q_{i,R})$, and $\mathbf m_i\equiv\mathbf Q_i \text{ mod}(2k)$. Since $c_L=c_R$, the physical spins $s_i$ are integers. Define the neutral spin $\hat{s}$ at fixed spin $s$ and charge $\mathbf{Q}$,
\begin{equation}
 \hat s=s-\frac{Q_{L}^2-Q_{R}^2}{4k}\,.
 \label{eq:sigma_epsilon_U1}
\end{equation}
Now, Fourier transforming to fixed left- and right-charge on each boundary selects specified charges in the primary-counting characters. Fourier transforming to fixed spins gives
\begin{equation}
\begin{aligned}
 \langle Z^P_{s_1,\textbf{Q}_1}(\beta_1)Z^P_{s_2,\textbf{Q}_2}(\beta_2)\rangle_c \simeq \frac{
 e^{-2\pi\sum_i\beta_i\left( \frac{Q_{i,L}^2+Q_{i,R}^2}{4k}-1/6\right)}}
 {2\pi^2\sqrt{\beta_1\beta_2}}
 \int_0^1dx_1\int_0^1dx_2\,
 e^{-2\pi i(\hat s_1x_1+\hat s_2x_2)}
 \mathcal M^{U(1)}_{\mathbf Q_1\mathbf Q_2}(\tau_1,\tau_2).
\end{aligned}
\label{eq:U1-full-sector-correlator}
\end{equation}

\begin{figure}[htbp]
\centering
\begin{tikzpicture}[x=1.1cm,y=0.85cm,>=Stealth]
 \draw[->,black] (-2.65,0)--(2.75,0) node[right] {$s$};
 \draw[->,black] (0,-0.45)--(0,4.55) node[above] {$E_{s,\mathbf Q}$};
 \foreach \x in {-2,-1,1,2}{\draw[black] (\x,0.06)--(\x,-0.06) node[below] {\small $\x$};}
 \draw[dashed,black,thick] (-2.4,2.233333)--(0,-0.166667)--(2.4,2.233333);
 \draw[cbBlue,very thick] (-2.4,4.233333)--(0.75,1.083333)--(2.4,2.733333);
 \fill[cbBlue] (0.75,1.083333) circle (1.5pt);
 \node[cbBlue,anchor=south] at (0.75,1.8) {$\left(\frac34,\frac{13\pi}{6}\right)$};
 \draw[cbBlue,very thick] (0.65,4.25)--(1.1,4.25);
 \node[anchor=west,black] at (1.15,4.25) {\small $(Q_L,Q_R)=(2,1)$};
 \draw[dashed,black,thick] (0.65,3.8)--(1.1,3.8);
 \node[anchor=west,black] at (1.15,3.8) {\small $(Q_L,Q_R)=(0,0)$};
\end{tikzpicture}
\caption{The threshold energy at fixed spin for $k=1$. The dashed line is the uncharged threshold $E_{s,\mathbf 0}=2\pi(|s|-1/6)$, and the solid blue line is~\eqref{eq:U1-paired-threshold} for $(Q_L,Q_R)=(2,1)$.}
\label{sf_fig}
\end{figure}
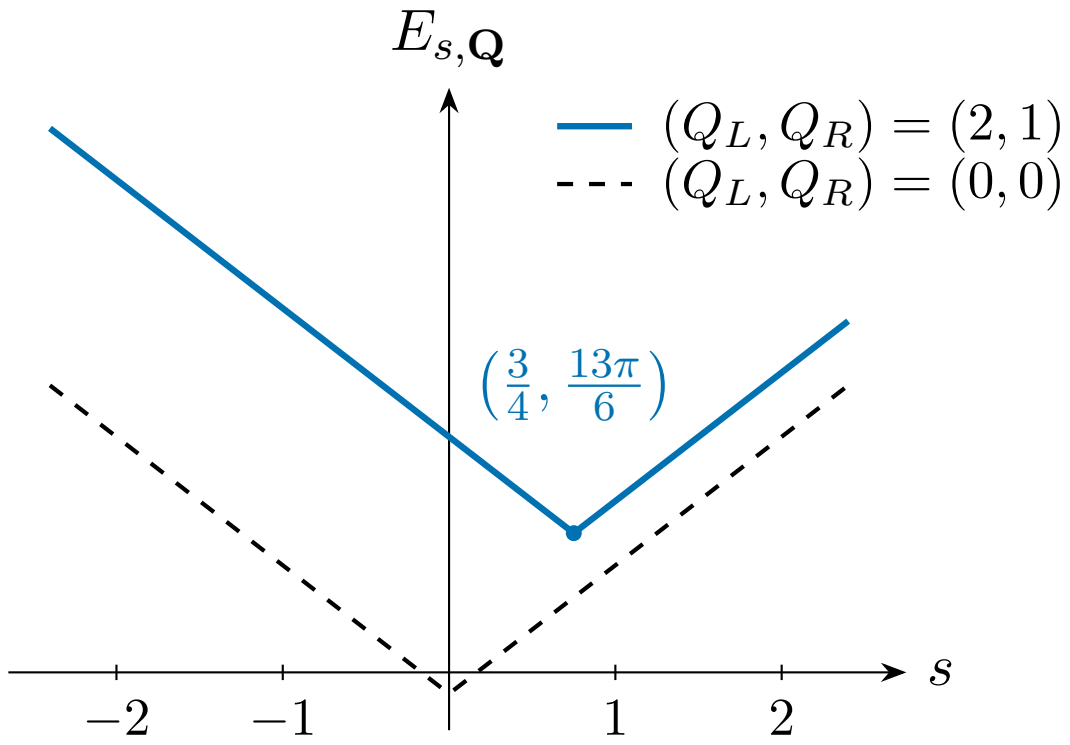

The $c=0$ part of the sum implicit in $\mathcal{M}$ accounts for the RMT ramp as in pure gravity and in our categorical examples in Section~\ref{S:holoCat}.  We now evaluate it. It includes $\gamma=T^n$, and their $-\mathbf1$ images. We have
\begin{equation}
U^{U(1)}_{\mathbf Q_1\mathbf Q_2}(T^n)
 =\delta_{\mathbf m_1\mathbf m_2}
 e^{2\pi i n(Q_{1,L}^2-Q_{1,R}^2)/(4k)},\qquad
 \mathcal W_{T^n}=\frac{\beta_1\beta_2}
 {|\tau_1+\tau_2+n|^2}\,.
\end{equation}
These character matrices $U^{U(1)}$ can also be obtained by writing out the primary-counting seed directly as (with $Z^P_{\rm grav}$ given in~\eqref{equ:CJ_primary_sec3})
\begin{equation}
\begin{aligned}
 Z_{\rm seed}^P(X_1;X_2)
 &=Z^P_{\rm grav}(\tau_1;\tau_2)\prod_{a=L,R}\Bigg[
 (q_{1,a}q_{2,a})^{-1/24}\\[-2pt]
 &\qquad\times\sum_{p_a,w_a\in\mathbb Z}
 q_{1,a}^{(p_a+kw_a)^2/(4k)}q_{2,a}^{(p_a-kw_a)^2/(4k)}
 e^{2\pi i[z_{1,a}(p_a+kw_a)+z_{2,a}(p_a-kw_a)]}\Bigg]\,.
\end{aligned}
\label{prim_seed}
\end{equation}
The $-\mathbf{1}$ effectively implements $\CRT$. Its images generate the $T^n$ terms with a flip in the sign of $\textbf{Q}_2$. By~\eqref{eq:full-modular-coefficient} we then have
\begin{equation}
    \mathcal{M}^{U(1)}_{\textbf{Q}_1\textbf{Q}_2} = \left( \delta_{\textbf{m}_1\textbf{m}_2}+ \delta_{\textbf{m}_1,-\textbf{m}_2}\right)\sum_{n=-\infty}^{\infty} \frac{\beta_1\beta_2 e^{2\pi i n\frac{Q_{1,L}^2-Q_{1,R}^2}{4k}}}{(x_1+x_2+n)^2+(\beta_1+\beta_2)^2} + (c>0)\,,
\end{equation}
and the transformation to fixed spin yields
\begin{equation}
 \langle Z^P_{s_1,\mathbf Q_1}(\beta_1)Z^P_{s_2,\mathbf Q_2}(\beta_2)\rangle_c
 \simeq \frac{1}{2\pi}\frac{\sqrt{\beta_1\beta_2}}{\beta_1+\beta_2}\,
 \widetilde{\delta}_{12}\,
 e^{-\beta_1 E_{s_1,\mathbf Q_1}-\beta_2E_{s_2,\mathbf{Q}_2}} + (c>0)\,,
\label{ramp_U(1)_bulk}
\end{equation}
where the remainder accounts for the $c>0$ terms. The BTZ threshold energy $E_{s,\textbf{Q}}$ at spin $s$ and charge $\textbf{Q}$ is 
\begin{equation}
\begin{aligned}
 \frac{E_{s,\mathbf Q}}{2\pi}
 &=\left|s-\frac{Q_L^2-Q_R^2}{4k}\right|
   +\frac{Q_L^2+Q_R^2}{4k}-\frac16=\begin{cases}
 s+\dfrac{Q_R^2}{2k}-\dfrac16,
   &s\geq\dfrac{Q_L^2-Q_R^2}{4k},\\[4pt]
 -s+\dfrac{Q_L^2}{2k}-\dfrac16,
   &s<\dfrac{Q_L^2-Q_R^2}{4k}.
 \end{cases}
\end{aligned}
\label{eq:U1-paired-threshold}
\end{equation}
and
\begin{equation}
\widetilde{\delta}_{12}
 =\delta_{\hat s_1,\hat s_2}
 \left(\delta_{\mathbf m_1,\mathbf m_2}+\delta_{\mathbf m_1,-\mathbf m_2}\right)\,.
\label{SF_delta_bulk}
\end{equation}
In Fig.~\ref{sf_fig}, we compare the uncharged threshold energy $E_{s,\mathbf{0}}$ with the  one shifted by charge $(Q_L,Q_R)=(2,1)$ for $k=1$.

Eq.~\eqref{ramp_U(1)_bulk} is the main result of this Subsection. It describes long-range repulsion at fixed spin and charge, precisely realizing the expectations described in Subsection~\ref{sec:U1-CFT} as we now explain. 

The result~\eqref{ramp_U(1)_bulk} implies a linear ramp in the spin- and charge-resolved SFF when the neutral spins coincide, and the charges either coincide or are opposite, at least $(\text{mod }2k)$. That is, there are correlations between sectors of identical neutral spin but different charge, for example where the left-charges differ by $2k$. This repulsion between different charge sectors arises because each $U(1)_k$ algebra is extended by the spectral-flow operators $\mathcal U_{\pm1}$ of~\eqref{equ:SF_vertex_opt}, with charges $\pm2k$, so that primary states of the extended algebra are labeled by $m_a=Q_a\bmod2k$.

Resolved with respect to the extended algebra the SFF has the tensor structure
\begin{equation}
    \langle Z^P_{s_1,\mathbf{m}_1}(\beta_1)Z^P_{s_2,\mathbf{m}_2}(\beta_2) \rangle_c \propto \delta_{\hat{s}_1,\hat{s}_2}\left( \delta_{m_{1,L},m_{2,L}}\delta_{m_{1,R},m_{2,R}} + \delta_{m_{1,L},-m_{2,L}}\delta_{m_{1,R},-m_{2,R}}\right) + (c>0)\,,
\end{equation}
which is what we anticipate from a RMT ansatz with $\CRT$, so that there is level repulsion between sectors of identical neutral spin and charges, or identical neutral spin and opposite charges.

Let us unpack the consequences of $\CRT$ in detail. Since $\CRT$ acts as $(m_L,m_R)\to (-m_L,-m_R)$, a RMT ansatz would predict that a sector not fixed by $\CRT$ exhibit GUE self-statistics, and repel against its conjugate sector. This is exactly what we find above from the torus wormhole~\eqref{ramp_U(1)_bulk}, with the self-SFF and conjugate-SFF giving linear ramps with the GUE value for the slope $\frac{1}{2\pi}$. 

The $\CRT$-invariant sectors have $m_L$ and $m_R$ equal to $0$ or $k$. In Section~\ref{sec:U1-CFT} we argued that on the multiplicity space of fixed $(m_L,m_R)$ we have
\begin{equation}
    \CRT^2_{M} = (-1)^{k(\delta_{m_L,k}+\delta_{m_R,k})}\,,
\end{equation}
with the corresponding symmetry classes summarized in Table~\ref{table:U1_classes}.
The RMT prediction is then that the $(0,k)$ and $(k,0)$ sectors have GSE statistics for odd $k$ and GOE statistics for even $k$, while $(0,0)$ and $(k,k)$ have GOE statistics for all $k$. The torus wormhole~\eqref{ramp_U(1)_bulk} gives the GOE slope $\frac{1}{\pi}$ for all self-conjugate sectors. However, when $\CRT^2_M=-1$, the neutral states come in Kramers pairs, and so RMT prediction is that, in those cases, the self-SFF has slope $\frac{1}{4\pi} \times 2^2 = \frac{1}{\pi}$, where the $\frac{1}{4\pi}$ is the GSE prediction and $2^2$ comes from the Kramers doubling. This coincides with the GOE slope, and also matches the torus wormhole answer~\eqref{ramp_U(1)_bulk}. 

So as in our examples with discrete or categorical global symmetry, the torus wormhole amplitude recovers the RMT prediction for long-range repulsion, although one must go beyond the wormhole to distinguish whether the underlying self-conjugate sectors are of GOE- or GSE-type.

So far we have focused on the RMT contribution encoded in the $c=0$ part of the modular sum. The $c>0$ terms contribute corrections to the SFF at fixed spin and charge, which we discuss in Subsection~\ref{sec:jacobi-corrections}. 

\subsection{Example: $SU(2)$ global symmetry}
\label{sec:nonabelian-holography}
In this section, we work out the  torus wormhole kernel with nonabelian symmetry  $G=SU(2)$ and extract its predictions for spectral statistics.
\paragraph{Wormhole seed.}

The canonical $SU(2)_k$ Chern-Simons kernel for a single chiral half reads from \eqref{eq:CS-kernel-general} as

\begin{equation}
\mathcal K_{\rm SU(2)_k}(\tau_1,z_1;\tau_2,z_2)
 =\sum_{j=0,\frac12,\ldots,\frac{k}{2}}
 \chi_j^k(\tau_1,z_1)\chi_j^k(\tau_2,z_2),
\label{su(2)_seed}
\end{equation}
where $\chi_j^k$ is the affine character~\eqref{su(2)-char}. 
Combining the gravitational seed with $SU(2)_k\times SU(2)_{-k}$ kernel, the full wormhole seed  \eqref{eq:PI-seed-conversion} is therefore
\begin{equation}
\begin{aligned}
 Z_{\rm seed}^{\mathrm{can}}(X_1;X_2)
 &=Z_{\rm grav}(\tau_1;\tau_2)
 \prod_{a=L,R}\mathcal K_{\rm SU(2)_k}
 (\tau_{1,a},z_{1,a};\tau_{2,a},z_{2,a})\\
 &=Z_{\rm grav}(\tau_1;\tau_2)
 \sum_{j_L,j_R=0,\frac12,\ldots,\frac{k}{2}}
 \prod_{i=1}^2\chi_{j_L}^k(\tau_i,z_{i,L})
                    \chi_{j_R}^k(-\bar\tau_i,z_{i,R}).
\end{aligned}
\label{eq:SU2-paired-seed}
\end{equation}

The modular property of this amplitude can again be checked using the $S$ and $T$ matrices of $SU(2)_k$ characters, as in the $U(1)$ example. In the present Cartan normalization, the Jacobi index for $SU(2)_k$ is $k/4$, so the seed transforms under simultaneous modular transformations as
\begin{equation}
\begin{aligned}
 Z_{\rm seed}^{\mathrm{can}}(\gamma\cdot X_1;\gamma^\sharp\cdot X_2)
 &=e^{\Phi^{SU(2)}_\gamma(X_1)+\Phi^{SU(2)}_{\gamma^\sharp}(X_2)}
 Z_{\rm seed}^{\mathrm{can}}(X_1;X_2), \\
  \Phi^{SU(2)}_\gamma(X)
 &=\frac{\pi i k c}{2}
 \left(\frac{z_L^2}{c\tau+d}-\frac{z_R^2}{c\bar\tau+d}\right).
\end{aligned}
\end{equation}
The path integral version of the seed differs from above canonical seed by  $\mathcal G_{SU(2)}(X_1)\mathcal G_{SU(2)}(X_2)$ from~\eqref{eq:canonical-PI-conversion}, which cancels the $\Phi^{SU(2)}_\gamma(X)$ factors and leads to simultaneous modular invariance.

For the elliptic transformations, one needs to separate the genuine $SU(2)$ large gauge transformations from center twists. The characters satisfy
\begin{equation}
\begin{aligned}
 \chi_j^k(\tau,z+2u\tau+2v)
 &=e^{-2\pi i k(u^2\tau+uz)}\chi_j^k(\tau,z),
 &&u,v\in\mathbb Z,\\
 \chi_j^k(\tau,z+1)&=(-1)^{2j}\chi_j^k(\tau,z),\\
 \chi_j^k(\tau,z+\tau)
 &=e^{-2\pi i k(\tau/4+z/2)}\chi_{\frac{k}{2}-j}^k(\tau,z).
\end{aligned}
\label{eq:SU2-elliptic}
\end{equation}
Only the first line corresponds to a valid large gauge transformation of $SU(2)$ gauge field, which gives an independent elliptic covariance to the wormhole seed on each boundary for either chirality.
This is because, with $z$ coupled to $J^3_0$, a
shift $z\to z+n\tau+m$ is a single-valued $SU(2)$ gauge transformation only if
$e^{2\pi i nJ^3_0}=e^{2\pi i mJ^3_0}=1$, i.e.\ only if $n,m$ lie in the cocharacter
lattice of $SU(2)$, which is $2\mathbb Z$  as in~\eqref{eq:cocharacter-lattice_SU2}.
 The unit shifts instead implement the center element $e^{2\pi iJ^3_0}=-1$ or produce the central twist sector. In particular, the third line in \eqref{eq:SU2-elliptic} does not identify $j$ with $k/2-j$ in the untwisted $SU(2)$ theory.\footnote{ A quotient by the center changes this statement. For the bosonic $SO(3)_k=SU(2)_k/\mathbb{Z}_2$
 at $k\in4\mathbb Z$, the extended characters of the non-fixed orbits are
$\chi_j^k+\chi_{k/2-j}^k$ with integer $j<k/4$, while the fixed point $j=k/4$ requires additional
resolution. Here we retain the simply connected gauge group $SU(2)_L\times SU(2)_R$.}

As in the $U(1)$ example, the twisted $SU(2)$ seed follows by applying fractional spectral flow separately for each chirality and on each boundary. Using the convention of~\eqref{su(2)_J3_SF}--\eqref{su(2)_Jpm_SF}, the flowed character is
\begin{equation}
\chi_j^{k,\zeta}(\tau,z)
 =e^{2\pi i k(\tau\zeta^2/4+z\zeta/2)}
 \chi_j^k(\tau,-z-\zeta\tau).
 \label{eq:SU2_twist_character}
\end{equation}
Replacing each character in~\eqref{eq:SU2-paired-seed} by its flowed counterpart gives the twisted seed amplitude. Again the twisted seed contains no new information and we focus on the untwisted case below.

A final comment before the discussion of spectral data is that the $\zeta=0$ case of \eqref{eq:SU2_twist_character} gives $\chi_j^k(\tau,-z)=\chi_j^k(\tau,z)$. This sign
flip of the source $z$ is the action of $-\mathbf 1\in SL(2,\mathbb Z)$. Thus $\CRT$ maps each $SU(2)$ representation to itself, and each modular image and its $-\mathbf1$ partner contribute equally, giving a factor of two in the full modular sum. From the boundary point of view, this is a manifestation of the $su(2)$ Weyl symmetry.

\paragraph{Spectral data and the RMT contribution.}
We now proceed to  extract the spectral data  from the full amplitude~\eqref{eq:full-spectral-projection} for $G=SU(2)$. The primary label in this case becomes $\Lambda_i=\mathbf j_i=(j_{i,L},j_{i,R})$ with $i=1,2$, and each spin $j=0,\frac{1}{2},...,\frac{k}{2}$. Removing the affine descendants from the $SU(2)$ character \eqref{su(2)-char} gives
\begin{align}
    \chi_j^{k,P}(\tau,z)=q^{a_j}\chi_j(z),\qquad
 \chi_j(z)=\sum_{m=-j}^j e^{2\pi i zm}
 =\frac{\sin((2j+1)\pi z)}{\sin(\pi z)},
\end{align}
where
\begin{align}
     a_j=h_j-\frac{c^{\mathrm{Sug}}}{24},\qquad
 h_j=\frac{j(j+1)}{k+2},\qquad c^{\mathrm{Sug}}=\frac{3k}{k+2}.
\end{align}
The primary-counting character $\Xi_{\Lambda_i}^P$ in~\eqref{eq:full-spectral-projection}  can therefore be written as
\begin{align}
    \Xi_{\Lambda_i}^P= \prod_{a=L,R}q_{i,a}^{a_{j_{i,a}}}\chi_{j_{i,a}}(z_{i,a})
\end{align}
We thus obtain the complete primary-counting amplitude as
\begin{equation}
\begin{aligned}
 Z_{\rm wormhole}^{P}(X_1;X_2)
 =\frac{|q_1q_2|^{-1/12}}{2\pi^2\sqrt{y_1y_2}}
 \sum_{\mathbf j_1,\mathbf j_2} ~\prod_{i=1}^2\prod_{a=L,R}q_{i,a}^{a_{j_{i,a}}}\chi_{j_{i,a}}(z_{i,a}) \times  \mathcal M^{SU(2)}_{\mathbf j_1\mathbf j_2}(\tau_1,\tau_2),
\end{aligned}
\label{eq:SU2-full-primary-amplitude}
\end{equation}

To extract the spectral data, we first project \eqref{eq:SU2-full-primary-amplitude} onto the spin-$j$ representation of $SU(2)$ at each boundary, for each chirality, using
\begin{equation}
\int_0^1 dz\,2\sin^2(\pi z)\chi_j(z)\chi_{j'}(z)=\delta_{jj'}.
\end{equation}
This projection counts each  spin-$j$ primary multiplet once. To count all states in the multiplet,  the canonical trace at boundary $i$ should be multiplied by a degeneracy factor $D_{\mathbf j_i}=(2j_{i,L}+1)(2j_{i,R}+1)$.  Fourier transforming in $x_i=\operatorname{Re}\tau_i$ to fixed spin $s_i$ sector and defining
\begin{equation}
\begin{aligned}
 \sigma_{\mathbf j}&=h_{j_L}-h_{j_R},\qquad
 \epsilon_{\mathbf j}=h_{j_L}+h_{j_R},\\
 \hat s_i&=s_i-\sigma_{\mathbf j_i},\qquad
 v_k=\frac{c^{\mathrm{Sug}}+1}{12}.
 \label{eq:SU2_parameter}
\end{aligned}
\end{equation}
we obtain
\begin{equation}
\begin{aligned}
\langle Z^P_{s_1,\mathbf j_1}(\beta_1) Z^P_{s_2,\mathbf j_2}(\beta_2)\rangle_c \simeq
\frac{D_{\mathbf j_1}D_{\mathbf j_2}}
 {2\pi^2\sqrt{\beta_1\beta_2}}\,
 e^{-2\pi\sum_i\beta_i(\epsilon_{\mathbf j_i}-v_k)}
 \int_0^1 dx_1\int_0^1 dx_2\,
 e^{-2\pi i(\hat s_1x_1+\hat s_2x_2)}
 \mathcal M^{SU(2)}_{\mathbf j_1\mathbf j_2}(\tau_1,\tau_2).
\end{aligned}
\label{eq:SU2-full-sector-correlator}
\end{equation}
This is the torus-wormhole prediction for the $SU(2)$ spectral correlator in each
sector. 

To extract the ramp contribution, we focus on  the  $c=0$ terms in the modular sum  in $\mathcal M^{SU(2)}_{\mathbf j_1\mathbf j_2}$. The $T$-matrix of $SU(2)$ character gives
\begin{equation}
U^{SU(2)}_{\mathbf j_1\mathbf j_2}(\pm T^n)
 =\delta_{\mathbf j_1\mathbf j_2}\,
 e^{2\pi i n(h_{j_{1,L}}-h_{j_{1,R}})}.
\end{equation}
Note that
the two signs of $\pm T^n$ give identical contributions. This again reflects the fact that   every $SU(2)$ representation is self-conjugate.
Unfolding the $T^n$ sum in~\eqref{eq:SU2-full-sector-correlator} therefore gives
\begin{equation}
 \langle Z^P_{s_1,\mathbf j_1}(\beta_1) Z^P_{s_2,\mathbf j_2}(\beta_2)\rangle_c \simeq\delta_{s_1s_2}\delta_{j_{1,L}j_{2,L}}\delta_{j_{1,R}j_{2,R}}\,
 \frac{D_{\mathbf j_1}^2}{\pi}\,
 \frac{\sqrt{\beta_1\beta_2}}{\beta_1+\beta_2}e^{-\beta_1 E_{s_1,\mathbf j_1}-\beta_2 E_{s_2,\mathbf j_2}}+(c>0),
\label{eq:SU(2)_RMT_result}
\end{equation}
where the $(c>0)$ represents the remaining terms in the modular sum, and $E_{s,\bf{j}}$ is the BTZ threshold energy  at spin $s$ and $SU(2)$ spin $\bf{j}$, given by
\begin{equation}
\begin{aligned}
 \frac{E_{s,\mathbf j}}{2\pi}
 &=|s-h_{j_L}+h_{j_R}|+h_{j_L}+h_{j_R}
   -\frac{1+c^{\mathrm{Sug}}}{12}\\
 &=|s-a_{j_L}+a_{j_R}|+a_{j_L}+a_{j_R}-\frac{1}{12}.
\end{aligned}
\label{eq:SU2-paired-threshold}
\end{equation}
This thus reproduces the diagonal boundary prediction \eqref{eq:su2-diagonal-prediction} of the ramp contribution.

As every pair $(j_L,j_R)$ is self-conjugate, the symmetry class is determined by the antiunitary on the multiplicity space. Assuming $\CRT^2=1$ on the full Hilbert space, $\CRT^2_M$ is determined as in Table \ref{table:SU2_classes}, where sectors with $j_L+j_R\in \mathbb{Z}$ exhibit GOE statistics, while those with $j_L+j_R\in \mathbb{Z}+\frac{1}{2}$ exhibit GSE statistics.  After dividing out the multiplicity factor $D_{\mathbf j}^2$, both classes have the common $1/\pi$ ramp coefficient. The
interpretation is the same as in Subsection~\ref{sec:U1-holography}: the GSE trace counts both members of each Kramers pair, which supplies a factor of four relative to the
coefficient $1/(4\pi)$ obtained by counting each pair once.

\subsection{Modular corrections}
\label{sec:jacobi-corrections}

In this section, we compute the modular corrections to the spectral data of $U(1)$ and $SU(2)$ from  the full canonical wormhole amplitude~\eqref{eq:full-spectral-projection}. This requires us to first write out the modular sum in  $\mathcal M^G_{\Lambda_1\Lambda_2}$ explicitly and then project onto fixed-spin and fixed-representation sectors. The $c>0$ terms  in the modular sum give off-diagonal corrections to the spectral correlator. As in pure gravity, each $c>0$ term is again suppressed by $1/\beta$ at low temperature and $O(1/T)$ at large Lorentzian time.

\paragraph{The modular sum.}
The $SL(2,\mathbb{Z})$ modular sum can be decomposed into a sum over a set of representatives of $PSL(2,\mathbb{Z})$ with $c\geq 0$, plus their  $-\mathbf 1$ images, which implement the $\CRT$.  We denote by $\bar\Lambda=(\bar\lambda_L,\bar\lambda_R)$ the conjugate representation pair under the $-\mathbf 1$ and combine each modular image with its $-\mathbf 1$  partner as
\begin{equation}
V^G_{\Lambda_1\Lambda_2}(\gamma)
 :=U^G_{\Lambda_1\Lambda_2}(\gamma)
   +U^G_{\Lambda_1\bar\Lambda_2}(\gamma).
\label{eq:continuous-central-pair}
\end{equation}
For $U(1)$,  this sums  both charges  $(m_L,m_R)$ and $(-m_L,-m_R)$ on the second boundary. For $SU(2)$, it gives $V^{SU(2)}=2U^{SU(2)}$. 

Now write $\gamma_{cd}=\left(\begin{smallmatrix}a&b\\c&d\end{smallmatrix}\right)$ with integer entries satisfying $ad-bc=1$. The Bruhat decomposition then rewrites ~\eqref{eq:full-modular-coefficient} as
\begin{equation}
\begin{aligned}
 \mathcal M^G_{\Lambda_1\Lambda_2}(\tau_1,\tau_2)
 &=\sum_{n\in\mathbb Z}V^G_{\Lambda_1\Lambda_2}(T^n)
       \mathcal W_{T^n}(\tau_1,\tau_2)\\
 &\quad+\sum_{c>0}\ \sum_{\substack{d\ ({\rm mod}\ c)\\(c,d)=1}}
       \sum_{j,\ell\in\mathbb Z}
       V^G_{\Lambda_1\Lambda_2}(T^j\gamma_{cd}T^\ell)
       \mathcal W_{T^j\gamma_{cd}T^\ell}(\tau_1,\tau_2).
\end{aligned}
\label{PS_decomp}
\end{equation}
The first line contains $c=0$ terms, which have been evaluated in previous $U(1)$ and $SU(2)$ examples and shown to match with diagonal RMT predictions. The second line includes all $c\ne0$ modular corrections and will give rise to off-diagonal spectral correlations. As in pure gravity case~\cite{Cotler:2020ugk}, this sum requires regularization.

\paragraph{Spectral data}
  Let $\alpha_i$ specify the charge or representation at boundary $i$. For $U(1)$, we have $\alpha_i=\mathbf Q_i$, with the extended primary label $\Lambda_i=\mathbf m_i=\mathbf Q_i$ mod $2k$ and the degeneracy factor $D_{\alpha_i}=1$. For $SU(2)$, we have $\alpha_i=\Lambda_i=\mathbf j_i$, with $D_{\alpha_i}=(2j_{i,L}+1)(2j_{i,R}+1)$. Writing $\tau_i=x_i+i\beta_i$, the projection onto fixed $\alpha_i$ and fixed neutral spin $r_i=\hat{s}_i$ at each boundary gives
  \begin{equation}
\begin{aligned}
 \langle Z^P_{s_1,\alpha_1}(\beta_1)Z^P_{s_2,\alpha_2}(\beta_2)\rangle_c
 &\simeq \mathcal P^G_{12}
   \int_0^1 dx_1\int_0^1 dx_2\,
  e(-r_1x_1-r_2x_2)
   \mathcal M^G_{\Lambda_1\Lambda_2}(\tau_1,\tau_2),
\end{aligned}
\label{WH_amp_result}
\end{equation}
with $e(x)=e^{2\pi ix}$ and the prefactor
\begin{align}
    \mathcal P^G_{12}
 &:=\frac{D_{\alpha_1}D_{\alpha_2}}{2\pi^2\sqrt{\beta_1\beta_2}}
     e^{-2\pi\sum_i\beta_i(\epsilon_{\alpha_i}-v_G)},
\end{align}
where $\epsilon_{\alpha_i}$ and $\sigma_{\alpha_i}$ are the sum and difference of the left- and right-moving Sugawara weights, respectively, and
\begin{align}
    v_G=\begin{cases}
  1/6,&G=U(1),\\
  (1+c^{\mathrm{Sug}})/12,&G=SU(2).
 \end{cases}
\label{eq:continuous-neutral-frequencies}
\end{align}
 In the $U(1)$ case, \eqref{WH_amp_result} initially counts primaries of the unextended current algebra. The restriction to extended primaries requires further procedures as described in Subsection~\ref{sec:U1-CFT}.

In both examples we considered, the $T$-matrix is diagonal with $U^G(T)_{\Lambda\Lambda'}=\delta_{\Lambda\Lambda'}e(\sigma_\Lambda)$. The $T^n$-translation factors in~\eqref{PS_decomp} thus obey
\begin{equation}
\begin{aligned}
 V^G_{\Lambda_1\Lambda_2}(T^j\gamma_{cd}T^\ell)
 &=e(j\sigma_{\Lambda_1}+\ell\sigma_{\Lambda_2})
   V^G_{\Lambda_1\Lambda_2}(\gamma_{cd}),\\
 \mathcal W_{T^j\gamma_{cd}T^\ell}(\tau_1,\tau_2)
 &=\mathcal W_{\gamma_{cd}}(\tau_1+j,\tau_2+\ell).
\end{aligned}
\label{eq:continuous-unfolding}
\end{equation}
The gravitational term $ W_{\gamma_{cd}}$ can be written as
\begin{align}
W_{\gamma_{cd}}(\tau_1,\tau_2)=   \frac{\mathrm{Im}(\tau_1) \mathrm{Im}(\tau_2)}{c^2\left|(\tau_1+\frac{a}{c})(\tau_2+\frac{d}{c})-\frac{1}{c^2}\right|^2}. 
\end{align}
Thus, shifting the integration variables $x_1$ and $x_2$ by $a/c$ and $d/c$ decouples $a$ and $d$ dependence in the $ W_{\gamma_{cd}}$ term. The resulting phases combine with the modular matrices into the vector-valued Kloosterman sum
\begin{equation}
\mathcal A^G_{\Lambda_1\Lambda_2}(r_1,r_2;c)
 :=\sum_{\substack{d\ ({\rm mod}\ c)\\(c,d)=1}}
 e\!\left(\frac{r_1a+r_2d}{c}\right)
 V^G_{\Lambda_1\Lambda_2}(\gamma_{cd}).
\label{Kloost_U(1)}
\end{equation}
Since the physical spins $s_i=r_i+\sigma_{\Lambda_i}$ are integers,\footnote{ Note that this also guarantees that the Kloosterman sum is well defined, as changing the modular transformation by $\gamma_{cd}\mapsto T^j\gamma_{cd}T^\ell$ simply multiplies the summand by $e(j(r_1+\sigma_{\Lambda_1})+\ell(r_2+\sigma_{\Lambda_2}))=1$.} further changing variables to $x_1+j$ and $x_2+\ell$ cancels the translation phase in \eqref{eq:continuous-unfolding} and unfolds the two Fourier integrals to $\mathbb{R}^2$.

We finally arrive at
\begin{equation}
\begin{aligned}
 \langle Z^P_{s_1,\alpha_1}(\beta_1)Z^P_{s_2,\alpha_2}(\beta_2)\rangle_c
 &=(c=0\text{ terms})+\mathcal P^G_{12}\sum_{c>0}
 \mathcal A^G_{\Lambda_1\Lambda_2}(r_1,r_2;c)
 \mathcal I_c(r_1,r_2;\beta_1,\beta_2).
\end{aligned}
\label{WH_amp_result_prime}
\end{equation}
The first term is the RMT contribution which for $G=U(1)$ is displayed in~\eqref{ramp_U(1)_bulk}, and for $G=SU(2)$ is displayed in~\eqref{eq:SU(2)_RMT_result}. The $c>0$ sums are modular corrections, with $\mathcal I_c$  the same integral that appears in pure gravity~\cite{Cotler:2020ugk}, now evaluated at the possibly fractional neutral spins.
\begin{equation}
\mathcal I_c(r_1,r_2;\beta_1,\beta_2)
 :=\int_{\mathbb R^2}dx_1\,dx_2\,
 e(-r_1x_1-r_2x_2)
 \frac{\beta_1\beta_2}
 {c^2\left|(x_1+i\beta_1)(x_2+i\beta_2)-c^{-2}\right|^2}\,,
\label{eq:continuous-gravitational-integral}
\end{equation}
Different representation sectors are correlated through the Kloosterman sum, while different spin sectors are correlated through $\mathcal I_c$.

To be slightly more explicit, we now give the Kloosterman sum expressions for $c=1$. For $U(1)$, using $S$-matrix $\rho_k(S)_{mn}=(2k)^{-1/2}e(-mn/(2k))$, the $c=1$ Kloosterman sum is
\begin{equation}
\mathcal A^{U(1)}_{\mathbf m_1\mathbf m_2}(r_1,r_2;1)
 =\frac1k\cos\!\left[\frac\pi k
 (m_{1,L}m_{2,L}-m_{1,R}m_{2,R})\right].
\label{eq:U1-c1-correction}
\end{equation}
For $SU(2)$, we have
\begin{equation}
\mathcal A^{SU(2)}_{\mathbf j_1\mathbf j_2}(r_1,r_2;1)
 =\frac{4}{k+2}\prod_{a=L,R}
 \sin\!\left[\frac{\pi(2j_{1,a}+1)(2j_{2,a}+1)}{k+2}\right].
\label{eq:SU2-c1-correction}
\end{equation}
Both expressions are generally nonzero.

\paragraph{Low temperature limit}

Performing the $x_i$ integral in  $\mathcal I_c$  gives~\cite{Cotler:2020ugk}
\begin{equation}
\begin{aligned}
 \mathcal I_c(r_1,r_2;\beta_1,\beta_2)
 &=\frac{\pi\beta_2}{c^2}e^{-2\pi\beta_1|r_1|}
   \int_{-\infty}^{\infty}\frac{dx}{x^2+B_c^2}\,
   e^{-2\pi i r_2x}
   \exp\!\left[-\frac{2\pi(|r_1|\beta_2+i r_1x)}
   {c^2(x^2+\beta_2^2)}\right],\\
 B_c^2&=\beta_2^2+\frac{\beta_2}{c^2\beta_1}.
\end{aligned}
\label{eq:continuous-one-dimensional-integral}
\end{equation}
The low temperature limit is defined as 
 $\beta_i=\beta b_i$ with $b_i>0$ fixed and $\beta\to\infty$. At fixed $c$ and fixed $r_i$, applying the same contour analysis of~\cite{Cotler:2020ugk} gives
\begin{equation}
\begin{aligned}
 \mathcal I_c(r_1,r_2;\beta_1,\beta_2)
 &=\frac{\pi^2}{c^2}
   e^{-2\pi(\beta_1|r_1|+\beta_2|r_2|)}
   \left[\mathcal J_c(r_1,r_2)+O(\beta^{-1})\right],\\
 \mathcal J_c(r_1,r_2)
 &:=\begin{cases}
 J_0\!\left(\dfrac{4\pi\sqrt{r_1r_2}}{c}\right),&r_1r_2>0,\\
 1,&r_1r_2\le0,
 \end{cases}
\end{aligned}
\label{eq:continuous-low-temperature}
\end{equation}
where $J_0$ is the Bessel function of the first kind. Consequently, each fixed-$c$ correction in~\eqref{WH_amp_result_prime} has the threshold factors $e^{-\sum_i\beta_i E_{s_i,\alpha_i}}$, with $E_{s_i,\alpha_i}=2\pi(|r_i|+\epsilon_{\alpha_i}-v_G)$, and is $O(\beta^{-1})$ suppressed compared to the $c=0$ ramp term in the low temperature limit. This suppression is the same as in pure gravity case.

\section{Discussion}
\label{S:discuss}

In this work we proposed categorical RMT, a random-matrix ansatz for chaotic local quantum systems on a circle with ordinary or non-invertible global symmetry. As summarized in the Introduction, the independent blocks are the irreducible representations of the tube algebra, which acts on the sum of all twisted Hilbert spaces. The symmetry fixes the exact degeneracies and identifications among these blocks, and ordinary RMT describes the rest. For a group this is ordinary symmetry-resolved RMT, applied twisted sector by twisted sector. For a non-invertible symmetry it says more. Hilbert spaces with different twists can share a block, and so share part of their spectrum, although no invertible symmetry relates them. The Fibonacci anyon chain shows exactly this, while the $S_3$ qutrit chain checks the rest of the framework. In a CFT the blocks are the primary spectra at fixed spin in each tube-algebra sector, and modular covariance adds correlations among blocks that categorical RMT would treat as independent, generalizing the correlations between spin sectors found in~\cite{Haehl:2023mhf,DiUbaldo:2023qli}.

We realized the long-range part of this structure holographically, in three-dimensional Einstein gravity coupled to the SymTFT that encodes the boundary symmetry. The relevant bulk amplitude is the torus wormhole of~\cite{Cotler:2020ugk}, decorated by a SymTFT cobordism. Its gravitational part gives the ramp of the ordinary torus wormhole, while the SymTFT determines which twisted Hilbert spaces share a tube-algebra sector. Boundary $\CRT$ rotates the torus by $\pi$, and in the bulk it acts on the SymTFT anyons by charge conjugation. It therefore fixes the self-dual sectors and exchanges the others in pairs with identical spectra. The seed wormhole together with its $\pi$-rotated image gives the categorical-RMT ramp of every sector. The other modular images add the correlations that modular covariance requires, so in a CFT distinct sectors are never exactly independent. The torus wormhole reaches the ramp but not the Heisenberg-time crossover to the plateau. It therefore probes long-range spectral correlations, not the short-range level repulsion that our lattice calculations probe.

The modular images also tell us when categorical RMT applies. They estimate the Thouless time, after which the ramp governs the spectral form factor, and so play the role for the torus wormhole that quasinormal modes play for the double cone~\cite{Chen:2023hra}. Consider primaries at fixed neutral spin $s\neq0$ on a circle of length $L$, in the spectral form factor at inverse temperature $\beta$ and time $T$. For $T\gg\beta$ and $T\gg L$, the images are smaller than the seed ramp by a factor of order $\beta L/T^2$. So when $\beta\lesssim L$ the ramp has taken over by $T\sim L$, the time light takes to cross the circle, and at low temperature, $\beta\gg L$, the images are suppressed by $L/\beta$ at all times. We have not computed the window $\beta\ll T\ll L$, so at high temperature $L$ is only an upper bound on the Thouless time. The estimates above are for pure gravity, but apply in these holographic models where the boundary has global symmetry, as the SymTFT changes only the Kloosterman sums, not the dependence on $\beta$, $L$ and $T$. The approach to the ramp is a power law in $T$, as for the hydrodynamic modes of~\cite{Chen:2023hra}, rather than the exponential decay set by gapped quasinormal modes. In the Cardy random matrix model of~\cite{Ambrosini:2026int}, the Thouless time is instead set by the regulator. It is the inverse energy width of the windows over which modular invariance is imposed, which the authors hold fixed so that the Thouless time is of order the size of the circle.

For continuous symmetries, the Sugawara construction splits the stress tensor into a current part, fixed by representation theory, and a neutral part, which describes the unconstrained spectral data. Spectral flow relates sectors of different charge, and relates each twisted Hilbert space to the untwisted one; for $U(1)_k$ the integer flows also extend the chiral algebra. Categorical RMT then applies to the neutral spectrum at fixed neutral spin, in each module of the extended chiral algebra. The bulk realizes the same split. Chern-Simons theory realizes the current-algebra characters and, through large gauge transformations, spectral flow, while gravity encodes the ramp of the neutral spectrum.

Some of these neutral sectors are symplectic even though the theory is bosonic. We consider theories with holomorphic and antiholomorphic currents. Then for $U(1)_k$ globaly symmetry with $k$ odd, this happens in the modules of charges $(m_L,m_R)=(k,0)$ or $(0,k)$ mod $2k$, and for  $SU(2)_k$ symmetry with $SU(2)$ spins $j_L$ and $j_R$, in the modules with half-integer $j_L+j_R$. On these $\CRT$ squares to $-1$ on the current-algebra part. Since $\CRT^2=+1$ in a bosonic theory, it must square to $-1$ on the neutral multiplicity space as well. The neutral Hamiltonian is then a GSE matrix, and each of its levels is exactly doubly degenerate. The torus wormhole cannot see this. It gives these sectors a ramp with the GOE slope $1/\pi$, which is also the GSE slope when the trace counts each Kramers pair twice. Telling them apart holographically requires going beyond the torus wormhole.

In our analysis of spectral statistics in CFTs with continuous global symmetries, we considered models with $G_L\times G_R$ vector-like symmetry. We did this so that torus wormhole amplitudes in Section~\ref{sec:continuous-holography}, which include a sum over $T$ images, admit a simple spectral interpretation. However it is straightforward to propose a categorical RMT ansatz for 2d CFTs with chiral global symmetry. The ``chiral tube algebras’' of~\cite{Benjamin:2026oqw}, generated by the modes of local and non-local chiral currents, seem useful 
in that case. Our proposal, in this language, reads: in the presence of extended chiral algebras, RMT describes the multiplicity space of each irreducible module of the full chiral tube algebra at fixed spin modulo the spectral flow (spectral flow maps the twisted modules of ~\cite{Benjamin:2026oqw} to untwisted ones, as they show explicitly for $SU(2)_1$). This is indeed consistent with the dictionary in Subsection~\ref{sec:continuous-RMT-statement}.

We close with two directions for future work.

\vspace{-.2cm}
\paragraph{Categorical ETH.}

Spectral statistics provide one diagnostic of quantum chaos. The eigenstate thermalization hypothesis provides another through the matrix elements of simple operators. This suggests a categorical version of ETH in which the tube algebra fixes exact selection rules and relations among matrix elements, while the remaining reduced matrix elements thermalize on the multiplicity spaces.

In forthcoming work~\cite{Jensen:2026cat}, we show that such an ansatz indeed describes chaotic lattice models. The same symmetry-preserving deformations used here to produce random matrix statistics also produce thermalized operator statistics once we resolve the matrix elements into the appropriate categorical sectors.

\vspace{-.2cm}
\paragraph{An effective theory of categorical spectral correlations.}

A nonlinear sigma model associated with causal symmetry breaking~\cite{Altland:2020ccq,Altland:2021rqn} gives an effective field theory explanation of ordinary RMT universality. Can this description be refined to account for systems with spatial locality? Such a theory should incorporate twisted Hilbert spaces and the topological operators that intertwine them from the outset.

For a CFT, such an effective theory must also reproduce the correlations imposed by modular covariance. Spatial locality brings in twisted sectors, and modular transformations relate their torus amplitudes, producing correlations between different spin and tube algebra sectors. These sectors, their intertwiners, and the resulting modular correlations should all appear naturally in an appropriately upgraded version of the sigma model. A successful theory should reproduce both the categorical-RMT seed and its modular completion, and explain why the modular corrections become subleading after the Thouless time. It is an open problem to find such a description.

\subsection*{Acknowledgments}

We would like to thank L.~Bottini, C.~Declamp, J.~Cotler, and G.~Turiaci for valuable conversations. The exact diagonalization numerics in Section~\ref{sec:categorical-RMT} were performed with the assistance of Claude Opus 4.8.  KJ and MR are supported in part by NSERC Subatomic Discovery Grants. CC acknowledges the support of the NSERC CGRS D. The work of SP was supported by the Department of Atomic Energy, Government of India, under Project Identification Number RTI-4012 and the Infosys Endowment for the study of the Quantum Structure of Spacetime. SP would also like to acknowledge his debt to the
people of India for their steady support of the study of basic science.

\appendix

\section{Highest-weight representations of $su(2)_k$ and spectral flow}

In this appendix, we briefly review the highest-weight representations of $su(2)_k$, derive the spectral flow of the $su(2)_k$ Virasoro generators, and show how various states transform under it. 

\subsection{Highest-weight representations}
\label{HWR_su(2)}

 To discuss the representation theory of $su(2)_k$,  it is more convenient to work in the basis
 \begin{align}
     J^3_{\text{new}}=\frac{1}{\sqrt{2}}J^3, \quad J^{\pm}=\frac{1}{\sqrt{2}}(J^1\pm i J^2).
     \label{equ:SU(2)_new_basis}
 \end{align}
In the following we drop the label ``new'' and simply write $J^3_{\text{new}}$ as $J^3$. In this basis, the Kac--Moody algebra reads
\begin{align}
\begin{split}
    &[J^3_m,J^3_n]=\frac{k}{2}m \delta_{m+n,0}\,, \\
    &[J^3_m,J^{\pm}_{n}]=\pm J^{\pm}_{m+n}\,, \\
    &[J^+_m,J^-_n]=km\delta_{m+n,0}+2J^3_{m+n}\,.
\end{split}
\end{align}
The physically relevant representations are highest-weight representations. A primary state $|j,m\rangle$, with spin $j\in \{0, \frac12,1,\dots,\frac k2\}$ and $J^3_0$ eigenvalue $m\in \{-j, -j+1,\dots,j\}$, satisfies
\begin{align} 
\begin{split}
    &J_0^{3} |j,m\rangle=m|j,m\rangle, \\
    & J^+_0|j,m\rangle= 
    \sqrt{(j-m)(j+m+1)}\,|j,m+1\rangle, \quad  J^+_0|j,j\rangle= 0, \\
    & J^-_0|j,m\rangle=  \sqrt{(j+m)(j-m+1)}\,|j,m-1\rangle, \quad  J^-_0|j,-j\rangle= 0, \\
    &J^a_n|j,m\rangle=0 \text{ for $n>0$}, \\
    &L_0|j,m\rangle=\frac{j(j+1)}{k+2}|j,m\rangle.
\end{split}
\end{align}
In order for the descendant states $(J_{-1}^+)^N |j,j\rangle$ to have a non-negative norm, the spin $j$ and the level $k$ must satisfy the integrability bound
\begin{align}
j\leq \frac{k}{2}.
\end{align}

\subsection{Spectral flow of $su(2)_k$ Virasoro generators}
\label{su(2)_vir_SF}

We take $r\geq 0$ and $w>0$. With $J^a$ in the original basis, the Sugawara construction gives
 \begin{align}
 \begin{split}
     L_{r}&=\frac{1}{2(k+2)}\sum_{n=-\infty}^{\infty} :J^a_n J^a_{r-n}: \\
     &=\frac{1}{2(k+2)}\sum_{n=-\infty}^{\infty} \left(:J_n^-J_{r-n}^{+}:+:J_n^+J_{r-n}^{-}:+2:J_n^3J_{r-n}^3 : \right) \\
     &=\frac{1}{2(k+2)}\Bigg{(} \sum_{n=r+1}^{\infty} \left(J_{r-n}^+ J_n^{-}+J_{r-n}^- J_n^+ \right) +\sum_{n=-\infty}^{r-1}\left(J_n^-J_{r-n}^{+} +J_n^+ J_{r-n}^- \right)    \\
     & \quad \quad \quad \quad \quad \quad \quad + J_0^+J_r^-+J_0^-J_r^+ +2\sum_{n=-\infty}^{\infty}\left(:J_{n}^3 J_{r-n}^3:\right) \Bigg{)}
\end{split}
 \end{align}
 Under the spectral flow $\sigma_w$, which is \eqref{su(2)_J3_SF}--\eqref{su(2)_Jpm_SF} at $\zeta=w\in\mathbb{Z}$,
 \begin{align}
     \sigma_w(J_r^3)=w\frac{k}{2}\delta_{r,0}-J_r^3, \quad \sigma_w(J_r^{\pm})=J_{r\pm w}^{\mp},
 \end{align}
the Virasoro generator transforms as
\begin{align}
    \sigma_w(L_{r})&=\frac{1}{2(k+2)}\Bigg{(} \sum_{n=r+1}^{\infty} \left(J_{r-n+w}^- J_{n-w}^{+}+J_{r-n-w}^+ J_{n+w}^- \right) +\sum_{n=-\infty}^{r-1}\left(J_{n-w}^+J_{r-n+w}^{-} +J_{n+w}^- J_{r-n-w}^+ \right)   \nonumber \\
     & \quad \quad \quad \quad  \quad + J_w^-J_{r-w}^+ +J_{-w}^+ J_{r+w}^- +\frac{k^2w^2}{2}\delta_{r,0}-2J_r^3 kw+2\sum_{n=-\infty}^{\infty}\left(:J_{n}^3 J_{r-n}^3:\right) \Bigg{)} \nonumber \\
     &=\frac{1}{2(k+2)} \Bigg{(} \sum_{n=r+w+1}^{\infty}(J_{r-n+w}^- J_{n-w}^{+})+\sum_{n=r+1}^{r+w}(J_{n-w}^{+}J_{r-n+w}^- -k(n-w)\delta_{r,0}-2J_r^3) \nonumber \\
      & \quad \quad \quad \quad \quad  + \sum_{n=r+1}^{\infty}(J_{r-n-w}^+ J_{n+w}^-) + \sum_{n=-\infty}^{r-1}(J_{n-w}^+J_{r-n+w}^{-} ) \nonumber \\
      &\quad \quad \quad \quad \quad +\sum_{n=-\infty}^{r-w-1}(J_{n+w}^- J_{r-n-w}^+)+\sum_{n=r-w}^{r-1}(J_{r-n-w}^+J_{n+w}^- -k(r-n-w)\delta_{r,0}-2J_r^3) \nonumber \\
       &\quad \quad \quad \quad \quad + (J_{r-w}^+J_w^- -k(r-w)\delta_{r,0}-2J_r^3) +J_{-w}^+ J_{r+w}^- \nonumber \\
        &\quad \quad \quad \quad \quad +\frac{k^2w^2}{2}\delta_{r,0}-2J_r^3 kw+2\sum_{n=-\infty}^{\infty}\left(:J_{n}^3 J_{r-n}^3:\right) \Bigg{)}
\end{align}
By some relabeling of the sum, we have 
\begin{align}
    \sigma_w(L_{r})&=\frac{1}{2(k+2)}\Bigg{(} \sum_{n=r+1}^{\infty}(J_{r-n}^- J_n^+) +\sum_{n=-\infty}^{r-1}(J_n^- J_{r-n}^+) +\sum_{n=-\infty}^{r}(J_n^+ J_{r-n}^-) +\sum_{n=r}^{\infty}(J_{r-n}^+J_{n}^-) \nonumber \\
    &\quad \quad \quad \quad \quad - \sum_{n=r+1}^{r+w}\left(k(n-w)\delta_{r,0}\right)
-\sum_{n=r-w}^{r-1}(k(r-n-w)\delta_{r,0})  \nonumber \\
&\quad \quad \quad \quad \quad-4wJ_r^3 - k(r-w)\delta_{r,0}-2J_r^3 +\frac{k^2w^2}{2}\delta_{r,0}-2J_r^3 kw +2\sum_{n=-\infty}^{\infty}\left(:J_{n}^3 J_{r-n}^3:\right) \Bigg{)}  \nonumber \\
&=\frac{1}{2(k+2)}\Bigg{(}\sum_{n=r+1}^{\infty} \left(J_{r-n}^+ J_n^{-}+J_{r-n}^- J_n^+ \right) +\sum_{n=-\infty}^{r-1}\left(J_n^-J_{r-n}^{+} +J_n^+ J_{r-n}^- \right) 
  \nonumber \\
&\quad \quad \quad \quad \quad +( J_{0}^- J_r^++2J_r^3)+J_0^+ J_r^- +2\sum_{n=-\infty}^{\infty}\left(:J_{n}^3 J_{r-n}^3:\right)  \nonumber \\
&\quad \quad \quad \quad \quad+\frac{kw^2}{2}(k+2)\delta_{r,0} -2wJ_r^3 (k+2) -2J_r^3\Bigg{)}\nonumber \\
&=L_{r}-wJ_r^3+\frac{kw^2}{4}\delta_{r,0}.
\end{align}
This verifies the transformation rule for $L_r$ with $r\geq0$. Since $\sigma_w$ commutes with Hermitian conjugation, the case $r<0$ follows from $L_{-r}=L_r^\dagger$, and the case $w<0$ follows by the same steps.

\subsection{Spectral flow of various states}
\label{SF_states}

For a primary state $|j,m\rangle$ in a highest-weight representation of $su(2)_k$, we denote its spectral-flowed image by $[|j,m\rangle]^w$, where $w\in\mathbb{Z}$ is the spectral flow parameter. The $su(2)$ generators act on this state as
\begin{align}
    J_n^a [|j,m\rangle]^w =[\sigma_w(J_n^a)|j,m\rangle]^w,
\end{align}
where $\sigma_w$ is the automorphism \eqref{su(2)_J3_SF}--\eqref{su(2)_Jpm_SF} at $\zeta=w$. We first consider the case $w=1$. For $[|j,m\rangle]^1$ to be primary, it must be annihilated by all positive modes of $J^a$. It is obvious that $J_{n>0}^3$ and $J_{n>0}^+$ annihilate $[|j,m\rangle]^1$. Requiring $J_{1}^{-}$ to annihilate $[|j,m\rangle]^1$ gives
\begin{align}
    J_{1}^{-}[|j,m\rangle]^1 =[J_0^+|j,m\rangle]^1=0,
\end{align}
which holds only when $m=j$. Therefore, $[|j,m\rangle]^1$ is primary only for $m=j$.

The primary state $[|j,j\rangle]^1$ has Cartan charge $\frac{k}{2}-j$ and is itself a highest-weight state, since
\begin{align}
    J_0^3[|j,j\rangle]^1 = \Big(\frac{k}{2}-j\Big) [|j,j\rangle]^1, \quad J_0^{+} [|j,j\rangle]^1=0.
\end{align}
$J_0^-$ acts on $[|j,j\rangle]^1$ as the usual lowering operator,
\begin{align}
    J_0^3 J_0^- [|j,j\rangle]^1 =\Big[\Big(\frac{k}{2}-J_0^3\Big)J_{-1}^+ |j,j\rangle\Big]^1=\Big(\frac{k}{2}-j-1\Big)J_0^- [|j,j\rangle]^1.
    \label{su(2)_SF_des_prim}
\end{align}
Its conformal weight is $\frac{j(j+1)}{k+2}-j+\frac{k}{4}=\frac{(\frac k2-j)(\frac k2-j+1)}{k+2}$, the weight of spin $\frac k2-j$. We therefore identify
\begin{align}
     [|j,j\rangle]^1=\Big|\frac{k}{2}-j,\frac{k}{2}-j\Big\rangle.
\end{align}
Applying spectral flow once more, we obtain $[|\frac{k}{2}-j,\frac{k}{2}-j\rangle]^1=|j,j\rangle$. Thus, spectral flow acts as a $\mathbb{Z}_2$ transformation exchanging highest-weight primary states.

For general states, the action of spectral flow is much less constrained. For example, the primary state $|j,m\neq j\rangle$ is mapped to a descendant state, since $[|j,m\neq j\rangle]^1$ is not annihilated by $J_1^-$. The descendant state $J_{-1}^+|j,j\rangle$ is mapped to the primary state $|\frac{k}{2}-j,\frac{k}{2}-j-1\rangle$, as shown in \eqref{su(2)_SF_des_prim}. Finally, the descendant $J_{-1}^{-}|j,j\rangle$ is mapped to another descendant state, since $[J_{-1}^{-}|j,j\rangle]^1$ is not annihilated by $J_{2}^-$: indeed $J_2^-[J_{-1}^{-}|j,j\rangle]^1=[J_1^+J_{-1}^-|j,j\rangle]^1=(k+2j)[|j,j\rangle]^1$.

\section{Vector-valued modular functions}
\label{app:modular-sums}

The modular completions in Sections~\ref{S:holoCat} and~\ref{sec:continuous-holography}
rely on the same unfolding technique to extract spectral data from vector-valued
Poincar\'e series. In this appendix, we review the general construction of a
vector-valued modular function as a Poincar\'e series and derive its Fourier
decomposition. The resulting expression for the Fourier coefficients is a vector-valued
analogue of the pre-Kuznetsov formula~\cite{Iwaniec2002SpectralMO}.

We use the following notation:
\begin{itemize}
  \item $e(x)=e^{2\pi ix}$.
  \item $\tau=x+iy\in\mathbb H$, where $\mathbb H$ is the upper half-plane.
  \item $\Gamma=SL(2,\mathbb Z)$, and $\Gamma_\infty=\{T^n:n\in\mathbb Z\}$ is the
  translation subgroup generated by
  $T=\left(\begin{smallmatrix}1&1\\0&1\end{smallmatrix}\right)$. We decompose
  \begin{equation}
    \Gamma=\Gamma^+\sqcup(-\mathbf 1)\Gamma^+,\qquad
    \Gamma^+=\left\{\begin{pmatrix}a&b\\c&d\end{pmatrix}\in\Gamma:
    c>0\ \text{or}\ (c=0,\ d=1)\right\},
  \end{equation}
  so that $\Gamma^+$ is a set of representatives of
  $PSL(2,\mathbb Z)=\Gamma/\{\pm\mathbf 1\}$.
  \item Repeated indices are not summed over unless an explicit $\sum$ is written.
\end{itemize}

\subsection{Vector-valued modular functions and Poincar\'e series}
\label{app:VVMF}

Let $\varrho:\Gamma\to U(p)$ be a $p$-dimensional unitary representation. We define the
weight-zero slash operator $|_\varrho\gamma$, acting on vector-valued functions
$\vec Z:\mathbb H\to\mathbb C^p$, by
\begin{equation}
  (\vec Z|_\varrho\gamma)(\tau):=\varrho(\gamma)^{-1}\vec Z(\gamma\cdot\tau).
\end{equation}
A vector-valued modular function is a function $\vec Z$ that is invariant under this
slash operator for all $\gamma\in\Gamma$,
\begin{equation}
  \vec Z|_\varrho\gamma=\vec Z
  \ \Longleftrightarrow\
  \vec Z(\gamma\cdot\tau)=\varrho(\gamma)\vec Z(\tau).
\end{equation}
We do not assume holomorphy. In the applications here, $\varrho(T)$ has finite order, so we work in a basis in which it is diagonal,
\begin{equation}
  \varrho(T)^\mu{}_\nu=\delta^\mu{}_\nu\,e(s_\mu),
  \qquad s_\mu\in\mathbb Q,\quad 0\le s_\mu<1,
\end{equation}
so that the $\mu$-th component of $\vec Z$ has Fourier frequencies
$r_\mu\in s_\mu+\mathbb Z$.

\paragraph{Poincar\'e series.}
Write $r_\nu^m=s_\nu+m$ with $m\in\mathbb Z$, and let $E_{r_\nu^m}:\mathbb H\to\mathbb C$
be a scalar seed function of the form
\begin{equation}
  E_{r_\nu^m}(x+iy)=e(r_\nu^m x)\,E_{r_\nu^m}(iy).
\end{equation}
The associated Poincar\'e series is
\begin{equation}
  \vec P_m^\nu(\tau,E)
  =\sum_{\gamma\in\Gamma_\infty\backslash\Gamma}
  E_{r_\nu^m}(\gamma\cdot\tau)\,\varrho(\gamma)^{-1}\vec e_\nu,
  \label{eq:poincare_series_vec}
\end{equation}
where $\vec e_\nu$ is the $\nu$-th standard basis vector of $\mathbb C^p$. The summand
depends only on the coset $\Gamma_\infty\gamma$, since replacing $\gamma\to T\gamma$
multiplies it by $e(r_\nu^m-s_\nu)=1$. Moreover, $\vec P_m^\nu$ is a vector-valued
modular function. To see this, set $\gamma'=\gamma M$, and we have 
\begin{equation}
\begin{aligned}
  \vec P_m^\nu(M\cdot\tau,E)
  &=\sum_{\gamma'\in\Gamma_\infty\backslash\Gamma}
  E_{r_\nu^m}(\gamma'\cdot\tau)\,
  \varrho(\gamma'M^{-1})^{-1}\vec e_\nu\\
  &=\varrho(M)\,\vec P_m^\nu(\tau,E).
\end{aligned}
\label{eq:PS_VMF}
\end{equation}
The sum converges absolutely if $|E_{r_\nu^m}(\tau)|=\mathcal O(y^{1+\delta})$ as $y\to0$
for some $\delta>0$, which we will assume in the following discussion. In pure gravity, the wormhole seed does not satisfy this criterion. So applying our result there requires regularization and continuation of the gravity seed~\cite{Cotler:2020ugk}. 

\subsection{Fourier coefficients and the pre-Kuznetsov formula}

We now compute the Fourier coefficients of the Poincar\'e
series~\eqref{eq:poincare_series_vec}. Let $P_m^{+,\mu\nu}(\tau,E)$ denote the $\mu$-th
component of $\vec P_m^\nu(\tau,E)$ with the sum restricted to
$\Gamma_\infty\backslash\Gamma^+$. Its $n$-th Fourier coefficient is
\begin{equation}
  P_{nm}^{+,\mu\nu}(iy,E)
  :=\int_0^1dx\,e(-r_\mu^n x)\,P_m^{+,\mu\nu}(x+iy,E),
  \qquad r_\mu^n=s_\mu+n.
\end{equation}
To evaluate it, we use the Bruhat decomposition
\begin{equation}
  \Gamma^+
  =\Gamma_\infty\ \sqcup\
  \bigsqcup_{c>0}\ \bigsqcup_{\substack{d\ ({\rm mod}\ c)\\(c,d)=1}}
  \Gamma_\infty\gamma_{cd}\Gamma_\infty,
  \label{eq:Bruhat_decomp}
\end{equation}
where $\gamma_{cd}=\left(\begin{smallmatrix}a&b\\c&d\end{smallmatrix}\right)$ is any
fixed element of $\Gamma^+$ with bottom row $(c,d)$. The cosets of $\Gamma_\infty$ in
$\Gamma_\infty\gamma_{cd}\Gamma_\infty$ are represented by $\gamma_{cd}T^\ell$,
$\ell\in\mathbb Z$, which act on modular parameter as
\begin{equation}
  \gamma_{cd}T^\ell\cdot\tau=\frac ac-\frac{1}{c^2(\tau+\ell+d/c)}.
\end{equation}
Unfolding the sum over $\ell$ into the $x$-integral, which extends it to
$x\in\mathbb R$, and then shifting $x\to x-d/c$, we obtain
\begin{equation}
\begin{aligned}
  P_{nm}^{+,\mu\nu}(iy,E)
  &=\delta^{\mu\nu}\delta_{nm}E_{r_\nu^m}(iy)
    +\sum_{c>0}A^{\mu\nu}(r_\mu^n,r_\nu^m;c)\,
      \mathcal B_c(r_\mu^n,r_\nu^m;y),\\
  \mathcal B_c(r_\mu^n,r_\nu^m;y)
  &:=\int_{\mathbb R}dx\,e(-r_\mu^n x)\,
  E_{r_\nu^m}\!\left(-\frac1{c^2(x+iy)}\right),
\end{aligned}
\label{eq:PS_Fourier}
\end{equation}
where $A^{\mu\nu}$ is the vector-valued Kloosterman sum used
in~\eqref{eq:Kloost_VMF_sec3},
\begin{equation}
  A^{\mu\nu}(r_\mu^n,r_\nu^m;c)
  :=\sum_{\substack{d\ ({\rm mod}\ c)\\(c,d)=1}}
  e\!\left(\frac{r_\mu^n d+r_\nu^m a}{c}\right)
  \bigl(\varrho(\gamma_{cd})^{-1}\bigr)^{\mu\nu}.
  \label{eq:Kloost_VMF}
\end{equation}
The representative $\gamma_{cd}$ is defined only up to
$\gamma_{cd}\to T^j\gamma_{cd}T^\ell$. This multiplies the summand
in~\eqref{eq:Kloost_VMF} by $e\bigl(j(r_\nu^m-s_\nu)+\ell(r_\mu^n-s_\mu)\bigr)=1$, so
$A^{\mu\nu}$ is well defined.

The Fourier coefficients of the full sum over $\Gamma_\infty\backslash\Gamma$ are
obtained by adding the images under $-\mathbf 1$. Let $C=\varrho(-\mathbf 1)$. Since
$-\mathbf 1$ is central and squares to $\mathbf 1$, $C$ commutes with every
$\varrho(\gamma)$ and $C^2=\mathbf 1$. Hence $\varrho(-\gamma)^{-1}=C\varrho(\gamma)^{-1}$,
and the full coefficient is
\begin{equation}
\begin{aligned}
  P_{nm}^{\mu\nu}(iy,E)
  &=\sum_\kappa(\delta^{\mu\kappa}+C^{\mu\kappa})\,
    P_{nm}^{+,\kappa\nu}(iy,E)\\
  &=(\delta^{\mu\nu}+C^{\mu\nu})\,\delta_{nm}E_{r_\nu^m}(iy)\\
  &\quad+\sum_{c>0}\Bigl[
  A^{\mu\nu}(r_\mu^n,r_\nu^m;c)
  +\sum_\kappa C^{\mu\kappa}A^{\kappa\nu}(r_\mu^n,r_\nu^m;c)
  \Bigr]\mathcal B_c(r_\mu^n,r_\nu^m;y).
\end{aligned}
\label{equ:vec_modular_correction_SL}
\end{equation}
Because $C$ commutes with $\varrho(T)$, $C^{\mu\kappa}$ vanishes unless
$s_\kappa=s_\mu$, so the $C$-dependent terms indeed carry frequency $r_\mu^n$. In the
applications of Sections~\ref{S:holoCat} and~\ref{sec:continuous-holography}, $C$ is
charge conjugation, so the $-\mathbf 1$ images supply the contribution of the conjugate
sector and double that of a self-conjugate sector.
Equation~\eqref{equ:vec_modular_correction_SL} is the Kloosterman-sum side of a
vector-valued analogue of the pre-Kuznetsov formula~\cite{Iwaniec2002SpectralMO}, the
precursor of the Kuznetsov trace formula.

\section{Punctured tori}

For completeness it is interesting to discuss wormholes connecting two punctured physical tori, corresponding to the correlations in the torus one-point function of operators. The construction of the SymTFT replaces the cobordism discussed thus far 
with a genuinely ribbon-decorated cobordism. If a bulk ribbon labeled by $x$ has an endpoint on each asymptotic torus, then the boundary surfaces are marked tori, and the TQFT assigns a map

\begin{equation}
K_{M,x}:\cH_{\mathrm{top}}(T^2;x)_{\mathrm{in}}
\longrightarrow
\cH_{\mathrm{top}}(T^2;x)_{\mathrm{out}},
\label{eq:punctured-kernel}
\end{equation}
for self-dual $x$. More generally, the incoming and outgoing puncture labels are related by duality according to the boundary orientations. Here $\cH_{\mathrm{top}}(T^2;x)$ is the state space of a torus with one marked point labeled by $x$. In a modular tensor category $\mathcal B$ it admits the standard decomposition
\begin{equation}
\cH_{\mathrm{top}}(T^2;x)
\simeq
\bigoplus_{a\in\Irr(\mathcal B)}
\Hom_{\mathcal B}(a,x\otimes a),
\label{eq:punctured-torus-space}
\end{equation}
where $\mathcal B=Z(\cC)$ is the modular tensor category of bulk lines. A basis vector is prepared by an auxiliary solid torus whose boundary carries the $x$-puncture and whose interior contains an admissible ribbon graph: an $a$-labeled core ribbon joined to the boundary $x$-leg by a morphism $\alpha\in\Hom_{\mathcal B}(a,x\otimes a)$. Thus an open ribbon connecting the two asymptotic boundaries does not define an operator on $\cH_{\mathrm{top}}(T^2)$. Instead, it changes the state spaces to the punctured spaces in \eqref{eq:punctured-kernel}.

For the product cobordism $T^2\times I$ with a straight, untwisted $x$-ribbon joining the marked points, functoriality gives
\begin{equation}
K_{T^2\times I,x}=\mathbf 1_{\cH_{\mathrm{top}}(T^2;x)}.
\label{eq:punctured-cylinder-identity}
\end{equation}

This computes a correlator of torus one-point functions,
\begin{equation}
\left\langle Z_a^{(x,\alpha)}(\tau_1)
Z_b^{(x^\vee,\beta)}(\tau_2)\right\rangle,
\label{eq:puncturedcorr}
\end{equation}
not the ordinary \(\langle Z_aZ_b\rangle\).

We can also have a closed bulk ribbon that does not puncture the asymptotic tori. In general, the dictionary is given by the following table:
\begin{center}
\renewcommand{\arraystretch}{1.35}
\begin{tabular}{p{0.29\textwidth}p{0.31\textwidth}p{0.31\textwidth}}
\toprule
Bulk/topological input & State spaces & Boundary observable \\
\midrule
No ribbon endpoints & \(\cH(T^2)\to\cH(T^2)\) & \(\langle Z_aZ_b\rangle\) \\
Closed bulk ribbon \(L_x\) & \(\cH(T^2)\to\cH(T^2)\) & \(\langle Z_aZ_bW_x(L_x)\rangle\) \\
Line wrapping Euclidean time & defect Hilbert space \(\cH_a\) & \(Z_a=\operatorname{Tr}_{\cH_a}(q^{L_0-c/24}\bar q^{\bar L_0-\bar c/24})\) \\
Line wrapping the spatial cycle & an endomorphism component \(D_x^{(a)}:\cH_a\to\cH_a\) & \(Z_{a;x}=\operatorname{Tr}_{\cH_a}(D_x^{(a)}q^{L_0-c/24}\bar q^{\bar L_0-\bar c/24})\) \\
One endpoint on each torus & \(\cH(T^2;x)\to\cH(T^2;x)\) & punctured-torus correlator \(\langle Z_a^{(x,\alpha)}Z_b^{(x^\vee,\beta)}\rangle\) \\
\bottomrule
\end{tabular}
\end{center}

A key point is that the tube-sector label \(\nu\), the boundary twist label \(a\), and a literal inserted ribbon label \(x\) are different data. A nonzero contribution from sector \(\nu\) to \(\langle Z_aZ_b\rangle\) does not imply that a \(\nu\)-ribbon stretches between the two asymptotic boundaries.

\bibliography{sample}
\bibliographystyle{JHEP}
\end{document}